\pdfoutput=1

\documentclass[11pt,twoside,a4paper]{book}  

\usepackage{afterpage} 
\usepackage{graphicx}
\usepackage{subfig}
\usepackage{color}
\usepackage{setspace}
\usepackage[czech,english]{babel}
\usepackage[latin2]{inputenc}

\usepackage{bbold}
\usepackage{amsmath}
\usepackage{amssymb}
\usepackage{amsthm}
\input{Qcircuit.tex}

\usepackage{a4wide}
\def\ifndef#1{\expandafter\ifx\csname#1\endcsname\relax}

\newcommand\cobrx[2]
 {\begin{center}
  \ifndef{pdfpagewidth} 
    \includegraphics*[width=#1]{#2.eps}\else
    \includegraphics*[width=#1]{#2.pdf}
   \fi
  \end{center}}

\newcommand\coverpagestarts
{\pagenumbering{roman}\thispagestyle{empty}
\begin{center}
\large\sffamily
\University\\
\Faculty\\
\Department\\
\vglue 10mm
\cobrx{50mm}{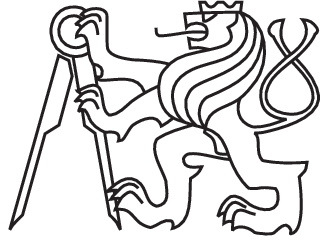}
\vglue 30mm
{\large\bf \DissTitle}\\
\bigskip
{\normalsize\rmfamily by}\\
\bigskip
{\large\it \FirstandFamilyName}\\
\vfill
{\normalsize\rmfamily
{A doctoral thesis submitted to}\\
\smallskip
{the {\Faculty}, {\University},}\\
{in partial fulfilment of the requirements for the degree of Doctor.}}\\
\vglue 20mm
{Ph.D. Programme: \PhDProgram}\\
{Branch of study: \PhDBranch}\\
\bigskip
\Month\ \Year
\end{center}
\newpage
\rule{0pt}{0pt}
\thispagestyle{empty}
\newpage
\rule{0pt}{0pt}
\vfill
\begin{description}
\item[Thesis Supervisor:]\ \\
\Supervisor\\
\SupervisorAffiliation
\end{description}
\ifndef{CoSupervisor}\relax\else\bigskip
\begin{description}
\item[Thesis Co-Supervisor:]\ \\
\CoSupervisor\\
\CoSupervisorAffiliation\\
\end{description}
\vglue 1cm
\fi
Copyright {\copyright} {\Year} by {\FirstandFamilyName}
}

\newcommand\mainbodystarts{\pagestyle{headings}\pagenumbering{arabic}\setcounter{page}{1}}

\theoremstyle{plain}
\newtheorem{theorem}{Theorem}[section]
\newtheorem{lemma}[theorem]{Lemma}
\newtheorem{proposition}[theorem]{Proposition}
\newtheorem*{corollary}{Corollary}
\newtheorem*{Church-Turing-thesis}{Church-Turing hypothesis}
\newtheorem{postulate}{Postulate}
\theoremstyle{definition}
\newtheorem{definition}{Definition}[section]
\newtheorem{example}{Example}

\theoremstyle{remark}
\newtheorem*{note}{Note}

\newcommand{\bcen}{\begin{center}}
\newcommand{\ecen}{\end{center}}
\newcommand{\blem}{\begin{lemma}\sl}
\newcommand{\elem}{\end{lemma}\rm}
\newcommand{\bnote}{\begin{note}\rm}
\newcommand{\enote}{\end{note}}
\newcommand{\bcor}{\begin{corollary}\sl}
\newcommand{\ecor}{\end{corollary}\rm}
\newcommand{\bdefi}{\begin{definition}\rm}
\newcommand{\edefi}{\end{definition}}
\newcommand{\btheo}{\begin{theorem}\sl}
\newcommand{\etheo}{\end{theorem}\rm}
\newcommand{\bprop}{\begin{proposition}\sl}
\newcommand{\eprop}{\end{proposition}\rm}
\newcommand{\bexam}{\begin{example}\rm}
\newcommand{\eexam}{\end{example}}
\newcommand{\bfig}{\begin{figure}\begin{center}}
\newcommand{\efig}{\end{center}\end{figure}}
\newcommand{\btab}{\begin{table}\begin{center}}
\newcommand{\etab}{\end{center}\end{table}}
\newcommand{\benum}{\begin{enumerate}}
\newcommand{\eenum}{\end{enumerate}}
\newcommand{\bitem}{\begin{itemize}}
\newcommand{\eitem}{\end{itemize}}
\newcommand{\bflushr}{\begin{flushright}}
\newcommand{\eflushr}{\end{flushright}}

\def\sq{\hfill\hbox{\rlap{$\sqcap$}$\sqcup$}\vskip \bigskipamount}

\def\sc#1{Section~\ref{#1}}

\newcommand{\abs}[1]{\ensuremath{\left\lvert#1\right\rvert}}
\newcommand{\norm}[1]{\ensuremath{\left\lVert#1\right\rVert}}
\newcommand{\braket}[2]{\ensuremath{\langle#1\vert#2\rangle}}
\newcommand{\ketbra}[2]{\ensuremath{\rvert#1\rangle\!\langle#2\rvert}}
\renewcommand{\bra}[1]{\ensuremath{\langle#1\rvert}} 
\renewcommand{\ket}[1]{\ensuremath{\lvert#1\rangle}} 

\DeclareMathOperator{\tr}{tr}	

\DeclareMathOperator{\diag}{diag}
\DeclareMathOperator{\betainc}{betainc}

\newcommand{\ap}{\text{'}}
\newcommand{\nicebox}[1]{\begin{center}\framebox{\parbox{0.94\textwidth}{#1}}\end{center}}

\newcommand\DissTitle{Quantum computing, phase estimation and applications}
\newcommand\FirstandFamilyName{Miroslav Dob\v{s}\'i\v{c}ek}

\newcommand\Month{January}
\newcommand\Year{2008}
\newcommand\Supervisor{Josef Kol\'a\v{r}}
\newcommand\SupervisorAffiliation{
Department of Computer Science and Engineering\\
Faculty of Electrical Engineering\\
Czech Technical University in Prague\\
Karlovo n\'{a}m\v{e}st\'{\i}~13\\
121 35 Prague 2\\
Czech Republic}
\newcommand\CoSupervisor{R\'obert L\'orencz}
\newcommand\CoSupervisorAffiliation{
Department of Computer Science and Engineering\\
Faculty of Electrical Engineering\\
Czech Technical University in Prague\\
Karlovo n\'{a}m\v{e}st\'{\i}~13\\
121 35 Prague 2\\
Czech Republic}

\newcommand\PhDProgram{Electrical Engineering and Information Technology}
\newcommand\PhDBranch{Information Science and Computer Engineering}
\newcommand\Department{Department of Computer Science and Engineering}
\newcommand\Faculty{Faculty of Electrical Engineering}
\newcommand\University{Czech Technical University in Prague}

\begin{document}

\coverpagestarts

\newpage

\chapter*{Abstract}


Recently, the field of unconventional computing has witnessed
a huge research effort to solve the problem of the assumed
power of computers operating purely according to the laws of
quantum physics. Quantum computing can be seen as a special
intermediate case between digital and real analog computing.
Importantly, there is a threshold theorem for error correction,
as opposed to the pure analog case. Alternatively, quantum
computing can be viewed as generalized probabilistic computing,
where non-negative real probabilities are replaced with complex
amplitudes. The main new resources are quantum mechanical
phenomena such as state superposition, interference and
entanglement. Superposition together with interference provide
a special kind of parallelism, while entanglement, especially
when spatially shared, supports unique means of communication.

One of the most important theoretical result is a proof by
Bernstein and Vazirani (1993) that there is an oracle
relative to which there is a language that can be efficiently
accepted by a quantum Turing machine, but cannot be efficiently
accepted by a bounded-error probabilistic Turing machine. The
problem which was considered is called Recursive Fourier Sampling
and the proposed quantum algorithm gives a quasipolynomial
speedup, $O(n)$ versus $O(n^{\log{n}})$. Next, Abrams and Lloyd
showed that a quantum computer can efficiently simulate many-body
quantum systems having a local Hamiltonian. An additional 
(informal) evidence of the assumed power of a quantum computer
is a bounded-error quantum polynomial time algorithm for large
integer factoring.
The Abrams-Lloyd algorithm is potentially the most useful
quantum algorithm known so far, and if a quantum computer is ever
built, it will revolutionize quantum chemical calculations. Thus
there is a growing consensus regarding investments into the
experimental quantum computing. 

In this thesis, attention is paid to small experimental testbed
applications with respect to the quantum phase estimation algorithm,
the core approach for finding energy eigenvalues. An iterative 
scheme for quantum phase estimation (IPEA) is derived from the 
Kitaev phase estimation, a study of robustness of the IPEA utilized
as a few-qubit testbed application is performed, and an improved 
protocol for phase reference alignment is presented. 
Additionally, a short overview of quantum cryptography is given,
with a particular focus on quantum steganography and authentication.

The work was supported by SSF Nanodev Consortium, by the European
Commission through the IST-015708 EuroSQIP project and by the Czech
Technical University through the grant CTU0507213.


\noindent{\bf Keywords:}
~quantum computing, quantum phase estimation, iterative quantum
protocols, quantum cryptography, quantum steganography




\newpage

\begin{center}\end{center}
\newcounter{MyPublRefCount}
\setcounter{MyPublRefCount}{0}
\def\bibauthor#1{\addtocounter{MyPublRefCount}{1}\bibitem[A.\theMyPublRefCount]{#1}}

\itemsep=0pt

\chapter*{Publications of the author}

Parts of the thesis have been presented in the following
journal papers and presentations:

\section*{Journal papers}

\bitem
\bibauthor{DobsicekPEA07}
Miroslav Dobšíček,
G\"{o}ran Johansson,
Vitaly Shumeiko,
and G\"{o}ran Wendin.
\newblock Arbitrary accuracy iterative quantum phase
          estimation algorithm using a single ancillary
	  qubit: A two-qubit benchmark.
\newblock {\em Physical Review A}, 76:030306(R), 2007.
\eitem

\bitem
\bibauthor{DobsicekAuth07}
Miroslav Dobšíček.
\newblock Simulation on Quantum Authentication.
\newblock {\em Physics of Particles and Nuclei Letters},
          4 (2):158-161, 2007.
\eitem

\section*{Presentation list}

\bitem
\bibauthor{Johansson-Dobsicek-barcelona}
G\"{o}ran Johansson, Lars Tornberg, Miroslav Dobšíček,
Margareta Wallquist, Vitaly Shumeiko, and G\"{o}ran Wendin.
\newblock Superconducting Qubits and Few Qubit Experiments.
Invited talk.
\newblock In {\em International Conference on Quantum
Information Processing and Communication}, 15-19th of
October 2007, Barcelona, Spain, 2007.
\eitem

\bitem
\bibauthor{Dobsicek-Johansson-london}
Miroslav Dobšíček, G\"{o}ran Johansson, Vitaly Shumeiko,
and  G\"{o}ran Wendin.
\newblock Arbitrary accuracy iterative phase estimation
algorithm as a two qubit benchmark. Contributed poster.
\newblock In {\em 7th QIPC Cluster Meeting}, 13-14th of
October 2006, London, UK, 2006.
\eitem

\bitem
\bibauthor{Dobsicek-ctu-06}
Miroslav Dobšíček, Josef Kolář, and Róbert Lórencz.
\newblock A theoretic-framework for quantum steganography.
Contributed poster.
\newblock In {\em CTU Workshop 2006}, Prague, Czech
Republic, 2006.
\eitem

\bitem
\bibauthor{Dobsicek-cryptofest-05}
Miroslav Dobšíček.
\newblock Commercial products for quantum cryptography
(in Czech). Contributed talk.
\newblock In {\em Cryptofest 2005}, Prague, Czech Republic,
2005.
\eitem

\section*{Other publications}

\bitem
\bibauthor{Dobsicek-stego}
Miroslav Dobšíček.
\newblock Extended steganographic system. Contributed poster.
\newblock In {\em 8th International Student Conference on
Electrical Engineering}, Prague, Czech Republic, 2004.
\eitem

\bitem
\bibauthor{Dobsicek-linux}
Miroslav Dobšíček and Radim Ballner.
\newblock Linux - security and exploits (in Czech). Book.
\newblock Kopp Publishing, České Budějovice, 2004.
\eitem

\section*{Citations}

\bitem
\item Paper \cite{DobsicekPEA07} has been cited in:
\benum
\item Ignacio Garc\'{i}a-Mata and Dima L. Shepelyansky.
\newblock Quantum phase estimation algorithm in presence of
static imperfections.
\newblock Technical report, arXiv:0711.1756, submitted to
the European Physical Journal, 2007.

\item Liu Xiu-Mei, Luo Jun, and Sun Xian-Ping.
\newblock Experimental realization of arbitrary accuracy
iterative phase estimation algorithms on ensemble quantum computers.
\newblock {\em Chinese Physics Letters}, 24:3316--3319, 2007.
\eenum
\eitem

\bitem
\item Paper \cite{Dobsicek-stego} has been cited in:
\benum
\item Nameer N. El.Emam.
\newblock Hiding a large amount of data with high security
using steganography algorithm.
\newblock {\em Journal of Computer Science},
3 (4):223--232, 2007.

\item Nameer N. El.Emam.
\newblock Embedding a large amount of information using high
secure neural based steganography algorithm.
\newblock {\em International Journal of Signal Processing},
4 (1):95--106, 2007.
\eenum
\eitem

\newpage

\chapter*{Acknowledgements}

Many people have contributed in one or another way to my PhD
thesis. I would like to express my gratitude to all of them.

First of all, I want to give special thanks to my PhD 
supervisor Josef Kolář. Without his constant help, advises
and support I would never be able to finish my thesis. Also,
he provided me with numerous opportunities for travelling,
meeting new people and professional advances. Thank you a lot.
Many thanks
goes to my co-supervisor Róbert Lórencz. I am very thankful
to him for introducing me to his group of Applied Numerics
and Cryptography, and for being such a supportive colleague
and friend. It was thanks to him I talked with a top class
cryptographer such as Tomáš Rosa and meet new friends Jiří
Buček, Tomáš Zahradnický and Tomáš Brabec in the group.

In the same way, I want to deeply thank Vitaly Shumeiko
and G\"{o}ran Wendin from the Chalmers University of 
Technology in Sweden, where I spent almost a half of my
PhD studies. Vitaly and G\"oran gave me a great opportunity
to work with them in the Laboratory of Applied Quantum 
Physics. The laboratory provides a very flexible environment
and the research performed there, regarding  superconducting
quantum computing, is one of the best in the world. It was a
great honor, pleasure and fun to be a member of the lab. I
gratefully acknowledge a financial support from the
laboratory.

I am especially thankful to G\"oran Johansson for helping
me with my research. His teaching, personal passion for 
science and easy going attitude have been the main reasons
for my achievements during my stay at the lab. Thanks 
G\"oran! Also, I want to thank Mikael Fogelstr\"{o}m,
Tomas L\"{o}fwander, Margareta Wallquist, Jonas Sk\"{o}ldberg,
Lars Tornberg, Jens Michelsen and Cecilia Holmquist for providing
a very friendly environment. I had a pleasure to share a room with
Cissi and Jens. I enjoyed it very much and it was fun. Additionally,
I thank Lars for taking me out bouldering on the wonderful island
H\"{o}n\"{o}.

I wish to thank Štěpán Holub from the Charles University in Prague
who introduced me to the field of quantum computing, Antoni Lozano
from Universitat Polit\`{e}cnica de Catalunya in Barcelona who
supervised me during my Erasmus stay at Barcelona and become a friend,
\v{C}estm\'{i}r Burd\'{i}k, Miloslav Znojil and Vladimir Gerdt who
I had stimulating discussions with during my short stay at the Joint
Institute for Nuclear Research in Dubna, Russia.

On the private side, I want to thank the friends I have met
in Barcelona. Thanks Luisa and Juan for letting my live in your
apartment and teaching me Spanish, thanks Fabien for being a great
friend and bringing me back to computer hacking, thanks Roxane for
your smile and designing a wedding invitation for me and my 
girlfriend Hanka. I am going to ask you for this favor soon :-)
Thanks Anna for being at Barcelona at the same time I was.
It was fun.

Very special thanks go to the small Czech community in G\"{o}teborg.
Mainly to Petr, Dominik, Sandro, Vojta, Jana B., Filip with Jana M.
and their kids, Honza, Tom\'{a}\v{s}, Zden\v{e}k
and Lenka. Additionally, to my friends in Prague who I regretfully now meet
quite rarely. I am greeting Jiříček with Karolínka, Robi with Milča,
Radek, Zornička, Radim, Mates, Canibal, Slunce, Helča and Polárníci.
You all were and still are an important part of my life. Thank you
for being my friends.

Finally, my greatest thanks to the people who are my family,
my parents, my brother and sister and my girlfriend Hanka. I am
very much indebted for your love and great support. Life is fine
as you all are here.

\newpage



\tableofcontents

\newpage

\listoffigures

\newpage


\newpage

\mainbodystarts



 \chapter{Introduction}
 \label{chap:introduction}


\section{Computation and information processing}
Computation and information processing in general always
played an important role in  human society. We can find
mechanical devices performing computation in ancient
astronomy, the Greek Antikythera mechanism (150 BCE) is
the oldest known mechanical analog computer, we can see
basic information processing such as storing data in a
written form on the walls of Egyptian pyramids. Somewhat
more advanced information processing in a modern sense is
historically connected to solving equations. Arabic
mathematician Muhammad ben Musa al-Khwarizmi (825 CE) seems
to be the first who used a concept of repeatedly applying a
set of reduction rules in order to reduce the complexity of
the equation at hand. The reduction (al-jebr) later gave
name to the algebra and the concept itself is now called an
algorithm in honor of al-Khwarizmi.

Later on, advances in math and technology gave birth to the
question of which problems/functions in principle can be
calculated, and to which extent can a mechanical device
(a machine) support an automatic problem reduction according
to a given algorithm. These questions date back to Gottfried
Leibnitz (1646 - 1716) who after having successfully
constructed a mechanical calculating device dreamt of a
formal language for a machine that could manipulate symbols
and determine the truth values of mathematical statements.

In 1936, Alan Turing developed an abstract notation of a
programmable universal computer, a Turing machine, in order
to address Hilbert's Entscheidunsgproblem (German for
'decision problem').
Using this abstraction, he was able to prove that there is
no mechanical procedure which can be used to decide
arbitrary statements in mathematics. In particular, Turing
reduced the halting problem to the Entscheidunsgproblem.
The term 'mechanical procedure' does not directly refer to
digital or analog implementation, however, classical
mechanics was assumed implicitly.

Also Turing's result directly spawns a short proof of
G\"odel Incompleteness Theorem, which is otherwise very long.
In modern terms, the original G\"odel's proof (1931) consists
mainly of defining a sort of computational model (a~G\"odel
numbering) and the work of Turing is heavily influenced by
that. A few months before Turing's result, Alonzo Church
(adviser of Turing) finished his work on lambda calculus
and proved that any function of positive integers can be
mechanically computed only if it is recursive. Soon it
turned out that lambda calculus and Turing machine
abstraction are equivalent (so are all new
non-hypercomputation proposals) and the authors formulated
the Church-Turing hypothesis:
\nicebox{
\begin{center}{\it
 'Any function "naturally to be regarded as computable"
 is computable by a Turing machine'.
}\end{center}}

The Church-Turing hypothesis cannot be mathematically proven
and therefore it is sometimes proposed as a definition or
as a physical law using a modern reformulation by
David Deutsch (1985):
\nicebox{
\begin{center}{\it
 'Every finitely realizable physical system can be
 perfectly simulated by a universal model computing machine
 operating by finite means'.}
\end{center}}

A few years later, after Church and Turing's theoretical
breakthrough, the use of digital electronics largely
invented by Claude Shannon and the stored program
architecture proposed by John von Neumann gave birth
to the first real-life computing devices with all the
capabilities of a Turing machine. The discovery of the
transistor led to a rapid progress in technology and
the digital computer era started.

It may be surprising that while previous computing devices
were analog ones (based on mechanical or hydraulic principles),
electronic analog computers were beaten by electronic
digital computers. The reasons are purely technical.
Practical computing with analog electronics is limited by
the range over which the variables might vary, and,
moreover, there is no theorem saying that noise can ever be
efficiently suppressed. Today, electronic analog computing
is still studied in the field of artificial neural networks
with limited precision real weights. The main point is that
some problems map more naturally to an analog computer than
to a digital computer and therefore the computation might be
faster due to reduced mapping costs.

It is worth to mention here, that an analog computer does
not implicitly assume a scale-independent continuous
phenomenon for a practical computation.  In fact, a
hypothetical device using real numbers with infinite
precision which are harnessable for computation would allow
for non-Turing (non-recursive) functions to be computed.
This is called a hypercomputation. A hypercomputer would not
only violate the Church-Turing hypothesis, but it would need
to exploit a real continuous phenomenon, which we know is
largely ruled out by quantum physics. The difference between
digital computing and real analog computing might also be
seen as the difference between computable numbers and
algebraic systems vs. real numbers and differential systems.

On the theoretical ground, it was found that it is valuable
to classify algorithms and their complexity according to
time and space demands with respect to the size of the
input. Several prominent complexity classes refer to
problems which are solvable by a Turing machine in
polynomial time, $P$, problems solvable in polynomial space,
$PSPACE$, problems solvable in exponential time, $EXP$, and
problems for which, if the answer is 'yes', then there is a
polynomial time proof, $NP$. It follows that
$P \subseteq NP \subseteq PSPACE \subseteq EXP$.

The empirical practice reveals that the vast majority of
important problems with polynomial time/space demands have
their complexity bounded by polynomials of low degree,
while problems with exponential
time demands have the base not close to one. It seems there
is a sharp difference among problems contained in $NP$,
fast efficiently solvable problems on one hand and
practically intractable problems on the other one. The
contrast of $O(n^2)$ and $O(2^n)$ for $n>100$ is
overwhelming and therefore the structure of the $NP$ class
is of utmost importance.

In the early 1970's, Cook, Karp and Levin proved an existence
of $NP$-complete problems, a subset of problems in $NP$, that
are at least as difficult as any other problems in $NP$, and
they actually showed that the decision version of many
practical problems such as the Boolean satisfiability problem,
the Traveling salesperson problem or the Clique problem are
$NP$-complete. So far, we have no indication that
$NP$-complete problems can be solved in polynomial time
instead of exponential time on a Turing machine, and yet we
are not able to prove or disprove the relation $P \neq NP$.
Currently, the $P \neq NP$ relation is one of the six
Millennium Prize Problems established by the Clay Mathematics
Institute.

Clearly, we know problems contained in $NP$ for
which we did not find a polynomial runtime algorithm nor
we proved them to be $NP$-complete. This is partly due to
our lack of knowledge, and, moreover, if $P \neq NP$, then
the Ladner's Theorem (1975) says that there must be
intermediate problems between $P$ and $NP$-complete.
Next to several artificial problems developed just for the
purpose of being $NP$-intermediate, natural problems such as
primality testing, factoring of large integers and graph
isomorphism were on the list together until recently.

\subsection*{Probabilistic computing}
It was always considered that relaxing from deterministic
computing to probabilistic computing might have a positive
impact on problems solving. This expectation is connected
to relatively fast converging iterative processes which when
supplied with a 'right' guess do produce the correct
answer most of the time. The right
guess is given by empirical sampling over similar
cases. In this view, the art of designing a probabilistic
algorithm consists of replacing a right guess by a random
guess while keeping the convergence of the process.
Probabilistic algorithms that always return the
correct answer within an expected (finite) runtime is
called a Las Vegas simulation. If the expected runtime
is bounded by a polynomial we talk about a zero-error
probabilistic polynomial-time problems, class $ZPP$.
With a modification to stop
after a certain number of steps and output an approximate
(possibly false) result we get a Monte Carlo simulation.
If the error probability can be bounded and
polynomial-runtime is kept the corresponding class is
$BPP$. The most general class of probabilistic polynomial
time problems is $PP$. It follows that $P\subseteq ZPP
\subseteq BPP \subseteq PP \subseteq PSPACE$.

Advances in complexity theory and probabilistic algorithms
led to so called Extended or Strong Church-Turing hypothesis
due to Bernstein and Vazirani (1993):
\nicebox{
\begin{center}{\it  
 'Any "reasonable" model of computation can be
 efficiently simulated on a probabilistic Turing machine'
}\end{center}}       

The message of this hypothesis is three-fold. First, it
is practically oriented. It refers to reasonable models
of computation in order to stand aside from hypothetical
hypercomputers and classical extremely massive parallel
systems where the space and/or energy requirements grow
exponentially with the size of the input. An example of
such model is the original Adleman DNA computer proposal
(1994).
Secondly, it allows a Turing machine to access a random
number generator. This expresses the assumption that
probabilistic algorithms might be stronger than
deterministic ones. This assumption was supported mainly
by a very fast probabilistic primality testing while
deterministic testing needed exponential time.
Third, the word 'efficiently'
means that all reasonable computational devices regardless
of their physical implementation are equivalent up to
polynomial-time reductions. In other words, a problem
classification is independent of machines while it is
true that an architecture reflecting the inner structure
of the problem may yield a polynomial speed-up compared to
other architectures.

One puzzling problem is the question of where the power
of probabilistic algorithms comes from. Practice reveals
that pseudo-random number generators are sufficient for
most probabilistic algorithms in order to be successful.
Additionally, classical mechanics which is still the main
framework for computational theory does not allow for
true random number generators. In fact, only practically
unpredictable numbers can be generated. For this reason it
is now believed that $P$ might be equal to $BPP$.

The assumption $P=BPP$ started
a study branch called 'derandomization'. The random essence
is at first reduced to a one call of a random number
generator, and then replaced by a deterministic process.
In 2002, one remarkable success
of derandomization was a deterministic polynomial-time
primality testing algorithm (the AKS test) developed by
Agrawal, Kayal and Saxena.
Primality testing nicely reflects advances in
theoretical computer science. In 1975, the test was shown
to be in $NP$, later in 1992, it fell to $ZPP$, and in 2002,
it went to $P$.
The AKS primality test is however much slower than
the probabilistic one and therefore it is used only as a
part of mathematical proofs.

This approach reflects problem solving in practice.
Problem classification alone is sometime not the right
figure of merit for its practical tractability. It only
considers worst-case times, i.e. it gives us information
that the problem never grows faster than a certain upper
bound. In fact, relaxation to average cases (application
specific) can reveal nice inner structure which allows
for practically fast heuristics, random sampling,
evolution techniques and so on. These methods are often
of advantage in general.

Some $P$ and $NP$-intermediate problems let alone
$NP$-complete ones, do not show enough structure which
can be exploited for practical speed-up on average. On the
other hand they seem to naturally map on unconventional
computing devices such as biological, chemical and quantum
systems. Actually, these problems are usually
problems from their respective fields. This mirrors the
fact that information processing is always a physical
process and some processes in their natural form are faster
compared to others.
In the future, we might end up with a computer equipped
with several co-processors which operate by different means
than digital electronics. Another tantalizing prospect
related to unconventional computers is that they might
exhibit even faster than polynomial speed-up for specific
problems and thus be a threat to the Strong Church-Turing
hypothesis. The most serious candidate that appears not to
obey the Strong Church-Turing hypothesis is a yet to be
built quantum computer.

\begin{figure}[h!]
 \includegraphics[angle=-90,scale=0.7]
 {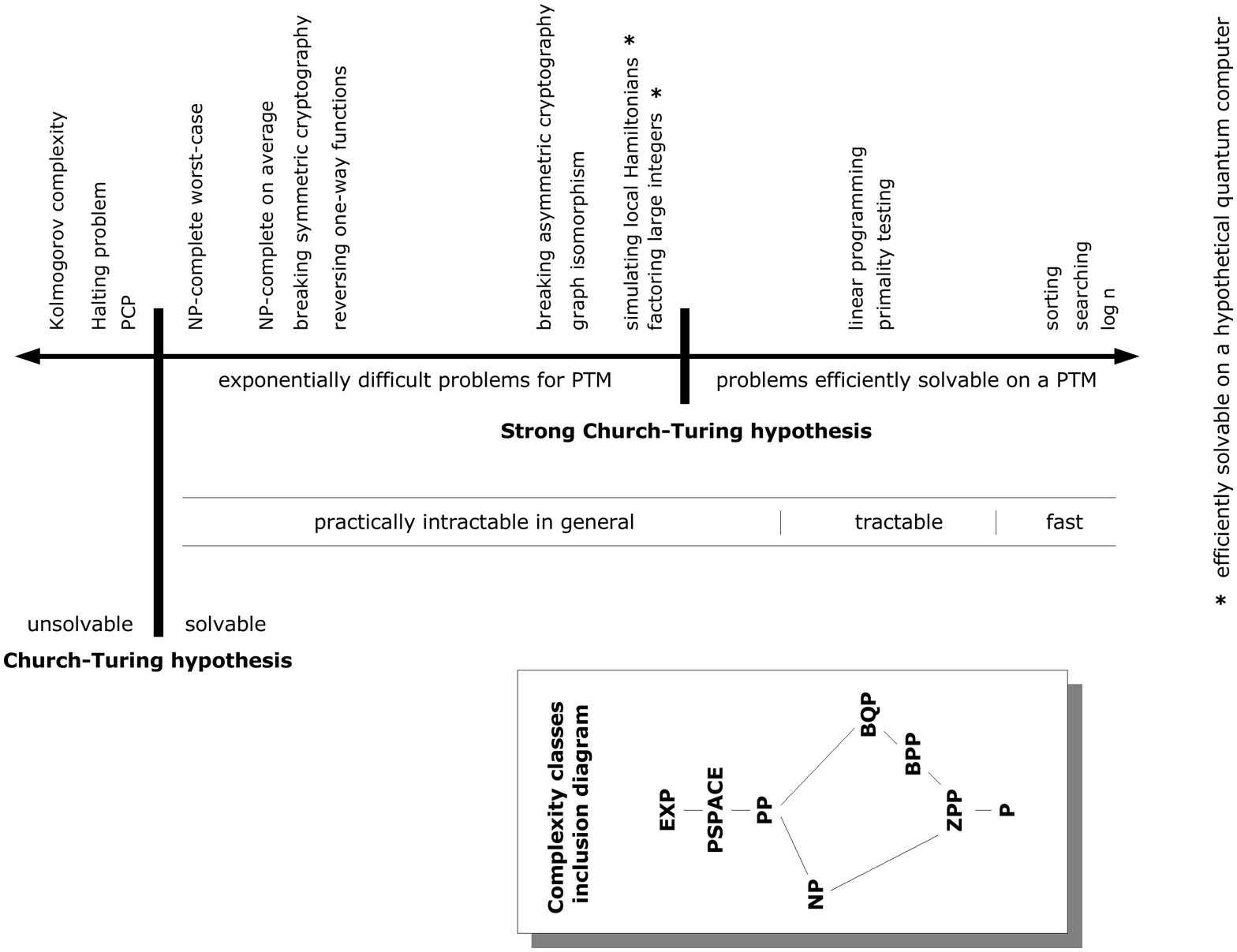}
 \caption[Key points in the structure of decision problems]
         {Currently known key points in the structure of
          decision problems.}
 \label{structure-decision-problems}
\end{figure}

Figure~\ref{structure-decision-problems} summarizes
currently known key points in the structure of decision
problems and indicates the assumed power of quantum
computers. PTM stands for a probabilistic Turing machine.

\newpage

\subsection*{Quantum computing} 
Classical information processing is developed in the
framework of classical mechanics. Since classical physics
was superseded by quantum physics in many ways, it is
natural to explore the potential of information processing
in the framework of quantum mechanics. In 1982, Richard
P. Feynman conjectured that quantum physics in general can
be simulated on a probabilistic Turing machine only with an
exponential slowdown and speculated about a universal
quantum mechanical system capable of an efficient
simulation. At about the same time, it was proved by
Bennett, Fredkin and Toffoli that a universal
computation can be done reversibly. This was an important
step since processes in quantum physics are reversible.
An abstract model for quantum computing, a quantum
Turing machine, was created by David Deutsch (1985) as a
tool for constructing a test of the many-world interpretation
of quantum physics. Quaintly enough, the test now hinges
on the question whether quantum information processing
allows for strong artificial intelligence. The resemblance
to the work of Turing is striking. They both developed an
abstract toolkit with a great theoretical value on its own
for further reasoning, and the reasoning at the end concerns
artificial intelligence. Theoretical quantum information
processing seems to have good prospects for deeper
understanding of connections between computation, nature
and life.

States in a classical computer are represented as strings
from a finite alphabet, typically binary strings,
while states of a quantum computer are
represented by unit vectors in a finite dimensional
complex Hilbert space. Complex amplitudes of these
vectors are allowed to change continuously, but they cannot
be directly observed. Therefore quantum computing can be
seen as a special intermediate case between digital and real
analog computing. Importantly, there is a threshold theorem
for quantum error correction, as opposed to the pure analog
case.
Alternatively, quantum computing can be seen as generalized
probabilistic computing, where non-negative real
probabilities are replaced with complex amplitudes.
The main new resources are quantum mechanical phenomena
such as state superposition, interference and entanglement.
Superposition together with interference
provide a special kind of parallelism, while entanglement,
especially when spatially shared, supports unique means of
communication.

The first evidence (promise problems) of the assumed quantum
computer power was given by D. Deutsch (1985), Deutsch and
Jozsa (1992), Bernstein and Vazirani (1993), and D. Simon
(1994).
Although these promise problems concerning global properties
of functions were somewhat artificial, the corresponding
quantum circuits composed from a set of universal gates
gave an exponential speed-up over the classical deterministic
solutions.
In 1994, Peter W. Shor published a bounded-error quantum
polynomial time algorithm  for large integer
factoring and discrete logarithm calculations.
The corresponding complexity class is called $BQP$.
The factoring problem
is believed to be classically intractable and currently the
RSA public-key cryptosystem is based on its intractability.
Later in 1999, Abrams and Lloyd described a $BQP$ algorithm
for finding energy eigenvalues and eigenvectors of quantum
physical systems governed by a local
Hamiltonian. Again, this is exponentially faster than many
{\it ab initio} methods now used in physics and chemistry
problems where local Hamiltonian takes place.
It turned out that all of the above mentioned problems
are instances of a hidden subgroup problem (HSP) over
Abelian groups, and the problem-solution approach can be
explained in terms of the quantum phase estimation algorithm
(PEA). A key ingredient in this algorithm is a quantum
Fourier transform (QFT) for which we actually know a
quadratic size circuit in the case of Abelian groups. 
Efficient solution for the HSP over non-Abelian groups is
an open problem. It is known that an efficient circuit for
the HSP over the symmetric group would lead to an efficient
algorithm for the graph isomorphism problem. The HSP over
the dihedral group is in the same manner connected to the
shortest vector in a lattice problem. Recently, there has
been progress in solving certain non-Abelian HSP utilizing
transforms beyond QFT. A discovery of an efficient circuit
for the  Clebsch-Gordan transform over the Heisenberg group
led to an efficient solution for the HSP over the Heisenberg
group. The Clebsch-Gordan transform also led to a 
discovery of an efficient circuit for the Schur transform
which might possibly open up for new quantum
algorithms.

The question whether a quantum computer can solve
efficiently $NP$-complete problems remains to be answered
as well as for a classical computer. In 1996, L. Grover
presented a sort of no-go for quantum exponential speed-up for
completely unstructured problems. He developed an algorithm
for unsorted database search which is provably optimal and
yields only a square root speed-up. In particular, a desired
item from an unsorted database of size $N$ can be searched
in time $O(\sqrt{N})$.

From the practical realization point of view, there is no
known viable technology for a scalable quantum computer yet.
Experiments have been performed with up to ten quantum
bits. Problems are caused mainly due to environmental
noise.
Other problems arise from entanglement which makes the
border line between a central processing unit and memory
not that sharp as it is in a classical computer. One
possible approach toward solving this problem is a
distributed quantum computer. Here, relatively easy to
stabilize small cores are interconnected with quantum and
classical channels, and by exploiting quantum state
teleportation and non-local control they cooperate on the
problem solving.

Besides quantum computing, there is also a discipline called
quantum cryptography. The protocol BB84 developed by Bennett
and Brassard in 1984 solves the problem of secret-key
distribution in an unconditionally secure way. This is
assured for free by the quantum physics laws themselves.
The first commercial products with the BB84 are already
available. They work as point-to-point systems over a fiber
optic line, up to 100km of length, at rates approximately
100kbits/sec. Another commercially available product in the
form of a standard PCI card or an USB device is a quantum
random number generator. The bit rate is over 16Mbits/sec.
In general, special properties of a quantum physical system
such as the no-cloning property of an unknown quantum state
and the randomness in a wave function collapse allow for
many quantum enhanced cryptographic primitives. These 
primitives range from authentication to covered communication
(steganography) and more.










\section{Motivation and contribution of the thesis}
The Abrams-Lloyd algorithm for finding energy eigenvalues is
potentially the most useful quantum algorithm known so far,
and if a quantum computer is ever built, it will 
revolutionize quantum chemical calculations. Thus there is a
growing consensus regarding investments into the experimental
quantum computing. 


In this thesis, attention is paid to small experimental
testbed applications with respect to the quantum phase
estimation algorithm, the core approach for finding energy
eigenvalues. An iterative scheme for quantum phase estimation
(IPEA) is derived from the Kitaev phase estimation, a study of
robustness of the IPEA utilized as a few-qubit testbed
application is performed, and an improved protocol for phase
reference alignment is presented. The derivation of the IPEA
involves incorporating a classical postprocessing algorithm
into a quantum circuit. The resulting quantum circuit then
becomes more efficient. Such a result may be quite inspiring
in the further development of quantum algorithms as the usual
approach is to perform as much as possible of necessary
calculations classically, since quantum resources are
considered 'expensive'.

Additionally, a short
overview of a quantum cryptography is given, with a particular
focus on quantum steganography and authentication.


\section{Organization of the thesis}

The outline of this thesis is as follows. Chapter 2 defines
the terms used in the thesis. A rather large part of this
chapter pertains to Turing machines and portions of linear
algebra. Chapter 3 provides a summary on quantum computing
and introduces the quantum Turing machine and quantum 
circuit model. A special attention is paid to the thin 
borderline between quantum computing and classical computing.
Chapter 4 considers fast quantum algorithms, quantum phase
estimation and its iterative variants, and a few testbed
circuits. This is the main chapter of the thesis. Chapter 5
contains the overview on quantum cryptography. Finally,
Chapter 6 consists of conclusions and the direction for 
future work.









 \chapter{Basic definitions and postulates}
 \label{chap:basic-definitions}



%
%

\section{Turing machines}

A Turing machine is a symbol-manipulating device
equipped with a read/write head and possibly infinite
cell-divided tape. The head reads the current symbol on
the tape, changes its internal state accordingly, writes an
output symbol if any, and then moves to the neighboring
left or right cell.  A computation consists of periodical
repeating of these steps. At any moment, the computation
is completely described by a Turing machine configuration.

{\definition[Turing machine configuration]
 A configuration of a Turing machine is an ordered triple:
 \begin{itemize}
  \item the contents of the tape,
  \item the current state,
  \item the position of the head.\sq
 \end{itemize}
}

Despite the simplicity, the
Church-Turing hypothesis says that a Turing machine can
simulate the logic of any computer that could be possibly
constructed. A Turing machine is said to
be universal if it is able to simulate any other Turing
machine. The smallest universal Turing machine has
three symbols and two internal states \cite{Smith07}. No
smaller
universal Turing machine is possible. Other commonly
used computational abstractions are a cellular automata
and a gate model. Gates sets such as \{NAND, fanout\},
\{NOR, fanout\}, \{AND, NOT\} constitute universal
sets. The Toffoli gate alone is a universal gate for
reversible computing.
Formally, a Turing machine is defined as follows.

{\definition[Turing machine]
 A Turing machine is an ordered six-tuple 
 $M=(Q,\Sigma,b,q_0,F,\delta)$, where
 \begin{enumerate}
  \item $Q$ is a {\it finite} set of states,
  \item $\Sigma$ is a finite set of tape symbols
        (alphabet),
  \item $b \in \Sigma$ is the blank symbol,
  \item $q_0 \in Q$ is the initial state,
  \item $F \subseteq Q$ is the set of final or accepting
        states,
  \item $\delta: \Sigma \times Q \rightarrow
                 \Sigma \times Q \times \{L,R\}$
	is a transition function, where $L$ is shift left,
	$R$ is shift right.
 \end{enumerate}
}

The transition function 
$$
 \delta: (\text{current symbol, current state)}
 \longrightarrow
 (\text{output symbol, new state, shift left/right})
$$
is usually given as an action table or a diagram. Sometimes
it is written as
$$
 \delta: (\text{current symbol, current state,
 output symbol, new state, shift left/right})
 \longrightarrow
 \{0, 1\}
$$
in order to denote allowed transitions among configurations.

{\example[Turing machine action table] \label{turing-machine-example}
A simple
action table for a machine
$$M=(\{q_0,q_1,q_2,q_f\},\,\{\ap0\ap,\ap1\ap,b\},\,b,\,q_0,\,\{q_f\},\,\delta)$$
calculating the XOR function over input bits $a$ and $b$ is
shown in the Figure~\ref{transition-function}.
The result is stored at the place of the second input bit.

\begin{figure}[h!]\begin{center}
 \begin{tabular}{cc||c}
  $a$ & $b$ & XOR \\ \hline \hline
  0 & 0 & 0 \\ \hline
  0 & 1 & 1 \\ \hline
  1 & 0 & 1 \\ \hline
  1 & 1 & 0 \\
 \end{tabular}\hspace{3cm}
 \begin{tabular}{c||c|c}
  $\delta$ & '0' & '1' \\ \hline\hline
  $q_0$    & $(\ap0\ap,q_1,L)$   & $(\ap0\ap,q_2,L)$ \\ \hline 
  $q_1$    & $(\ap0\ap,q_f,L)$   & $(\ap1\ap,q_f,L)$ \\ \hline
  $q_2$    & $(\ap1\ap,q_f,L)$   & $(\ap0\ap,q_f,L)$ \\
 \end{tabular}
 \caption[Turing machine action table]
         {The XOR function and corresponding Turing machine
          action table.}
 \label{transition-function}
\end{center}\end{figure}
}

The transition function $\delta$ makes a given machine $M$
to behave like a computer with a fixed program. In this
sense, only one function can be calculated. However, since
the action table itself can be encoded as a string, we can
say that a part of the tape is considered to be a program
description and the rest is the input:

$$\underset{\text{program}}
           {\underbrace{\ap0101011 \ldots}}
  \underset{\text{input data}}
           {\underbrace{1101 \ldots 010\ap}}
  \overset{\delta}{\longrightarrow}
  \underset{\text{program}}
           {\underbrace{\ap0101011 \ldots}}
  \underset{\text{output data}}
           {\underbrace{0011 \ldots 100\ap}}\,.
$$

The Turing machine from the 
Example~\ref{turing-machine-example}
can be interpreted as recognizing two programs:
\begin{verbatim}
program '0': perform identity on the input bit,
program '1': negate the input bit.
\end{verbatim}

Due to string manipulation nature, Turing machines (and
various automata) are often studied in terms of languages
they accept.

{\definition[Accepting a language]
 A Turing machine accepts a language $L$ iff
 \begin{itemize}
  \item for all strings $x \in L$: the machine halts in a
   final state,
  \item for all strings $x \notin L$: the machine halts
   in a non-final state or the machine enters an infinite
   loop.\sq
 \end{itemize}
}

The Turing machine from the 
Example~\ref{turing-machine-example}
halts in a non-final state for the blank symbol $b$
(no program or input bit is given).

{\definition[Time complexity] Let $M$ be a Turing machine
and $f(n)$ a function $f: \mathbb{N}\rightarrow\mathbb{N}$.
We say that $M$ has a time complexity $f(n)$ if for each
input of size $n$ the machine halts after at most $f(n)$
steps.
}

{\definition[Space complexity] Let $M$ be a Turing machine
and $f(n)$ a function $f: \mathbb{N} \rightarrow\mathbb{N}$.
We say that $M$ has a space complexity $f(n)$ if for each
input of size $n$ the machine utilizes at most $f(n)$ cells
before halting.
}

\subsection{Probabilistic Turing machines}

{\definition[Probabilistic Turing machine]
 A probabilistic Turing machine is a generalized Turing
 machine where the transition function does not output
 only a single triple (output symbol, new state, shift),
 but a probabilistic distribution of such triples.
 Formally, the transition function assigns to each
 possible transition a non-negative probability:
 \begin{equation}
   \delta: Q \times \Sigma \times Q \times \Sigma
   \times \{L, R\} \longrightarrow [0,1]
 \end{equation}
 in such a way that the set of transitions from one
 configuration $c_0$ to all its direct
 successor-configurations $c_1,\ldots,c_k$
 must fulfill the local probability condition
 \begin{equation}
  \label{LPC}
  \hspace{3.3cm}
  \sum_{i=1}^k p_i = 1, \hspace{3cm} 
  \text{(LPC)}
 \end{equation}
 where $p_i$ is the probability of transition from $c_0$
 to $c_i$.
}\sq

It may happen that several different paths $\ell$ of
computation lead to the same configuration $c_x$.
In such a case, this configuration has a probability
equal to the sum of probabilities of corresponding paths
\begin{equation}
 p(c_x)=\sum_{\ell}p_x^{(\ell)}\,.
\end{equation}
At any moment, the local probability condition assures
that probabilities of all distinct configurations sum to
one (global probability condition).
\begin{equation}
 \label{GPC}
 \hspace{3.1cm}\sum_{\forall c_i\neq c_j}p_i = 1.
 \hspace{2.8cm} \text{(GPC)}
\end{equation}

{\definition[Transition matrix] Let us have a Turing
 machine $M$. The transition matrix consists of elements
 $T_{i,j}$ defined as
 \begin{equation}
  T_{i,j} = \begin{cases}
   p_i & \text{if the machine goes from $c_j$ to $c_i$
               in one step with probability $p_i$},\\
   0 & \text{otherwise}.
  \end{cases}
 \end{equation}
}\sq

Transition matrix for a probabilistic Turing machine is
a stochastic matrix where each column sums to one (LPC).
We can
say that transition matrix $T$ maps one normalized (GPC)
linear
combination (superposition) of configurations to another
normalized linear combination of configurations. Thus
stochastic matrices preserve the $1$-norm 
(Eq. \ref{p-norm}, page \pageref{p-norm}). For every
column vector of configurations
$v$, $\norm{T\cdot v}_1=\norm{v}_1$.

Transitions allowed by stochastic matrices can also be
viewed in terms of constructive interference. The
constructive interference is to be understood as the
possibility for different computational paths to end up
in the same configuration and thus increase the total
probability of that configuration.
Probabilities of other configurations
are proportionally smaller. One has to program a
probabilistic Turing machine in such a way that the
probability of a desirable configuration (correct result)
is gradually enlarged, and probabilities of undesirable
configurations subsequently diminishes.
Destructive interference might allow for more progressive
separation of desirable and undesirable configurations, but
with a probabilistic Turing machine this is not possible
since we have only non-negative probabilities.

{\example[Transition matrix and configuration tree]
An example of stochastic transition matrix and a
configuration tree
starting in a configuration $c_1$ is shown in the
Figure~\ref{transition-matrix}. Local and global
probability conditions
are manifestly fulfilled at each level of the tree.
The computation halts in a configuration $c_4$ with
probability $p(c_4)=0.1+ 0.4\cdot 0.4+ 0.4\cdot
0.6\cdot1=0.5$ and in a configuration $c_5$ 
with probability $p(c_5)=0.5$.
\begin{figure}[h!]
 \begin{center}\begin{tabular}{cc}
 \begin{tabular}{c}
 $T=\left(
 \begin{array}{ccccc}
    0   & 0   & 0   & 0 & 0\\
    0.4 & 0   & 0   & 0 &0\\
    0   & 0.6 & 0   & 0 &0\\
    0.1 & 0.4 & 1   & 1 & 0\\
    0.5 & 0   & 0  &  0 & 1
 \end{array}\right)$
 \end{tabular}&
 \begin{tabular}{c}
 \includegraphics[scale=0.4]{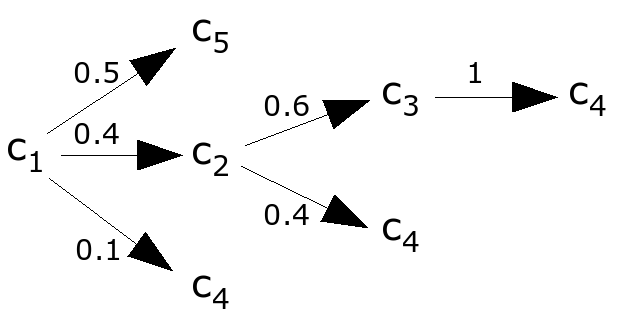} 
 \end{tabular}
 \end{tabular}
 \caption[Stochastic transition matrix]
         {An example of stochastic transition matrix and
          configuration tree with root $c_1$.}
 \label{transition-matrix}
 \end{center}
\end{figure}
}

{\definition[Accepting a language on a probabilistic
Turing machine]
 A probabilistic Turing machine PTM accepts a language $L$
 with completeness $c$ and soundness $s$ iff
 \begin{itemize}
  \item for all strings $x \in L$:  PTM accepts $x$ with
        probability $p > c$,
  \item for all strings $x \notin L$: PTM  accepts $x$ with
        probability $p \leq s$.\sq
 \end{itemize}
}

In order to study the power of a probabilistic Turing
machine, we need to restrict real probabilities from
the range $[0, 1]$ only to a certain 'universal' set of
probabilities. Assumption that all real numbers from
the range $[0,1]$ with infinite precision can be used,
would allow one to encode the hard problems right into
them. 
C. H. Papadimitriou in \cite{Papadimitriou94} showed
that for each
probabilistic Turing machine $M_1$ there exists a
probabilistic Turing machine $M_2$ with probabilities
restricted to the set $\{0,\frac{1}{2},1\}$ which
simulates $M_1$ with at most polynomial slowdown. The
probability $\frac{1}{2}$ shows a crucial dependence
on the fair coin flipping if we assume a
probabilistic Turing machine to be more powerful than
a deterministic Turing machine.

{\example[Probabilistic Turing machine with restricted
probabilities]
Let $M_1$ be a probabilistic Turing machine using
probabilities $0.4$ and $0.6$ at some part of the
computation. Let $M_2$ be a probabilistic Turing machine
with restricted probabilities $\{0, \frac{1}{2}, 1\}$.
In the Figure~\ref{restricted-probabilities},
$M_2$ approximates $M_1$ with precision
$\varepsilon \geq 1/2^3$.
\begin{figure}[h!]
 \begin{center}
 \includegraphics[scale=0.9]
 {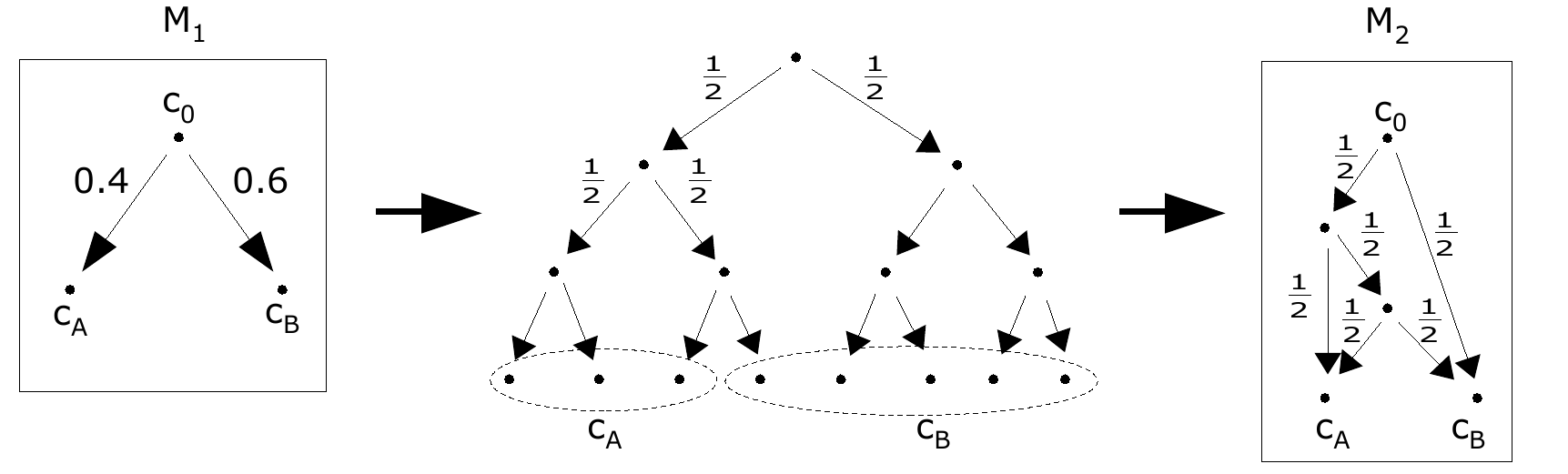}
 \caption[Restricted set of probabilities]
         {Probabilistic Turing machine with restricted
          probabilities.}
 \label{restricted-probabilities}
 \end{center}
\end{figure}
}

\subsection{Nondeterministic Turing machines}

{\definition[Nondeterministic Turing machine]
 Nondeterministic Turing machine is a hypothetical kind
 of a probabilistic Turing machine. It explores all the
 possible computational paths at parallel, and if the
 solution exists, then with non-zero probability the right
 path is actually taken. Equivalently, we can say that
 the machine goes along the right computational path, and
 if there is a fork, it nondeterministically chooses with
 non-zero probability the right path again. 
 Nondeterministic Turing Machine NTM accepts a language $L$
 iff
 \begin{itemize}
  \item for all $x \in L$: NTM accepts $x$ with probability
        $p>0$,
  \item for all $x \notin L$: NTM accepts $x$ with probability
        $p=0$.
 \end{itemize}
}\sq

Modern computers are usually not described as direct
instances of a Turing machine. Their architecture is
well described by abstract models as 'stored program'
Random Access Machine (RAM) and synchronous/asynchronous
Parallel Random Access Machine (PRAM). These models
simplify algorithm analysis mainly due to grouping tape
cells to fixed size 'words', which are uniquely indexed
and can be thus directly accessed in a unit time as opposed
to a Turing machine.

\section{Complexity classes}
Algorithms and problems are categorized into complexity
classes.

{\definition[Polynomial-time, P]
 P is the class of decision problems deterministically
 solvable in polynomial time on a Turing machine.
 A language $L$ is in a class P if
 \begin{itemize}
  \item for all $x \in L$: the machine accepts $x$ with
        probability $p=1$,
  \item for all $x \notin L$: the machine accepts $x$ with
        probability $p=0$,
 \end{itemize}
 and the time complexity is upper-bounded by some
 polynomial $poly(x)$.
}

{\definition[Nondeterministic polynomial-time, NP]
 NP is the class of decision problems nondeterministically
 solvable in polynomial time on a Turing machine.
 A language $L$ is in a class NP if
 \begin{itemize}
  \item for all $x \in L$: the machine accepts $x$ with
        probability $p>0$,
  \item for all $x \notin L$: the machine accepts $x$ with
        probability $p=0$,
 \end{itemize}
 and the time complexity is upper-bounded by some
 polynomial $poly(x)$.

 Equivalently, NP is the class of decision problems such
 that, if the answer is 'yes', then there is a deterministic
 polynomial time proof of this fact. 
}

{\definition[Zero-error probabilistic polynomial-time, ZPP]
 ZPP is the class of decision problems probabilistically
 solvable in expected polynomial time on a Turing
 machine with zero error probability.
 A language $L$ is in a class ZPP if
 \begin{itemize}
  \item for all $x \in L$: the machine accepts $x$ with
        probability $p=1$,
  \item for all $x \notin L$: the machine accepts $x$ with
        probability $p=0$,
 \end{itemize}
 and the time complexity is upper-bounded by some
 polynomial $poly(x)$.
}

{\definition[Bounded-error probabilistic polynomial-time, BPP]
 BPP is the class of decision problems probabilistically
 solvable in  polynomial time on a Turing machine with
 bounded error probability.
 A language $L$ is in a class BPP if
 \begin{itemize}
  \item for all $x \in L$: the machine accepts $x$ with
        probability $p>\frac{2}{3}$,
  \item for all $x \notin L$ the machine accepts $x$ with
        probability $p \leq \frac{1}{3}$,
 \end{itemize}
 and the time complexity is upper-bounded by some
 polynomial $poly(x)$.
 
 The numbers $\frac{2}{3}$ and $\frac{1}{3}$ may be replaced
 to any $a$ and $b$ such that $a=\frac{1}{2}+poly^{-1}(x)$ and 
 $b=\frac{1}{2}-poly^{-1}(x)$. The Chernoff bound \cite{Chernoff-bound}
 says that by repeating trials the error probability diminishes
 exponentially fast and the success probability is
 accordingly increased.
}

{\definition[Probabilistic polynomial-time, PP]
 PP is the class of decision problems probabilistically
 solvable in polynomial time on a Turing machine but the
 error probability need not to be bounded by a polynomial.
 Compared to the BPP class, the numbers $a$ and $b$ are
 allowed to scale as $\frac{1}{2}\pm \exp^{-1}(poly(x))$.
 A language $L$ is in a class PP if
 \begin{itemize}
  \item for all $x \in L$: the machine accepts $x$ with
        probability $p>1/2$,
  \item for all $ x \notin L$: the machine accepts $x$ with
        probability $p\leq 1/2$,
 \end{itemize}
 and the time complexity is upper-bounded by some
 polynomial $poly(x)$.
}

{\definition[Polynomial-space, PSPACE]
 PSPACE is the class of decision problems solvable
 on a Turing machine whose space complexity is
 upper-bounded by some polynomial with respect
 to the size of the input.
}

{\definition[Exponential-time, EXP]
 EXP is the class of decision problems deterministically
 solvable a Turing machine whose time complexity is
 upper-bounded by an exponential function $\exp{(poly(n))}$,
 where $poly(n)$ represents a polynomial with respect to
 the size of the input $n$.
}\sq

It follows that $P \subseteq ZPP \subseteq NP \subseteq PP
\subseteq PSPACE \subseteq EXP$ and $P \subseteq ZPP
\subseteq BPP \subseteq PP \subseteq PSPACE \subseteq EXP$.
The relation of BPP and NP is unknown. However, it is
believed that $P=BPP$ and $P\neq NP$. Detailed proofs and
further information can be found in \cite{Papadimitriou94}.

\section{Linear algebra}

Table~\ref{tab:dirac-notation} summarizes the Dirac
notation for
notions from linear algebra. This 'bra(c)ket' notation is
widely used within quantum mechanics. Another formalism
based on density matrices will be introduced later.

\begin{table}[h!]
\begin{center}
\begin{tabular}{c|l} \hline \hline
 {\bf Notation}    &
   {\bf Description} \\ \hline
 $\ket{\upsilon}$
   & Column vector. Known as 'ket'.\\
 
 $\bra{\upsilon}$
   & Row vector dual to \ket{\upsilon}.
   Known as 'bra'.\\
 
 $\braket{\psi}{\upsilon}$
   & Inner product of vectors \bra{\psi}
   and \ket{\upsilon}.\\
 
 $\ketbra{\psi}{\upsilon}$
   & Outer product of \ket{\psi}
   and \bra{\upsilon}.\\
 
 $\ket{\psi}\otimes\ket{\upsilon}$
   & Tensor product of \ket{\psi}
   and \ket{\upsilon}.\\
 
 $\ket{\psi}\ket{\upsilon}$
   & Abbreviated notation for tensor product
   of \ket{\psi} and \ket{\upsilon}.\\
 
 $\ket{\psi,\upsilon}$
   & Abbreviated notation for tensor product
   of \ket{\psi} and \ket{\upsilon}.\\
 
 $A^{\dagger}$
   & Adjoint operator of the $A$ matrix.
   $A^{\dagger}=\left(A^{T}\right)^{*}
               =\left(A^{*}\right)^{T}$.\\

 $\bra{\psi}A\ket{\upsilon}$
   & Inner product of \bra{\psi}
   and $A\ket{\upsilon}$.\\

 $\norm{\upsilon}$
   & Notation for a vector norm.\\ \hline \hline
\end{tabular}
\caption{Dirac notation.}
\label{tab:dirac-notation}
\end{center}
\end{table}

\subsection{Hilbert space}

{\definition[Vector space]
 A set $\mathcal{V}=\left(\mathcal{V},+,\cdot\,\right)$
 is called a vector space over a scalar field $\mathcal{F}$
 iff the operations
 $+: \mathcal{V} \times \mathcal{V} \rightarrow \mathcal{V}$
 (vector addition) and $ \cdot: \mathcal{F} \times \mathcal{V}
 \rightarrow \mathcal{V}$ (scalar multiplication) are defined 
 and
 \begin{enumerate}
  \item $(\mathcal{V},+)$ is a commutative group with a neutral
  element $\overrightarrow{0}$,\\[7pt]
  and for all $\upsilon, \psi \in \mathcal{V}, \quad
  \alpha,\beta \in \mathcal{F}$:\vspace{-4pt}
  \item $\alpha(\beta\ket{\upsilon})
         =(\alpha\beta)\ket{\upsilon}$,
  \item $(\alpha+\beta)\ket{\upsilon}
         =\alpha\ket{\upsilon}+\beta\ket{\upsilon}$,
  \item $\alpha(\ket{\upsilon}+\ket{\psi})
         =\alpha\ket{\upsilon}+\alpha\ket{\psi}$.
 \end{enumerate}
} 

{\definition[Vector norm]
 Let $\mathcal{V}$ be a vector space over a scalar field
 $\mathcal{F}$. A vector norm is a function $\norm{\cdot}:
 \mathcal{V} \rightarrow \mathbb{R^+}$ satisfying for all
 $\upsilon, \psi \in \mathcal{V},\quad
 \eta \in \mathcal{F}$:
 \begin{enumerate}
  \item $\norm{\upsilon}=0 \;\Leftrightarrow\;
        \upsilon=\overrightarrow{0}$,
  \item $\norm{\eta\upsilon}=\abs{\eta}\norm{\upsilon}$,
  \item $\norm{\,\ket{\psi}+\ket{\upsilon}\,}
        \leq \norm{\psi}
        + \norm{\upsilon}$.
	\hspace{6cm}\text{(triangle inequality)}
 \end{enumerate}
} 

{\definition[$p$-norm]
 Let $p \geq 0$ be a real number and 
 $\upsilon=(\upsilon_1,\upsilon_2,\ldots,\upsilon_n)$
 an $n$-dimensional vector. The $p$-norm is a vector norm
 defined as
 \begin{equation}
  \label{p-norm}
  \norm{\upsilon}_p =
  \left(\sum_{i=1}^n{\abs{\upsilon_i}^p}\right)^\frac{1}{p}.
 \end{equation}
 For $p=1$ we get the taxicab norm and for $p=2$ we get
 the Euclidean norm.
} 

{\definition[Inner product] Let $\mathcal{V}$ be a vector
 space over the field of complex numbers. An inner product
 over the vector space $\mathcal{V}$ is a function
 $\braket{\cdot}{\cdot}: \mathcal{V} \times \mathcal{V}
 \rightarrow \mathbb{C}$ satisfying for all
 $\upsilon, \psi, \lambda \in \mathcal{V}$:
 \begin{enumerate}
  \item $\braket{\upsilon}{\upsilon} \in \mathcal{R},
         \quad \braket{\upsilon}{\upsilon} \geq 0,
         \quad \braket{\upsilon}{\upsilon}=0
         \; \Leftrightarrow \;
         \ket{\upsilon}=\overrightarrow{0}$,
  \item $\braket{\upsilon}{\psi}
         =\braket{\psi}{\upsilon}^{*}$,
  \item $\bra{\upsilon}(\ket{\psi} + \ket{\lambda})
        =\braket{\upsilon}{\psi} + \braket{\upsilon}{\lambda}$.
 \end{enumerate}
 The inner product induces the $2$-norm
 $\norm{\upsilon}_2=\sqrt{\braket{\upsilon}{\upsilon}}$.
 A vector $\ket{\upsilon}$ is normalized or unit-vector under
 the $2$-norm iff
 $\norm{\upsilon}_2=1$.
} 

\nicebox{
{\note
For a real or complex vector space $\mathcal{V}$
with an inner product $\braket{\cdot}{\cdot}$ and norm
$\norm{\cdot}$ the following
inequality holds, $\upsilon, \psi \in \mathcal{V}$:
$$
 \hspace{5cm}
 \abs{\braket{\psi}{\upsilon}} \leq \norm{\psi}
 \cdot \norm{\upsilon}.
 \hspace{2cm} \text{(Schwarz inequality)}
$$
}} 

{\definition[Completeness] Let $\mathcal{V}$ be a vector
 space with the $2$-norm $\norm{\cdot}_2$ and 
 $\ket{\upsilon_{i}} \in \mathcal{V}$
 a sequence of vectors.
 \begin{itemize}
  \item $\ket{\upsilon_{i}}$ is a Cauchy sequence
   iff
   $$\,\forall \epsilon > 0 \;\; \exists N > 0 \text{ such
   that }
   \;\forall n,m > N : \norm{\,\ket{\upsilon_{n}}
   - \ket{\upsilon_{m}}\,}_2 < \epsilon.
   $$
  \item $\ket{\upsilon_{i}}$ is convergent
   iff 
   $$\;\exists \ket{\upsilon} \in \mathcal{V} \text{ such
   that }
   \,\forall \epsilon > 0 \;\; \exists N > 0 \;\;\,
   \forall n > N : \norm{\, \ket{\upsilon_{n}}
   - \ket{\upsilon}\,}_2 < \epsilon.
   $$
 \end{itemize}
 Space $\mathcal{V}$ is complete iff every Cauchy sequence
 converges.
} 

{\definition[Hilbert space] A Hilbert space $\mathcal{H}$
 is a vector space with an inner product
 $\braket{\cdot}{\cdot}$ that is complete under the induced
 norm $\norm{\upsilon}_2=\sqrt{\braket{\upsilon}{\upsilon}}$.
} 

\nicebox{
{\note
 Standard quantum computing is defined only for
 finite dimensional complex Hilbert spaces,
 $\mathcal{H}=\mathbb{C}^n$, and normalized (unit) vectors.
}}

{\definition[Kronecker delta function]
 The Kronecker delta function is a function of two
 variables (usually integers) defined as
 \begin{equation}
  \label{eq:delta_function}
  \delta_{ij}=\left\{\begin{array}{ll}1,
  & \mbox{ if $i=j$,}\\0,
  & \mbox{ otherwise. }\end{array}\right.
 \end{equation}
} 

{\definition[Basis vectors]
 An enumerable set of normalized vectors $\ket{e_{i}}$ forms
 an orthonormal basis of a Hilbert space $\mathcal{H}$ iff
 \begin{enumerate}
  \item $\forall\, i,j : \braket{e_{i}}{e_{j}}=\delta_{ij}$,
  \item any vector $\ket{\upsilon} \in \mathcal{H}$ can be
   written as $\ket{\upsilon}=\sum_{i}{\alpha_{i}\ket{e_{i}}}$.
 \end{enumerate}
} 

{\definition[Dimension of a Hilbert space]
 The dimension of a Hilbert space $\mathcal{H}$
 is the number of vectors of a basis of $\mathcal{H}$. We
 write
 \begin{equation}
   \dim{(\mathcal{H})}=\abs{\{\ket{e_i}\}}\,,
 \end{equation}
 where
 $\{\ket{e_i}\}$ denotes an orthonormal basis of $\mathcal{H}$.
} 

{\definition[Tensor product of two Hilbert spaces]
 \label{tensor-product-hilbert-spaces}
 Let $\mathcal{H}_{A}$ and $\mathcal{H}_{B}$ be Hilbert spaces
 with basis $\ket{e_{i}}$ and $\ket{f_{j}}$ respectively.
 A tensor product
 \begin{equation}
  \label{eq:spaces-tensor-product}
  \mathcal{H}_{AB}=\mathcal{H}_{A} \otimes \mathcal{H}_{B}
  =\left\{\ket{\upsilon} \otimes \ket{\psi} : \ket{\upsilon}
  \in \mathcal{H}_{A}, \ket{\psi} \in \mathcal{H}_{B}\right\}
 \end{equation}
 is also a Hilbert space
 \begin{itemize}
  \item with base $\ket{g_{k}}
   =\left\{\ket{e} \otimes \ket{f}: \ket{e} \in \ket{e_{i}},
   \ket{f} \in \ket{f_{j}} \right\}$ and
  \item inner product defined as
   $\langle\: a \otimes b \mid c \otimes d \:\rangle
   =\braket{a}{c}\braket{b}{d}$,\quad
   where $\ket{a},\ket{c} \in \mathcal{H}_{A}$\\
   and $\ket{b},\ket{d} \in \mathcal{H}_{B}$.
  \item $\dim{(\mathcal{H}_{AB})}
   =\dim{(\mathcal{H}_{A})}\, \cdot \,\dim{(\mathcal{H}_{B})}$.
 \end{itemize}
} 

\subsection{Linear operators}
{\definition[Linear operator]
 Let $\mathcal{V}$ be a vector space. Function
 $A : \mathcal{V} \rightarrow \mathcal{V}$ is a linear
 operator iff it is linear in its inputs, i.e.
 \begin{equation}
  \label{linear-operator}
  A \Bigl( \sum_{i}{\alpha_{i}\ket{e_{i}}}\Bigr)
  =\sum_{i}{\alpha_{i}\left(A\ket{e_{i}}\right)},
  \quad \ket{e_{i}} \in \mathcal{V}.
 \end{equation} 

 In $\mathbb{C}^{n}$ a linear operator $A$ can be expressed
 as an $n$-by-$n$ square matrix.
 \begin{equation}
  \label{linear_operator_matrix}
  A=\left( \begin{array}{ccc}
      a_{0,0}  &   \cdots& a_{0,n-1}\\
      \vdots   &   \ddots& \vdots   \\
      a_{n-1,0}&   \cdots& a_{n-1,n-1}
    \end{array} \right)
   = \sum_{i,j}{a_{ij}\ketbra{i}{j}},
     \quad a_{ij}=\bra{i}A\ket{j}.
 \end{equation}
} 

{\definition[Adjoint Operator]
 Let $A$ be a linear operator on a Hilbert space
 $\mathcal{H}$. A unique linear operator $A^{\dagger}$ on
 $\mathcal{H}$ satisfying
 \begin{equation}
  \label{adjoint_operator}
  \bra{\upsilon\,}A\,\psi\rangle 
  = \langle A^{\dagger}\, \upsilon\, \vert\, \psi \rangle,
 \end{equation}
 for all vectors $\upsilon,\psi \in \mathcal{H}$, is called
 an adjoint operator or Hermitian conjugate of the operator
 $A$. Additionally, we define
 \begin{equation}
  \label{ket_dagger}
  \ket{\upsilon}^{\dagger} \equiv \bra{\upsilon}.
 \end{equation}
} 

\nicebox{
{\note
 Let $A$ and $B$ be linear operators on a Hilbert space
 $\mathcal{H}$, $\ket{\upsilon} \in \mathcal{H}$. Then
 \begin{equation}
  \label{dagger_property_1}
  \left( AB \right)^{\dagger}=B^{\dagger}A^{\dagger},
 \end{equation}
 \begin{equation}
  \label{dagger_property_2}
  \left( A\ket{\upsilon} \right)^{\dagger}
  = \bra{\upsilon}A^{\dagger}.
 \end{equation}
}} 

{\definition A linear operator $A$ defined on a vector
 space $\mathcal{V}$ is called
 \begin{enumerate}
  \item identity operator $\mathbb{I}$ iff
        $A\ket{\upsilon}=\ket{\upsilon}$ for all vectors
	$\ket{\upsilon} \in \mathcal{V}$,
  \item zero operator $\mathbb{O}$ iff
        $A\ket{\upsilon}=\overrightarrow{0}$ for all
	vectors $\ket{\upsilon} \in \mathcal{V}$,
  \item normal iff $A^{\dagger}A=AA^{\dagger}$,
  \item self-adjoint or Hermitian iff $A^{\dagger}=A$,
  \item unitary iff $A^\dagger A = AA^{\dagger}=\mathbb{I}$,
  \item idempotent iff $A^{2}=A$,
  \item projection iff $A$ is self-adjoint and idempotent.
 \end{enumerate}
} 

{\theorem[Spectral decomposition]
 Any normal operator $A$ on a vector space $\mathcal{V}$
 is diagonal to some orthonormal basis for $\mathcal{V}$
 \begin{equation}
  \label{spectral_decomposition}
  A=\sum_{i}{\lambda_{i}\ketbra{i}{i}},
 \end{equation}
 where $\lambda_{i}$ are the eigenvalues of $A$,
 $\{\ket{i}\}$ is an orthonormal basis for
 $\mathcal{V}$, and each $\ket{i}$ is an eigenvector of $A$
 corresponding to eigenvalue $\lambda_{i}$. 
 {\em For proof see
 \cite{Nielsen-Chuang00} on page 72.}
} 

{\definition[Linear operator on a composed Hilbert space]
 Let $A$, $B$  be linear operators on $\mathcal{H}_{A}$,
 $\mathcal{H}_{B}$ respectively. The tensor product
 \begin{equation}
  A \otimes B = \left( \begin{array}{ccc}
     a_{1,1}B& \ldots& a_{1,m}B \\
       \vdots& \ddots& \vdots   \\
     a_{m,1}B& \ldots& a_{m,m}B
  \end{array} \right)
 \end{equation}
 is a linear operator on
 $\mathcal{H}_{A} \otimes \mathcal{H}_{B}$.
 \begin{equation}
  \label{operator_tensor_product}
  \left( A \otimes B \right)
  \left(\sum_{i,j}{\alpha_i\ket{e_{i}} \otimes
   \beta_j\ket{f_{j}}}\right)
  = \sum_{i,j}{\alpha_i \beta_j\bigl(A\ket{e_{i}} \otimes
   B\ket{f_{j}}\bigr)}.
\end{equation}
} 

\subsection{Operator functions}

{\definition[Operator function]
 Let $A=\sum_{i}{\lambda_{i}\ketbra{i}{i}}$ be a spectral
 decomposition for a normal operator $A$. An operator
 function $f$ on $A$ is defined as
 \begin{equation}
  \label{operator_function}
  f(A) = \sum_{i}{f(\lambda_{i})\,\ketbra{i}{i}}.
 \end{equation}
This allows us to define functions like square root,
logarithm or exponential for operators.
} 

{\definition[Trace of a matrix] The trace of a square matrix
 $A$ is defined as
 \begin{equation}
  \label{trace}
  \tr(A) = \sum_{i}{a_{ii}}.
 \end{equation} } 

\nicebox{
{\note Let $A$ and $B$ be linear operators and
 $\eta \in \mathbb{C}$. Then
 \begin{equation}
  \label{trace_cyclic_property}
  \bullet\hspace{2mm}\hspace{1mm}\tr(AB)=\tr(BA),
  \hspace{4.87cm} \mbox{(cyclic property)},
 \end{equation}
 \begin{equation}
  \label{trace_linear_property}
  \bullet\hspace{2mm}\begin{array}{l}
  \tr(A+B)=\tr(A)+\tr(B),\\
  \tr(\eta A)=\eta \, \tr(A),
  \end{array} \hspace{3.2cm} \mbox{(linear property)}.
 \end{equation}
}} 
From the cyclic property it follows that the trace of a
square matrix $A$ is invariant under the unitary similarity
transformation $A \rightarrow UAU^{\dagger}$.
\begin{equation}
  \tr\left(UAU^{\dagger}\right)
 =\tr\left(UU^{\dagger}A\right)
 =\tr\left(A\right)
\end{equation}
This property is important for the density operator formalism.
See Section \ref{sec:density-operator-formalism} on page
\pageref{sec:density-operator-formalism}.

\section{The postulates of quantum mechanics}
\label{def:postulates}

{\postulate[State space]
 Associated to any isolated physical system is a complex
 vector space with inner product (Hilbert space) known as
 the state space of the system. The state of the system 
 is completely described by its state vector, which is 
 a unit vector in the system's state space.
} 

{\postulate[Evolution] 
 The time evolution of the state of a closed quantum system
 is described by the Schr\"{o}dinger equation
 \begin{equation}
  \label{schrodinger-equation}
  i\hbar\frac{\partial}{\partial t}\ket{\upsilon}
  =H\ket{\upsilon},
 \end{equation}
 where $\hbar$ is the reduced Planck constant
 $\hbar \approx 1.05457\cdot 10^{-34}$Js and $H$ is a fixed
 self-adjoint operator (hermitian) known as the Hamiltonian
 of the closed system.
} 

In practice, it is convenient to absorb $\hbar$ into $H$,
effectively setting $\hbar=1$. The solution of the equation
\eqref{schrodinger-equation} constitutes a time evolution 
operator
\begin{equation}
 \label{SE-solution}
 U(\Delta t)=e^{-iH\Delta t}.
\end{equation}
The time evolution operator $U(\Delta t)$ is a unitary operator
since the Hamiltonian $H$ is a hermitian matrix, 
$H=H^{\dagger}$, and therefore the unitarity condition is
fulfilled,
$$
 U(\Delta t)U(\Delta t)^{\dagger}
 =e^{-iH\Delta t}\; e^{iH\Delta t}=I.
$$

We can reformulate the second postulate using unitary
evolution:

\textit{
 The time evolution of a closed quantum system from
 the state $\ket{\upsilon_1}$ at time $t_{1}$ to the state
 $\ket{\upsilon_2}$ at time $t_{2}$ is described by a
 unitary operator $U=U(t_{2}-t_{1})$,}
\begin{equation}
 \label{state_vector_evolution}
 \ket{\upsilon_2}=U\ket{\upsilon_1}.
\end{equation}

The correspondence between the discrete-time description
of dynamics using unitary operators and continuous time
description using Hamiltonians is one-to-one.

{\postulate[Quantum measurement]
 A quantum measurement is described by a collection
 $\{M_{m}\}$ of measurement operators. These are operators
 acting on the state space of the system being measured. 
 The index $m$ refers to the measurement outcome $r_m$ that
 may occur in the experiment. Measuring the system state
 $\ket{\upsilon}$ will give the result $r_m$ with probability
 \begin{equation}
  \label{state_vector_result_probability}
  p(r_m)=\bra{\upsilon}M_{m}^{\dagger}M_{m}\ket{\upsilon},
 \end{equation}
 and the state of the system reduces to the post-measurement
 state
 \begin{equation}
  \label{state_vector_collapse}
  \ket{\upsilon^{'}}
  =\frac{M_{m}\ket{\upsilon}}{\sqrt{p(r_m)}}.
 \end{equation}
 The operators $\{M_{m}\}$ satisfy the completeness equation
 \begin{equation}
  \sum_{m}{M_{m}^{\dagger}M_{m}}=I.
 \end{equation}
}

Using the completeness equation together with the normalization
condition, we can see that probabilities sum to one:
 \begin{equation}
  \label{measurement_operators_completness}
  1=\sum_{m}{p(r_m)}
   =\sum_{m}{\bra{\upsilon}M_{m}^{\dagger}M_{m}\ket{\upsilon}}
   .
\end{equation}

\nicebox{
{\note
 We say that states $\ket{\upsilon}$ and $\ket{\psi}$ are 
 equivalent, $\ket{\upsilon} \cong \ket{\psi}$, up to the
 global phase factor, iff
 $$
  \ket{\upsilon}=e^{i\Phi}\ket{\psi},
  \quad
  \Phi \in \mathbb{R}\,.
 $$
 The statistics of measurement
 predicted for these two states are the same!
 $$
  \label{global_phase}
  \bra{\upsilon}M_{m}^{\dagger}M_{m}\ket{\upsilon}
  =\bra{\psi}
   e^{-i\Phi}
   M_{m}^{\dagger}M_{m}
   e^{i\Phi}\ket{\psi}
  =\bra{\psi}
   M_{m}^{\dagger}M_{m}
   \ket{\psi}
 $$
}} 

The global phase must not be confused with the
relative phase. Let $\ket{0}, \ket{1}$ form a basis of
a Hilbert space $\mathcal{H}$. States
$\ket{\upsilon}=\alpha\ket{0}+\beta\ket{1}$ and
$\ket{\psi}=\alpha\ket{0}-\beta\ket{1}$ defined
on $\mathcal{H}$ do differ by a relative phase.

\subsubsection*{Projective measurement}
For many applications a special class of measurements
known as projective measurement is of importance.
A projective measurement is described by a self-adjoint
operator $M$, called an {\bf observable}, with a spectral
decomposition 
\begin{subequations}
\begin{equation}
 M=\sum_{m}{r_m\,P_{m}},
\end{equation}
where\\
\begin{equation}
 \sum_{m}{P_{m}}=I \qquad \mbox{ and }
 \qquad P_{m}P_{n}=\delta_{mn}P_{m}.
\end{equation}
\end{subequations}
$P_{m}$ is an orthogonal projector onto the eigenspace of
$M$ with eigenvalue $r_m$.
The eigenvalues $r_m$ correspond to possible outcomes of
the measurement. The probability of getting result $r_m$
and afterward state's collapse are given by
\begin{equation}
 \label{state_vector_projective_result_probability}
 p(r_m)=\bra{\upsilon}P_{m}\ket{\upsilon},
\end{equation}
\begin{equation}
 \label{state_vector_projective_collapse}
  \ket{\upsilon^{'}}
 =\frac{P_{m}\ket{\upsilon}}{\sqrt{p(r_m)}}.
\end{equation}

A projective measurement allows to express the measurement
\begin{center}
 {\it in an orthonormal basis $\ket{m}$}
\end{center}
simply by defining
\begin{center}
 $P_m=\ketbra{m}{m}$.
\end{center}

The corresponding observable $M$ is then given by a list of
projectors $P_m$. Actually, projective measurement is
equivalent to the general measurement postulate when
augmented with the ability to perform unitary
transformations. See \cite{Nielsen-Chuang00} for details.
Unless
stated otherwise, projective measurement is the default
measurement used in quantum computing.

\nicebox{
{\note The average value of a projective measurement is
\begin{equation}
 \label{state_vector_everage_value}
 E(M)=\sum_{m}{r_m \, p(r_m)} 
     =\sum_{m}{r_m\bra{\upsilon}P_{m}\ket{\upsilon}}
     =\bra{\upsilon}\left(\sum_{m}{r_m \, P_{m}}\right)
      \ket{\upsilon}=\bra{\upsilon}M\ket{\upsilon}.
\end{equation}
The average value $E(M)$ is often denoted by 
$\langle M \rangle$.
}} 

{\postulate[Composite systems]
 The state space $\mathcal{H}$ of a composite physical
 system is the tensor product of the state spaces
 $\mathcal{H}_{i}$ of its components,
 $$\mathcal{H}=\bigotimes_{i}\mathcal{H}_{i}\;.$$
 Moreover, if the subsystems are in the states
 $\ket{\upsilon_{i}} \in \mathcal{H}_{i}$, then the joint
 state $\ket{\upsilon} \in \mathcal{H}$ of the total
 system is
 $$\ket{\upsilon}=\bigotimes_{i}\ket{\upsilon_{i}}\,.$$
}

\nicebox{
{\note[Entanglement]
 Let $\mathcal{H}=\mathcal{H}_{A}
 \otimes \mathcal{H}_{B}$ be a Hilbert space. A joint state
 $\ket{\upsilon} \in \mathcal{H}$ that cannot be written as
 tensor product of some vectors 
 $\ket{\upsilon_{A}} \in \mathcal{H}_{A}$,
 $\ket{\upsilon_{B}} \in \mathcal{H}_{B}$
 is said to be entangled, otherwise we call this joint
 state a product state. The volume of entangled states
 is much higher than the volume of product states
 [Aubrun, Szarek].
}}

In entangled states, unitary operators and measurement
performed on one system affect the state of the second
system. In product states, these operations affect only
the state of the target component.

\section{Density operator formalism}
\label{sec:density-operator-formalism}
Next to the state vectors formalism, there exists an
alternative density operator (density matrix) formalism.
The postulates of quantum mechanics can be equivalently
written using density operators. Density operator
formalism is of advantage when describing individual
subsystems of a composite quantum system or dealing with
quantum systems whose state is not completely known.

{\definition[Density operator]
 Let a quantum system $\mathcal{S}$ with associated
 Hilbert space $\mathcal{H}$ be at a state
 $\ket{\upsilon_{i}} \in \mathcal{H}$ with probability
 $p_{i}$. The density operator for the system described
 by an ensemble $\{p_{i}, \ket{\upsilon_{i}}\}$ is defined
 as
 \begin{equation}
  \label{density_operator}
  \rho = \sum_{i}{p_{i}\ketbra{\upsilon_{i}}{\upsilon_{i}}}.
 \end{equation}
A density operator satisfies the conditions:
\begin{enumerate}
 \item $\tr(\rho)=1$,
 \item $\rho$ is a normal operator and its spectrum consists
 of positive real numbers.
\end{enumerate}\sq
} 

A quantum state represented by a density operator $\rho$
is said to be a {\bf pure state} iff
\begin{equation}
 \label{purity_criterion}
 \tr(\rho^{2})=1.
\end{equation}
Otherwise the state is said to be {\bf mixed}. For a pure
state
described by a state vector $\ket{\upsilon}$ the equation
(\ref{density_operator}) reduces to
\begin{equation}
 \rho=\ketbra{\upsilon}{\upsilon}.
\end{equation}

Using the density operator formalism, equation
(\ref{state_vector_evolution}) for temporal
unitary evolution of a closed quantum system has the form
\begin{equation}
 \label{density_operator_evolution}
 \rho'=U \rho\, U^{\dagger}.
\end{equation}
This can be easily seen from a transformation
\begin{equation}
 \sum_{i}{p_{i}\ketbra{\upsilon_{i}}{\upsilon_{i}}}
 \stackrel{\rm U}{\longrightarrow} \sum_{i}{p_{i}
 U \ketbra{\upsilon_{i}}{\upsilon_{i}} U^{\dagger}}.
\end{equation}
Equations (\ref{state_vector_result_probability}),
(\ref{state_vector_collapse}) for the $3^{rd}$ postulate
of quantum mechanics have the form
\begin{equation}
 \label{density_operator_result_probability}
 p(r_m)=\tr\left(M_{m}^{\dagger}M_{m}\rho\right),
\end{equation}
\begin{equation}
 \label{density_operator_collapse}
 \rho'=\frac{M_{m} \rho M_{m}^{\dagger}}{p(r_m)}.
\end{equation}
For composite systems described by the $4^{th}$ postulate
where the individual components are in the states
$\rho_{i}$, the joint state of the total system is
\begin{equation}
 \rho=\bigotimes_{i}\rho_{i}.
\end{equation}

\subsubsection*{The reduced density operator}
When we deal with a subsystem of a larger system
$\mathcal{S}$ whose state is described by a density
operator $\rho$, we need to find a function which will
provide the correct measurement statistics for this
subsystem. Such a function is a partial trace function
and the provided statistics is called a reduced density
operator. It can be shown that a partial trace is the
unique function with the above written property.

{\definition[Partial trace]
 Let $\rho_{A}$ and $\rho_{B}$ be a density operator
 of a system $A$ and $B$, respectively. The partial
 trace over system $B$ is defined by
 \begin{equation}
  \label{partial-trace}
  \tr_{B}\left( \rho_{A} \otimes \rho_{B} \right)
  = \rho_{A} \tr\left( \rho_{B} \right).
 \end{equation}
} 

{\definition[Reduced density operator]
 Let $\rho_{AB}$ be a density operator describing a composed
 state of physical systems $A$ and $B$. The reduced
 density operator $\rho_{A}$ for system A is defined by
\begin{equation}
 \label{reduced-density-operator}
 \rho_{A} = \tr_{B}\left( \rho_{AB} \right).
\end{equation}
} 

{\example[Tracing out]
Let $\rho_{AB}$ be a four-by-four density operator
describing a composed state of two two-dimensional systems
$A$ and $B$. According to the definition, the reduced
density operator $\rho_A$ and $\rho_B$ are calculated by
partial trace over system $B$ and $A$, respectively. 
There are two practical approaches how to perform this
tracing out.

One approach follows directly from the definition of the
partial trace. To be notationally clear, with respect to
the state labelling, let us write
\begin{align*}
 \rho_A &
 =\sum_{\phi,\phi' \in \{0,1\}}
  \alpha_{\phi,\phi'}\ketbra{\phi}{\phi'},\\
 \rho_B &
 =\sum_{\psi,\psi' \in \{0,1\}}
  \beta_{\psi,\psi'}\ketbra{\psi}{\psi'},
\end{align*}
then
\begin{align*}
 \rho_A \otimes \rho_B
 &=\sum_{\phi,\phi' \in \{0,1\}}
   \alpha_{\phi,\phi'} \, \ketbra{\phi}{\phi'}
   \otimes
   \rho_B \\
 &=\left(\begin{array}{cc}
  \alpha_{0,0}\,\rho_B & \alpha_{0,1}\,\rho_B \\
  \alpha_{1,0}\,\rho_B & \alpha_{1,1}\,\rho_B
  \end{array}\right) \\
 &=\sum_{\phi,\phi' \in \{0,1\}}
  \alpha_{\phi,\phi'} \, \ketbra{\phi}{\phi'}\;
  \otimes 
  \sum_{\psi,\psi' \in \{0,1\}}
  \beta_{\psi,\psi'}\ketbra{\psi}{\psi'} \\
 &=\sum_{\phi,\phi',\psi,\psi' \in \{0,1\}}
   \alpha_{\phi,\phi'}\beta_{\psi,\psi'}\,
   \ket{\phi}\ket{\psi}\bra{\phi'}\bra{\psi'}\\
 &=\left(\begin{array}{cccc}
   \alpha_{0,0}\beta_{0,0} & \alpha_{0,0}\beta_{0,1} &
   \alpha_{0,1}\beta_{0,0} & \alpha_{0,1}\beta_{0,1} \\
   \alpha_{0,0}\beta_{1,0} & \alpha_{0,0}\beta_{1,1} &
   \alpha_{0,1}\beta_{1,0} & \alpha_{0,1}\beta_{1,1} \\
   \alpha_{1,0}\beta_{0,0} & \alpha_{1,0}\beta_{0,1} &
   \alpha_{1,1}\beta_{0,0} & \alpha_{1,1}\beta_{0,1} \\
   \alpha_{1,0}\beta_{1,0} & \alpha_{1,0}\beta_{1,1} &
   \alpha_{1,1}\beta_{1,0} & \alpha_{1,1}\beta_{1,1}
   \end{array}\right) \\
 &=\sum_{\phi,\psi,\phi',\psi' \in \{0,1\}}
   x_{\phi,\psi,\phi',\psi'}\,
   \ket{\phi,\psi}\bra{\phi',\psi'}\\
 &=\sum_{\gamma,\gamma'=0}^3
   x_{\gamma,\gamma'} \ket{\gamma}\bra{\gamma'}\\
 &=\left(\begin{array}{cccc}
 x_{0,0} & x_{0,1} & x_{0,2} & x_{0,3} \\
 x_{1,0} & x_{1,1} & x_{1,2} & x_{1,3} \\
 x_{2,0} & x_{2,1} & x_{2,2} & x_{2,3} \\
 x_{3,0} & x_{3,1} & x_{3,2} & x_{3,3}
 \end{array}\right).
\end{align*}

Now we can see that the trace of the upper-left two-by-two
submatrix gives us 
$$
 \tr\left(\begin{array}{cc}
   x_{0,0} & x_{0,1} \\
   x_{1,0} & x_{1,1}
 \end{array}\right)
 = \tr(\alpha_{0,0}\;\rho_B)=\alpha_{0,0}=x_{0,0}+x_{1,1},
$$
since $\tr(\rho_B)$ must be equal to one. System $B$ is
traced out. Accordingly, we can take a submatrix
consisting of items on positions related to $\beta_{0,0}$
and trace out system $A$,
$$
 \tr\left(\begin{array}{cc}
 x_{0,0} & x_{0,2} \\
 x_{2,0} & x_{2,2}
 \end{array}\right)
 = \tr(\beta_{0,0}\;\rho_A)=\beta_{0,0}=x_{0,0}+x_{2,2}.
$$

Thus the reduced density operator for the system $A$ is
\begin{equation}
\begin{split}
\rho_A=\tr_B(\rho_{AB})
 =\tr_{B}\left(\begin{array}{cccc}\nonumber
 x_{0,0} & x_{0,1} & x_{0,2} & x_{0,3} \\
 x_{1,0} & x_{1,1} & x_{1,2} & x_{1,3} \\
 x_{2,0} & x_{2,1} & x_{2,2} & x_{2,3} \\
 x_{3,0} & x_{3,1} & x_{3,2} & x_{3,3}
 \end{array}\right)
 &=\left(\begin{array}{cc}
  \tr\left(\begin{array}{cc}
   x_{0,0} & x_{0,1} \\
   x_{1,0} & x_{1,1}
  \end{array}\right) &
  \tr\left(\begin{array}{cc}
   x_{0,2} & x_{0,3} \\
   x_{1,2} & x_{1,3}
  \end{array}\right) \\[5mm]
  \tr\left(\begin{array}{cc}
   x_{2,0} & x_{2,1} \\
   x_{3,0} & x_{3,1}
  \end{array}\right) &
  \tr\left(\begin{array}{cc}
   x_{2,2} & x_{2,3} \\
   x_{3,2} & x_{3,3}
  \end{array}\right)
 \end{array}\right)\\[5mm]
 &=\left(\begin{array}{cc}
   x_{0,0} + x_{1,1} \;& x_{0,2} + x_{1,3}\\[3mm]
   x_{2,0} + x_{3,1} \;& x_{2,2} + x_{3,3}
   \end{array}\right).
\end{split}
\end{equation}
The reduced density operator for the system $B$ is
\begin{equation}
\begin{split}
\rho_B=\tr_A(\rho_{AB})
 =\tr_A\left(\begin{array}{cccc}\nonumber
 x_{0,0} & x_{0,1} & x_{0,2} & x_{0,3} \\
 x_{1,0} & x_{1,1} & x_{1,2} & x_{1,3} \\
 x_{2,0} & x_{2,1} & x_{2,2} & x_{2,3} \\
 x_{3,0} & x_{3,1} & x_{3,2} & x_{3,3}
 \end{array}\right)
 &=\left(\begin{array}{cc}
   \tr\left(\begin{array}{cc}
   x_{0,0} & x_{0,2} \\
   x_{2,0} & x_{2,2}
  \end{array}\right) &
  \tr\left(\begin{array}{cc}
   x_{0,1} & x_{0,3} \\
   x_{2,1} & x_{2,3}
  \end{array}\right) \\[5mm]
  \tr\left(\begin{array}{cc}
   x_{1,0} &  x_{1,2} \\
   x_{3,0} & x_{3,2}
  \end{array}\right) &
  \tr\left(\begin{array}{cc}
   x_{1,1} & x_{1,3} \\
   x_{3,1} & x_{3,3}
  \end{array}\right)
 \end{array}\right)\\[5mm]
 &=\left(\begin{array}{cc}
   x_{0,0} + x_{2,2} \;& x_{0,1} + x_{2,3}\\[3mm]
   x_{1,0} + x_{3,2} \;& x_{1,1} + x_{3,3}
\end{array}\right).
\end{split}
\end{equation}

A different approach involves direct formulas for tracing
out. These formulas are derived from the trace
definition \eqref{trace} and the partial trace definition
\eqref{partial-trace},
\begin{equation}
 \rho_A = \tr_B(\rho_{AB})
 = \sum_{\phi,\phi' \in \{0,1\}}
 	             \ket{\phi}\;
 	             \Bigl( \sum_{\psi \in \{0,1\}}
                     \bra{\phi, \psi}
		     \;\rho_{AB}\;
		     \ket{\phi',\psi}
		     \Bigr)\;
		     \bra{\phi'},
\end{equation}
\begin{equation}
 \rho_B = \tr_A(\rho_{AB})
 = \sum_{\psi,\psi' \in \{0,1\}}
 	             \ket{\psi}\; 
 	             \Bigl( \sum_{\phi \in \{0,1\}}
                     \bra{\phi, \psi}
		     \;\rho_{AB}\;
		     \ket{\phi,\psi'}
		     \Bigr)\;
		     \bra{\psi'}.
\end{equation}
An insight into these formulas gives a simple rule
how to calculate a partial trace over a given system.
Let $\rho_{AB}$ be written in the form of
$$
 \rho_{AB}
 =\sum x_{\phi,\psi,\phi,\phi'}\,
  \ket{\phi,\psi}\bra{\phi',\psi'},
$$
then the partial trace over the $i$-th system (qubit) is
evaluated by replacing an outer product term
$\ket{x}\bra{y}$ for this system with a corresponding 
inner product $\braket{x}{y}$.

\vspace{1mm}
For example, let 
$\rho_{AB}
 =x_{0,1,0,0}\;\ket{01}\bra{00}
  \;\,+\;\,
  x_{0,1,1,1}\;\ket{01}\bra{11}
  \;\,+\;\, \cdots\;.$ Then,\\[3mm]
$\tr_B(\rho_{AB})
 =(\,x_{0,1,0,0}\;\ket{0}\bra{0}\;\underset{=0}{\underbrace{\braket{1}{0}}}
  \;\,+\;\,
  x_{0,1,1,1}\;\ket{0}\bra{1}\;\underset{=1}{\underbrace{\braket{1}{1}}}
  \;\,+\;\, \cdots )\;
 =\;x_{0,1,1,1}\; \ket{0}\bra{1} \;\,+\;\, \cdots\;,$

$\tr_A(\rho_{AB})
 =(\,x_{0,1,0,0}\;\ket{1}\bra{0}\;\underset{=1}{\underbrace{\braket{0}{0}}}
  \;\,+\;\, x_{0,1,1,1}\;\ket{1}\bra{1}\;\underset{=0}{\underbrace{\braket{0}{1}}}
  \;\,+\;\, \cdots )\;
 =\;x_{0,1,0,0}\; \ket{1}\bra{0} \;\,+\;\, \cdots\;.$\sq
} 

 \chapter{Quantum computing}
 \label{chap:quantum-computing}


\section{Quantum computing}                                





Quantum computing is an abstract concept of computation
defined within the framework of quantum mechanics. The
postulates of quantum mechanics relate the set of allowed
states to unit vectors in a complex Hilbert space, allowed
transitions are described by unitary matrices and the
measurement reveals complex amplitudes only
indirectly via the corresponding probabilities. For the
concept of quantum computing, it is convenient to adopt
these rules in the following way.

\subsection{Quantum states}
The smallest nontrivial quantum mechanical system is a
two-state system. In the context of quantum computing,
it is called a quantum bit, or a qubit for short.
Examples are two distinct polarization states of a photon,
or the two spin directions of an electron in a magnetic
field. The associated state space is a two-dimensional
complex Hilbert space $\mathcal{H}_2$ which is spanned by
an orthonormal basis
$\mathcal{B}=\{\ket{0}, \ket{1}\}$ labeled in some
canonical way. Labeling \ket{0} and \ket{1} refers to the
standard (computational) basis defined as:
\begin{equation}
 \hspace{5cm}
 \ket{0}=\left(\begin{array}{c}1\\0\end{array}\right),
 \quad
 \ket{1}=\left(\begin{array}{c}0\\1\end{array}\right).
 \hspace{1.3cm}\text{(standard basis)}
\end{equation}
An arbitrary qubit state $\ket{\upsilon}$ can be written as
a normalized linear combination (superposition) of basis
states
\begin{equation}
  \ket{\upsilon}=\alpha\ket{0} + \beta\ket{1} =
  \left(\begin{array}{c}\alpha\\ \beta\end{array}\right),
  \qquad \alpha,\beta \in \mathbb{C},
  \mbox{ with }\, \abs{\alpha}^{2}+\abs{\beta}^{2}=1.
  \hspace{1cm}\text{(qubit)}
\end{equation}
 The complex
 numbers $\alpha,\beta$ are so-called quantum mechanical
 amplitudes.
 The condition
 $$\abs{\alpha}^{2}+\abs{\beta}^{2}=1$$ follows
 from the $2$-norm unity condition $\norm{\upsilon}_2=1.$
 We have
 $$
  \norm{\upsilon}_2=\sqrt{\braket{\upsilon}{\upsilon}}=1
  \quad \Rightarrow \quad \braket{\upsilon}{\upsilon}=1,
 $$
 and for the standard basis we directly get
 $$
  \braket{\upsilon}{\upsilon}=\left(\alpha^*\;\beta^*\right)
  \left(\begin{array}{c}\alpha\\ \beta\end{array}\right)
  =\alpha\alpha^* + \beta\beta^*
  =\abs{\alpha}^2+\abs{\beta}^2=1\,.
 $$
 For any other orthonormal basis, e.g.
 \begin{equation}
  \hspace{3.6cm}
  \ket{0'}=\frac{1}{\sqrt{2}}
  \left(\begin{array}{r}1\\1\end{array}\right),\quad
  \ket{1'}=\frac{1}{\sqrt{2}}
  \left(\begin{array}{r}1\\-1\end{array}\right),
  \hspace{1.6cm} \text{(dual basis)}
 \end{equation}
 the same is true.

In principle,
it is also possible to work with three-state systems
(qutrits) or general $d$-state systems (qudits), but there
is no actual advantage of doing so, as well as for classical
trits/dits. For this reason, we can restrict ourselves to
qubits and larger systems composed out of qubits.
The joint state of a composed system is given by the tensor
product of its individual subsystem states. Let us have two
qubits $\ket{\upsilon},\ket{\psi} \in \mathcal{H}_{2}$,
$\ket{\upsilon}=\alpha\ket{0}+\beta\ket{1}$,
$\ket{\psi}=\gamma\ket{0}+\delta\ket{1}$.
Their tensor product is equal to
$$
 \ket{\upsilon}\ket{\psi}
 = \alpha\gamma\ket{0}\ket{0} + \alpha\delta\ket{0}\ket{1}
  +\beta\gamma \ket{1}\ket{0} + \beta\gamma\ket{1}\ket{1},
$$
and we can rewrite this equation in the form
$$
 \ket{\upsilon}\ket{\psi}
 =  \alpha_{00}\ket{00} + \alpha_{01}\ket{01}
  + \alpha_{10}\ket{10} + \alpha_{11}\ket{11}
 =  \sum_{i \in \{0, 1\}^{2}}{\alpha_{i}\ket{i}}.
$$
States $\ket{00}, \ket{01}, \ket{10}, \ket{11}$ are the
standard basis vectors of the Hilbert space
$\mathcal{H}_{2^2}$.
$$
 \ket{00}=\left(\begin{array}{r}1\\0\end{array}\right)
 \otimes \left(\begin{array}{r}1\\0\end{array}\right)
 = \left(\begin{array}{r}1\\0\\0\\0\end{array}\right), \quad
 \ket{01}=\left(\begin{array}{r}1\\0\end{array}\right)
 \otimes \left(\begin{array}{r}0\\1\end{array}\right)
 = \left(\begin{array}{r}0\\1\\0\\0\end{array}\right),
$$
$$
 \ket{10}=\left(\begin{array}{r}0\\1\end{array}\right)
 \otimes \left(\begin{array}{r}1\\0\end{array}\right)
 = \left(\begin{array}{r}0\\0\\1\\0\end{array}\right), \quad
 \ket{11}=\left(\begin{array}{r}0\\1\end{array}\right)
 \otimes \left(\begin{array}{r}0\\1\end{array}\right)
 = \left(\begin{array}{r}0\\0\\0\\1\end{array}\right).
$$
Labels $00, 01, 10, 11$ may be renamed to $0, 1, 2, 3$
because they can be easily seen as a binary representation
of these integers. Using this representation,
we write
$$
  \ket{\upsilon}\ket{\psi}=\alpha_{0}\ket{0} +
  \alpha_{1}\ket{1} + \alpha_{2}\ket{2} + \alpha_{3}\ket{3}
 =\sum_{i=0}^{3}{\alpha_{i}\ket{i}}.
$$
A general state of an $n$-qubit system (often called a
quantum register), $\ket{\psi} \in \mathcal{H}_{2^{n}}$,
is described as
\begin{equation}
 \hspace{2cm}
 \label{n-qubit}
 \ket{\psi}=\sum_{i \in \{0,1\}^{n}}\alpha_{i}\ket{i}
 \;\;=\;\;
 \sum_{i=0}^{2^{n}-1}{\alpha_{i}\ket{i}},
 \hspace{1.2cm}\text{(quantum register)}
\end{equation}
where $\sum_{i}{\abs{\alpha_{i}}^{2}}=1$ and $\{\ket{i}\}$
is the standard basis of an $2^{n}$-dimensional Hilbert
space $\mathcal{H}_{2^{n}}$.
Equation \eqref{n-qubit} shows  how a state space of
quantum system grows exponentially with its physical size
 - that is the number of the qubits.
It is due to the property of tensor product of Hilbert
spaces  (def. \ref{tensor-product-hilbert-spaces}) that
$$
  \dim(\mathcal{H}_{A} \otimes \mathcal{H}_{B})
 =\dim(\mathcal{H}_{A}) \cdot \dim(\mathcal{H}_{B})\,.
$$

\subsection{Measurement}
It may seem that an $n$-qubit register is capable
of storing exponentially more information than an $n$-bit
register. However, the Holevo theorem \cite{Holevo73}
states that one can retrieve faithfully only
$n$ \textbf{bits} from an $n$-qubit register. Using other
words, retrieving of classic information from a quantum
register requires measurement and by the $3^{rd}$ postulate,
we know that measurement destroys the superposition. In
particular, let $\mathcal{H}$ be a state space spanned
by basis $\mathcal{B}$, then the projective measurement
associated with some observable
$$
 M=\sum_m r_m P_m,
$$
where $P_m$ are orthogonal projectors
$$
 P_m=\ketbra{m}{m},
 \qquad \ket{m} \in \mathcal{B},
$$
projects the state $\ket{\psi} \in \mathcal{H}$,
$$
 \ket{\psi}=\sum_i{ \alpha_i \ket{i}},
 \qquad \ket{i} \in \mathcal{B},
$$
to the post-measurement state
$$
 \frac{P_m\ket{\psi}}{\sqrt{p(r_m)}}
 =\frac{1}{\abs{\alpha_m}}\sum_i{\alpha_i\ketbra{m}{m}i\rangle}
 =\frac{1}{\abs{\alpha_m}}\sum_{i}\alpha_m\ket{m}\delta_{mi}
 =\frac{\alpha_m}{\abs{\alpha_m}}\ket{m}\cong\ket{m},
$$
with probability
$$
 p(r_m)=\bra{\psi}P_m\ket{\psi}=
 \abs{\alpha_m}^2,
$$
where $r_m$ is the outcome of the measurement and
$\alpha_m (\abs{\alpha_m})^{-1}$ is an unobservable
global phase.
It is also possible to measure only a part of the register
at a time. For product states, the rest of the register
is kept untouched, for entangled states, the whole register
is affected.
{\example[Projective measurement]${\,}\quad{\;}$
\begin{description}
\item[Product state.]
Let $\ket{\psi} = \alpha\ket{000}+\beta\ket{001}+
\gamma\ket{010}+\delta\ket{011}\in
\mathcal{H}_{2^3}, \quad \alpha, \beta,
\gamma, \delta \neq 0$.
It is a product state since $\ket{\psi}=
\ket{0}\left(a\ket{0}+b\ket{1}\right)
\left(c\ket{0}+d\ket{1}\right)$ for some $a,b,c,d$
satisfying $\alpha=ac,\,
\beta~=~ad,\, \gamma~=~bc, \text{ and } \delta=bd$.
Measuring the middle qubit with respect to the
standard observable
$$
 M= \left(\begin{array}{rr}1&0\\0&-1\end{array}\right)
  = 1\cdot\left.\left(\begin{array}{rr}
    1&0\\0&0
    \end{array}\right)\right\vert_{P_0=\ketbra{0}{0}}
    \quad + \quad\; (-1)\cdot\left.\left(\begin{array}{rr}
    0&0\\0&1
    \end{array}\right)\right\vert_{P_1=\ketbra{1}{1}}
$$
yields the post-measurement state
$$
 \ket{\psi'}=\left\{\begin{array}{ll}
 c\ket{000}+d\ket{001},
 \quad &\mbox{ with probability } \; \abs{a}^2, \;\\
 &\mbox{ and the result of the measurement is } +1,\\
 c\ket{010}+d\ket{011},
 \quad &\mbox{ with probability } \; \abs{b}^2, \;\\
 &\mbox{ and the result of the measurement is } -1.
 \end{array}\right.
$$
The average value of this measurement is
$\langle M \rangle=\abs{a}^2-\abs{b}^2$. Results $+1$
and $-1$ are to be understood as observing $0$ and $1$,
respectively, in the measured register.

\item[Entangled state.]
Let $\ket{\psi}=\alpha\ket{00}+\beta\ket{11} \in
\mathcal{H}_{2^2}, \quad \alpha, \beta \neq 0.$
It is an entangled state since
$\ket{\psi}$ cannot be written as a tensor product of
two qubits
$$
  (a\ket{0}+b\ket{1})(c\ket{0}+d\ket{1})
 =ac\ket{00}+ad\ket{10}+bc\ket{10}+bd\ket{11},
$$
because
$$
 \left(\begin{array}{ccc}
  ac=\alpha & \Rightarrow & ac \neq 0\\
  bd=\beta  & \Rightarrow & bd \neq 0
 \end{array}\right)
 \Rightarrow
 \quad
  ad \neq 0, \quad
  bc \neq 0\; .
$$
Measuring the left qubit yields the post-measurement state
$$
 \ket{\psi'}=\left\{\begin{array}{ll}
 \ket{00},
 \quad \mbox{ with probability } \; \abs{\alpha}^2, \;
 \mbox{ and the result of the measurement is } +1,\\
 \ket{11},
 \quad \mbox{ with probability } \; \abs{\beta}^2,  \;
 \mbox{ and the result of the measurement is } -1.
 \end{array}\right.
$$
It can be seen that the measurement of the left qubit
uniquely determined the state of the right qubit.
\end{description}
}\sq
\subsubsection*{Classical vs. probabilistic vs. quantum register}
We can now compare a classical, probabilistic and quantum
register composed out of $n$ two-state systems.
\nicebox{
 \begin{center}
 \begin{tabular}{lrlll}
 Classical register:
 & $x$\;,\quad
 & $x \in \{0, 1\}^n$\;.\\[3mm] 
 Probabilistic register:
 & $\sum\limits_x p_x\, x$\;,\quad
 & $x \in \{0, 1\}^n$\;,\quad
 & $p_x \in \mathbb{R}^+$\;,\quad
 & $\sum\limits_x p_x = 1$\;.\\[3mm]
 Quantum register:
 & $\sum\limits_x \alpha_x \ket{x}$\;,\quad
 & $x \in \{0, 1\}^n$\;,\quad
 & $\alpha_x \in \mathbb{C}$\;,\quad
 & $\sum\limits_x \abs{\alpha_x}^2 = 1$\;.
 \end{tabular}
 \end{center}
}
A classical
$n$-bit register can store any string from the set
$\{0, 1\}^n$. Depending on the assigned semantic, these
strings can be interpreted as nonnegative integers
$\{0,\ldots, 2^n -1\}$, restricted set of real numbers, 
characters, and so on. The important point is that the
register contains only one such string at a time. This is
in contrast with a probabilistic register which
contains a probabilistic distribution of all possible
$n$-bit strings. These are exponentially many with respect
to the length $n$. However, only one string is observed
upon measurement. If we want to observe a different string,
we need to repeat the whole computation as the former
probabilistic distribution is no longer available nor it can
be cloned right before measurement for obvious reasons.
A quantum register can also contain a special distribution
of all possible $n$-bit strings. To each possible string is
assigned a complex amplitude and we talk about a quantum
superposition. After a measurement of all qubits, the
superposition is no longer available and we observe a single
string only. In this sense, a quantum register is very
similar to a probabilistic register.

\subsection{Evolution of quantum states}
However, a probabilistic and a quantum register differ in
the transitions they can undergo. An evolution of a quantum
register $\ket{\psi}$ is described
by a unitary matrix $U$, $\ket{\psi'}=U\ket{\psi}$. Since
for every unitary $U$ there exists its inversion
$U^{-1}=U^\dagger$, such an evolution is always reversible,
$U^{-1}\ket{\psi'}=U^\dagger U \ket{\psi}=\ket{\psi}$.
Models of classical computation are usually not reversible,
nevertheless, Bennett in \cite{Bennett73} showed the existence
of reversible universal Turing machines, and Fredkin and
Toffoli \cite{Fredkin-Toffoli82} showed a set of universal
classical
reversible gates for the circuit model. Therefore the
restriction to reversible computation should not be
considered as an obstacle.
In principle, for any computable function
$f: \{0, 1\}^n \rightarrow \{0, 1\}^m,$ 
there is a unitary matrix $U_f$ corresponding to the
reversible version of $f$. A quantum computer is at least
as powerful as a classical computer \cite{Benioff82}. 
The operator
$U_f$ for a nonreversible classical function $f$ is
typically implemented to perform a bijective mapping
\begin{equation}
 U_f : \ket{x,y} \longrightarrow \ket{x, y \oplus f(x)},
\end{equation}
where $x \in \{0, 1\}^n$, $y \in \{0, 1\}^m$ and $\oplus$
denotes the bitwise addition modulo 2 (equivalently, bitwise
exclusive disjunction). By setting the second register to
zero, $\ket{y}=\ket{0}=\ket{0^{(m)}}=\ket{0,\ldots,0}$,
we get a computation
\begin{equation}
 U_f\ket{x,0}=\ket{x,f(x)}.
\end{equation}
Classical reversible functions and quantum specific
transforms (inherently reversible) perform mapping
\begin{equation}
 U_f : \ket{x} \longrightarrow \ket{f(x)}.
\end{equation}

\begin{figure}[h!]
 \begin{center}
 \includegraphics[scale=0.8]{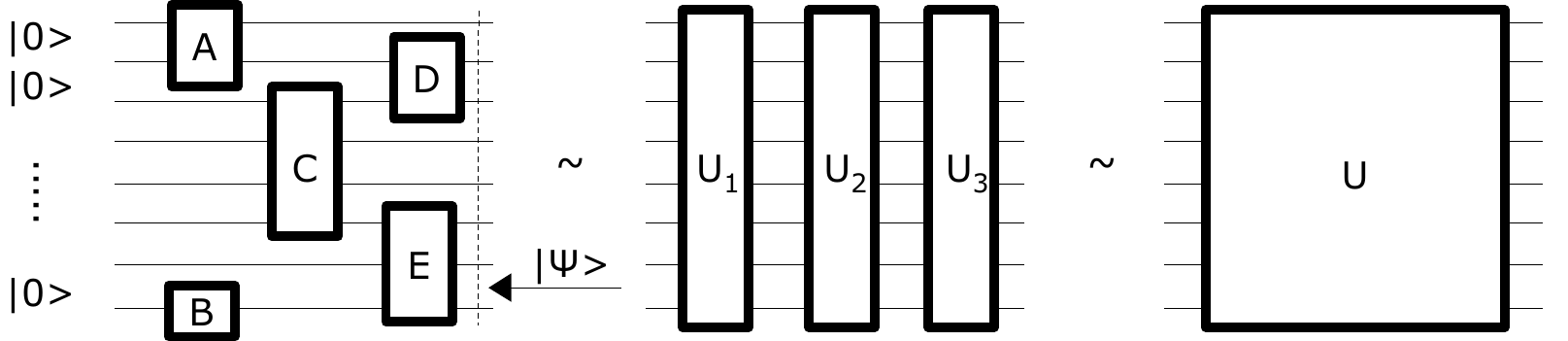}
 \caption[Graphical scheme of a quantum algorithm]
         {Graphical scheme of a quantum algorithm.}
 \label{quantum-diagram}
 \end{center}
\end{figure}

A quantum algorithm typically consists of several different
transforms applied to various quantum registers or
individual qubits. This can be graphically depicted as shown
in the Figure~\ref{quantum-diagram}. The figure
shows a procedure consisting from three evolutionary
steps $U_1, U_2, U_3$ applied over eight qubits to an
initial state $\ket{0}=\ket{0^{(8)}}$;
$\ket{\psi}=(U_3\cdot (U_2 \cdot (U_1\ket{0})))=U\ket{0}$.

The overall unitary operation at the first step has
the form $U_1=A^{(2)} \otimes \mathbb{I}^{(5)} \otimes
B^{(1)}$.
Analogically,
$U_2=\mathbb{I}^{(2)} \otimes C^{(4)} \otimes
\mathbb{I}^{(2)}, \text { and }
U_3=\mathbb{I}^{(1)}\otimes D^{(2)} \otimes
\mathbb{I}^{(2)} \otimes E^{(3)}$.
The upper index denotes the number of qubits over which
the transform is applied.

Examples of important unitary transforms over $n$ qubits
are the Hadamard transform and the discrete Fourier
transform. Note that $H^{(1)} = QFT^{(1)}$.
\begin{equation}
 \hspace{3cm}
 H^{(n)}:\quad
 \frac{1}{\sqrt{2^n}}
 \sum_{x, y=0}^{2^n - 1}
 (-1)^{x y} \ket{y}\bra{x}
 \hspace{1.89cm}
 \text{ (Hadamard transform) }
\end{equation}

\begin{equation}
 \hspace{.4cm}
 QFT^{(n)}:\quad
 \frac{1}{\sqrt{2^n}}
 \sum_{x, y=0}^{2^n -1 }
 e^{2 \pi i x y / 2^n} \ket{y}\bra{x}
 \hspace{1.5cm}
 \text{ (Fourier transform) }
 \label{qft-over-n-qubits}
\end{equation}

It is easy to see that these transforms are indeed unitary,
i.e. for the Hadamard transform, we have
\begin{align}
 H^{(n)} H^{(n)\dagger} 
 & = \frac{1}{\sqrt{2^n}}\frac{1}{\sqrt{2^n}}
 \sum_{x, y=0}^{2^n - 1}
 \underset{= 1}{\underbrace{(-1)^{xy} (-1)^{xy}}}
 \ket{y}\braket{x}{x}\bra{y} \nonumber \\[2mm] \nonumber
 &=\frac{1}{2^n}\sum_{ y=0}^{2^n -1}
 \ket{y}
 \underset{ =2^n}
          {\Bigl(
	   \underbrace{\sum_{x=0}^{2^n -1}\braket{x}{x}}
	   \Bigr)}
 \bra{y}
 = \sum_{ y=0}^{2^n -1 } \ketbra{y}{y} = \mathbb{I}^{(n)}.
\end{align}

The Hadamard transform is often used to create an equally
weighted superposition of standard basis states
\begin{align}
 H^{(n)}\ket{x}
 &=\frac{1}{\sqrt{2^n}}\sum_{y=0}^{2^n -1}
 (-1)^{xy}\ket{y},\qquad \ket{x},\ket{y} \in \mathcal{B},
 \\
 H^{(n)}\ket{0}
 &=\frac{1}{\sqrt{2^n}}\sum_{y=0}^{2^n -1}
 \ket{y}
  =\frac{1}{\sqrt{2^n}}
  (\ket{0}+\ket{1}+\ket{2}+\cdots+\ket{2^n-1}).
\end{align}

\paragraph{Quantum parallelism}
Since unitary operators are linear operators, they are
linear in its inputs. Let us prepare two quantum registers
in the following state
$$
 \ket{\psi}=(H^{(n)} \otimes \mathbb{I}^{(m)})
 \ket{\underset{n}{\underbrace{0,\ldots,0}} ,
      \underset{m}{\underbrace{0,\ldots,0}}}
 =\frac{1}{\sqrt{2^n}}\sum_{x=0}^{2^n -1}\ket{x,0},
$$
then a single application of an operator $U_f$ results in
a quantum parallel calculation of values $f(x)$ for all
$x \in \{0,\ldots, 2^n -1\}$.
\begin{equation}
 \ket{\psi'}=U_f\ket{\psi}
 =U_f\Bigl(\frac{1}{\sqrt{2^n}}\sum_{x=0}^{2^n -1}\ket{x,0}\Bigr)
 =\frac{1}{\sqrt{2^n}}\sum_{x=0}^{2^n -1}U_f\ket{x,0}
 =\frac{1}{\sqrt{2^n}}\sum_{x=0}^{2^n -1}\ket{x,f(x)}
\end{equation}
However, this quantum parallelism alone is not very useful.
A subsequent measurement would reveal exactly one pair
$(x, f(x))$
chosen at random. Such a process can be perfectly simulated
by a probabilistic computation. One idea might be to clone
the final superposition $\ket{\psi'}$ so that we can perform
additional
measurements, but the 'No cloning theorem' prohibits unknown
state cloning.

{\theorem[No cloning theorem]
 An unknown quantum state cannot be cloned. Namely, there is
 no unitary transformation performing evolution
\begin{equation}
 \ket{\chi}\ket{0}
 \stackrel{\rm U}{\longrightarrow}\ket{\chi}\ket{\chi}
 \end{equation}
 for any state $\ket{\chi}$.
} 

{\bf Proof.} Let us have two different orthonormal states
$\ket{\psi}$ and $\ket{\upsilon}$, and
$\ket{\chi}
  =\frac{1}{\sqrt{2}}(\ket{\psi} + \ket{\upsilon})$.
Assuming a 'cloning' U exists, we have
$U(\ket{\psi}\ket{0})=\ket{\psi}\ket{\psi}, \;\;
U(\ket{\upsilon}\ket{0})=\ket{\upsilon}\ket{\upsilon}$.

Additionally,
\begin{eqnarray}
 U(\ket{\chi}\ket{0})
 & = &
 U\left(\frac{1}{\sqrt{2}}
 (\ket{\psi} + \ket{\upsilon})\ket{0}\right)
 =\frac{1}{\sqrt{2}}
 (U(\ket{\psi}\ket{0}) + U(\ket{\upsilon}\ket{0}))
 =\frac{1}{\sqrt{2}}
 (\ket{\psi}\ket{\psi} + \ket{\upsilon}\ket{\upsilon})
 \nonumber \\
 & \neq &
 \ket{\chi}\ket{\chi}
 =\frac{1}{2}
 (\ket{\psi}\ket{\psi} + \ket{\psi}\ket{\upsilon}
 + \ket{\upsilon}\ket{\psi} + \ket{\psi}\ket{\psi}).
 \nonumber
\end{eqnarray}\sq
Note that the 'No cloning theorem' trivially holds for a
probabilistic register too.
Figure~\ref{probabilistic-cloning-matrix} shows a
hypothetical cloning matrix for a probabilistic one bit
register. The matrix is not a valid stochastic matrix since
its columns do not sum to one.
\begin{figure}[h!]
$$
 \left(\begin{array}{cccc}
 1& 0& .& .\\
 0& 1& .& .\\
 1& 0& .& .\\
 0& 1& .& .\\
 \end{array}\right)
 \cdot
 \left(\begin{array}{c}
 p_1 \\ p_2 \\ 0 \\ 0
 \end{array}\right)
 =
 \left(\begin{array}{c}
 p_1 \\ p_2 \\ p_1 \\ p_2
 \end{array}
 \right)
$$
\caption[Hypothetical cloning matrix]
        {A hypothetical cloning matrix for a probabilistic
         one bit register.}
\label{probabilistic-cloning-matrix}
\end{figure}

The only possible way to make use of quantum parallelism
is to employ quantum interference. Using interference
phenomena, we can modify the superposition, for example,
so that only a pair $(x, f(x))$ satisfying $f(x)=1$ is
measured with high probability. So far, we have no
indication that interference as present in quantum
mechanics can modify the superposition quickly enough so
that $NP$-complete problems can be solved in polynomial
time. Surprisingly, Abrams and Lloyd in \cite{Abrams-Lloyd98}
showed that
even small non-linearities, if allowed by quantum physics,
would enable for efficient $NP$-complete problems solving.

{\example[Grover amplitude amplification algorithm]
 Let us have a computable function
 $f: \{0, 1\}^n \rightarrow
 \{0, 1\}$ and a corresponding unitary operator $U_f$.
 We are looking for a string $x$ such
 that $f(x)=1$. An example of such problem is a Boolean
 satisfiability problem or a database search. An unsorted
 database search of size $N$ is especially interesting
 problem since there is a complete lack of inner structure
 to be exploited. Classically, we need to check all
 $N$ items, one by one, in order to find the right one in
 the worst case. On a quantum computer, we can use a
 different approach leading to a square root speed-up.

 Let $n=\lceil \log_2 N \rceil$. Using a weighted
 superposition of basis states over $n$ qubits,
 we can search through all
 $N$ item with a single query to $U_f$
 $$
  U_f  \Bigl( \frac{1}{\sqrt{2^n}}\sum_{x=0}^{2^n -1}
  \ket{x}\otimes\ket{0} \Bigr) =
  \frac{1}{\sqrt{2^n}}\sum_{x=0}^{2^n -1}
  \ket{x}\otimes\ket{0\oplus f(x)}\,.
 $$
 A small modification to the initial state of the second
 helps to change the computation in such a way that the
 values $f(x)$ are conveniently transfered to the
 amplitudes.
 \begin{align}
  U_f \Bigl(\frac{1}{\sqrt{2^n}}\sum_{x=0}^{2^n -1}
  \ket{x}\otimes\frac{1}{\sqrt{2}}(\ket{0}-\ket{1})\Bigr)
  = & \;
  \frac{1}{\sqrt{2^{n}}\sqrt{2}}\Bigl(\sum_{x=0}^{2^n -1}
  \ket{x}\ket{0\oplus f(x)}
  -
  \sum_{x=0}^{2^n -1}\ket{x}\ket{1\oplus f(x)}\Bigr)
  \nonumber \\ \nonumber
  = & \;
  \frac{1}{\sqrt{2^n}}\sum_{x=0}^{2^n -1}(-1)^{f(x)}
  \ket{x}\otimes\frac{1}{\sqrt{2}}(\ket{0}-\ket{1})\,.
 \end{align}
 Now, amplitude $-1/\sqrt{2^n}$ marks the item(s)
 with $f(x)=1$ and amplitude $1/\sqrt{2^n}$ marks the rest
 of items. The amplitude of the desired item can now be
 amplified using a transform known as inversion about mean.
 \begin{equation}
  \hspace{5cm}
   H^{(n)}\left(2\ketbra{0}{0}-\mathbb{I}^{(n)}\right)H^{(n)}
  \hspace{0.6cm}
  \text{(inversion about mean)}
 \end{equation}

 Inversion about mean applied to a general superposition
 produces transform
 \begin{align}
  &
  \Bigl(H^{(n)}
  \bigl(2\ketbra{0}{0}-\mathbb{I}^{(n)}\bigr)
  H^{(n)}\Bigr)
  \sum_{x=0}^{2^n -1} \alpha_x \ket{x}
  \; = \;
  \Bigl(\frac{2}{\sqrt{2^n}\sqrt{2^n}}
  \sum_{j,k=0}^{2^n -1}\ketbra{j}{k}
  \; - \;
  \mathbb{I}^{(n)}\Bigr)
  \sum_{x=0}^{2^n -1}\alpha_x \ket{x}
  \nonumber
  \\
  \nonumber
  = & \;
  \frac{2}{2^n}\sum_{j,k=0}^{2^n -1}
  \ketbra{j}{k}\sum_{x=0}^{2^n -1} \alpha_x \ket{x}
  \;\; - \;\;
  \mathbb{I}^{(n)}\sum_{x=0}^{2^n -1} \alpha_x \ket{x}
  \; = \;
  2\sum_{j=0}^{2^n -1}
  \ket{j}\cdot\underset{\text{ mean }\langle \alpha \rangle}
  {\underbrace{
  \Bigl(\sum_{x=0}^{2^n -1}
  \alpha_x\Bigr)\cdot\frac{1}{2^n}}}
  \;\; - \;\;
  \sum_{x=0}^{2^n-1} \alpha_x \ket{x}
  \\ \nonumber
  = &
  \underset{\text{may be relabelled}}
  {\underbrace{\sum_{j=0}^{2^n -1}
  2\langle \alpha \rangle \ket{j}}}
  \;\; - \;\;
  \sum_{x=0}^{2^n -1} \alpha_x \ket{x}
  \; = \;
  \sum_{x=0}^{2^n -1}
  \Bigl( -\alpha_x + 2\langle \alpha \rangle \Bigr) \ket{x}
  \;.
 \end{align}

 Repeated sequential action of the operator $U_f$ and
 inversion about
 mean is shown in the Figure~\ref{amplitude-amplification}.
 \begin{figure}[h!]
 \begin{center}
 \includegraphics[scale=0.8]
 {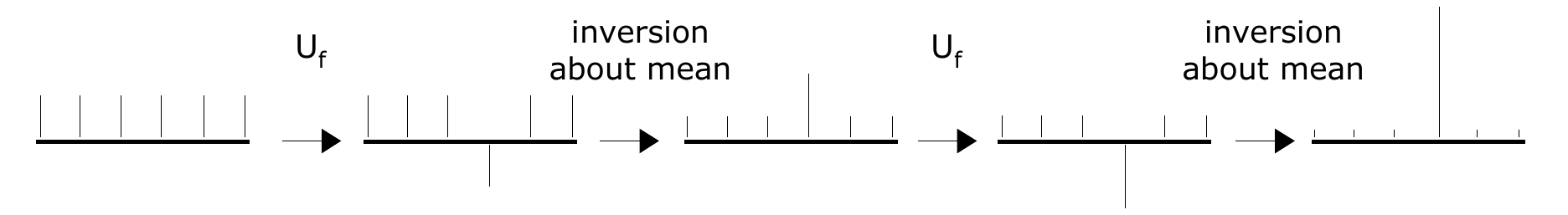}
 \end{center}
 \caption[Amplitude amplification by inversion about mean]
         {Amplitude amplification by inversion about mean.}
 \label{amplitude-amplification}
 \end{figure}

 It is clear that in order to stop the amplification at the
 right point one has to know at least approximately the
 number $M$ of items satisfying $f(x)=1$. The number of
 steps of amplification then grows like $O(\sqrt{N/M})$, see
 \cite{Grover97}. The number $M$ can be exactly determined
 by the counting algorithm \cite{Nielsen-Chuang00} if it is
 not roughly known by other means. 
 
 Recently, a similar type of amplification has been reported
 in energy transfer in plants during photosynthesis 
 \cite{Photosynthesis07}.
 \sq
}

A different example of a quantum algorithm is the period
finding algorithm. This algorithm is the only quantum part
of the Shor factoring algorithm \cite{Shor-factoring97}.
With respect to the
factoring, the task is to find the period $r$ of a function
$f_{k,N}(x)=k^x \pmod N$ for given coprimes $k$ and $N$.
The number $N$ corresponds to the number to be factored and
$k$ is chosen at random. Accidentally, if $k$ is not a
coprime to $N$, then $k$ is a factor of $N$. Once the period
$r$ is determined, a classical post-processing algorithm 
checks whether $r$ leads to a non-trivial factor of $N$.
This happens with high probability.

The period finding algorithm consists of three steps and
we need two $m$-qubit registers, where $m$ is chosen in the
order of $\Theta(\log(N^2))$. First,
we create a state with the period $r$ which we wish to
determine in the first register. We do it by exploiting
quantum parallelism and measurement on the second register.
The circuit implementing $U_f$ is essentially a reversible
version of a classical circuit for modular exponentiation
utilizing repeated squaring and requiring $O(m^3)$ universal
gates.

\begin{align}
 \sum_x \ket{x}\ket{0}
 \underset{\;\text{parallelism}\;}
          {{\overset{U_f}
	            {\;\longrightarrow\;}}}
 \sum_x \ket{x}\ket{k^x \hspace{-2mm}\mod N}
 \underset{\;\;\text{measurement}}
          {{\overset{\;\;\text{second register}}
	            {\;\longrightarrow\;}}} &
 \sum_{
  \{x' \; : \; y \,\equiv\, k^{x'} \hspace{-2mm}\mod N\}}
 \ket{x'}\ket{y}\nonumber \\ \nonumber
 &=\bigl( \ket{l} + \ket{l + r} + \ket{l+2r}+\cdots \bigr)
   \ket{y} \\ \nonumber
 &=\sum_j\ket{l+jr}\ket{y}
\end{align}

The result of the measurement is some non-negative integer
$y$. Now, the superposition in the first register starts
with a label $\ket{l}$, where $l$ is equal to the smallest
$x$ such that $y\equiv k^x\hspace{-1.4mm}\pmod N$.
We call $l$ an offset, $l \leq r$.
Note that in general $2^m$ is not an exact multiple of
the unknown $r$ and therefore  the amount of terms in the
sum $\sum_j \ket{l+jr}$ is not $2^m/r$ exactly. However, if
$2^m$ is sufficiently large, then the small decay of the
periodicity at $x$ close to $2^m$ will have a negligible
effect. Thus
$$
 \abs{\{j\}=\{0,1,2,\ldots\}} \approxeq \frac{2^m}{r}\,.
$$
Second, we apply the quantum Fourier transform to the
first register. Quantum Fourier transform over 
$\mathbb{Z}_{2^m}$ requires $O(m^2)$ gates. The Fourier
transform maps functions with period $r$ to functions
having non-zero values only at the amplitudes of the
frequency $1/r$. 

Thus we get rid of the offset and
neglecting states with negligible amplitudes the number
of terms in the superposition is strikingly decreased.
It is now proportional to $r$. There are exactly $r$ states
with amplitude peaks and other states with smaller
amplitudes are clustered around these peaks.
Therefore, a subsequent measurement on the first register
produces with high probability an $m$-bit non-negative
integer
$$
 a \approxeq j_a\frac{2^m}{r},
$$
where $j_a \in \{0, 1, 2, \ldots r-1\}$.
Third, we take a rational number $a/2^m$ and using a
continued fraction expansion we find a convergent which
resembles $j_a/r$. Thus we can extract $r$. The continued
fraction expansion can be computed using $O(m^3)$ gates
for elementary arithmetic.
$$
 \frac{a}{2^m}=0.\underset{m-\text{bits}}
                        {\underbrace{
			   a_{m-1} a_{m} \ldots a_0}}
 = \cfrac{1}{c_1 + \cfrac{1}{c_2 + \cfrac{1}{c_3 + \cdots}}}
 \Rightarrow \frac{j_a}{r}
$$

Choosing $m$ in the order of $\Theta(\log(N^2))$ ensures that
$r$ is revealed with high probability. A continued fraction
expansion is a very important part of this algorithm.

\section{Quantum Turing machines}
The first formal definition of a quantum Turing machine,
viewed as a quantum physical analogue of a probabilistic
Turing
machine, was given by D. Deutsch \cite{Deutsch85} in 1985.
The
existence of a fully universal quantum Turing machine which
can simulate any other quantum Turing machine with at most
polynomial slowdown was proven by Bernstein and Vazirani
\cite{Bernstein-Vazirani97} in 1993. Additionally, they
proved that there is an
oracle relative to which there is a language that can be
efficiently accepted by a quantum Turing machine but
cannot be efficiently accepted by a bounded-error
probabilistic Turing machine.  The problem which was
considered is called Recursive Fourier Sampling and the
proposed quantum algorithm gives a quasipolynomial speedup,
$O(n)$ versus $O(n^{\log{n}})$. This provides the first
{\it formal} evidence that a quantum Turing machine is
more powerful than a probabilistic Turing machine. 

Informal evidence is present in promise problems such
as the Deutsch-Jozsa problem \cite{Deutsch-Jozsa92}, the
Simon problem \cite{Simon97} or Shor's factoring algorithm
\cite{Shor-factoring97}. However, these
problems are only assumed to be hard, it is not proven.
For example, the Deutsch-Jozsa problem was later shown to be
in the complexity class BPP \cite{Bernstein-Vazirani97} and
thus efficiently
solvable on a probabilistic Turing machine. In this case,
the quantum algorithm for the Deutsch-Jozsa problem only
gives an advantage of deterministic approach instead of
a probabilistic one. Regarding the Simon promise problem,
its complexity was elaborated from the former expected
quantum polynomial time to the exact quantum polynomial time
\cite{Brassard-Hoyer97}. The Simon promise problem was the
first 'solid' problem where a quantum algorithm yields an
exponential speedup.

{\definition[Quantum Turing machine]
 A quantum Turing machine is a generalized Turing machine.
 The machine has a {\it finite} control 
 (finite set of states $Q$), possibly infinite tape, and a
 read/write head that recognizes a finite alphabet $\Sigma$.
 A transition function assigns to each transition a complex
 amplitude

 \begin{equation}
  \delta: Q \times \Sigma \times Q \times \Sigma \times
  \{L, R\} \rightarrow \mathbb{C}
 \end{equation}
 and the corresponding finite transition matrix is unitary.
 A particular configuration is observed with
 a probability equal to the squared absolute value of the
 amplitude associated with that configuration.
 \sq
}

The restriction to unitary matrices can be derived from the
need of preserving the inner product as only unit vectors
represent valid quantum states. A unitary matrix is the only
finite dimensional square matrix that preserves the inner
product.
The property of a unitary matrix $U^\dagger U=\mathbb{I}$
is equivalent to the assertion that column vectors
$e_i$ of $U$ are normalized and orthogonal;
\begin{center}
 for all
$i, j: \braket{e_i}{e_j}=\delta_{ij}$.
\end{center}
The normalization under the $2$-norm, $\norm{e_i}_2=1$,
is to be compared with the local probability condition
(Eq.~\ref{LPC}) for a probabilistic Turing machine. 
Columns of a stochastic matrix are normalized under the 
$1$-norm and for a probabilistic Turing machine this
ensures that a normalized linear combination of 
configurations is mapped to another normalized linear
combination of configurations (Eq~\ref{GPC}). However,
for a quantum Turing machine having normalized columns
of transition matrix is not sufficient to ensure that
normalized states are mapped again to normalized states.
This is due to the interference phenomena.

It may happen that several different paths $\ell$
of  computation lead to the same configuration $c_x$. In
such a case, this configuration has an amplitude equal to
the sum of amplitudes of corresponding paths.
The configuration $c_x$ is then observed with probability
\begin{equation}
 \hspace{2cm}
 \label{sum-of-amplitudes}
 p(c_x)=\abs{\sum_{\ell}\alpha_x^{(\ell)}}^2,
 \qquad
 \alpha_{x}^{(\ell)} \in \mathbb{C}.
\end{equation}

We talk about a {\bf constructive interference} if
\begin{equation}
 \label{complex-constructive-interference}
 \abs{\sum_{\ell}\alpha_x^{(\ell)}}^2
 \geq
 \sum_{\ell}\abs{\alpha_{x}^{(\ell)}}^2\,,
\end{equation}
and about a {\bf destructive interference} if
\begin{equation}
 \abs{\sum_{\ell}\alpha_x^{(\ell)}}^2
 <
 \sum_{\ell}\abs{\alpha_{x}^{(\ell)}}^2\,.
\end{equation}

For a probabilistic Turing machine,
$\alpha_{x}^{(\ell)} \in \mathbb{R}^+$,
we have only constructive interference in the form of
\begin{equation}
 \abs{\sum_{\ell}\alpha_x^{(\ell)}}
 =
 \sum_{\ell}\abs{\alpha_{x}^{(\ell)}},
\end{equation}
which can be seen as a special case of the
equation \eqref{complex-constructive-interference},
up to the used $p$-norm.

{\example[Destructive interference] Let us have a quantum
Turing machine whose transitions are described by a unitary
matrix $T$ and the computation starts from the
configuration $c_0$. A three-step evolution is shown in the
Figure~\ref{quantum-tree}. Observe that at the second level
we have non-zero probability to observe configuration
$c_1$. At the third level, the probability of observing
$c_1$ is zero due to the destructive interference.
\begin{figure}[h!]
 \begin{center}
 \begin{tabular}{cc}
  \begin{tabular}{c}
   $T=\frac{1}{\sqrt{2}}
   \left(\begin{array}{cr}
      1 & 1\\ 1 & -1
   \end{array}\right)$
 \end{tabular}&
 \begin{tabular}{c}
 \includegraphics[scale=0.9]{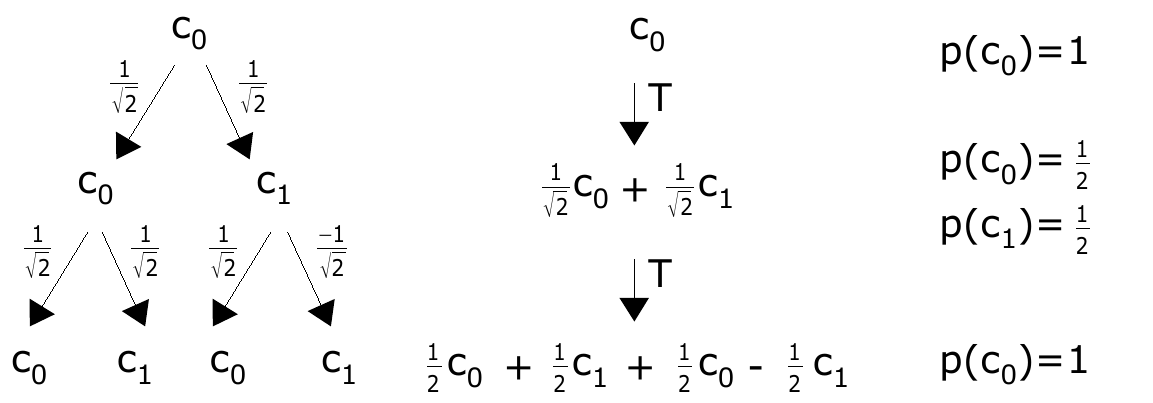}
 \end{tabular}
 \end{tabular}
 \end{center}
 \caption[The power of destructive interference]
         {The power of destructive interference.}
 \label{quantum-tree}
\end{figure}
}

{\definition[Accepting a language on a quantum Turing
 machine]
 A quantum Turing machine QTM accepts a language $L$
 with completeness $c$ and soundness $s$ iff there is
 a configuration
 $c(x)=(x',q_f,m)$, where $q_f \in F$, $m \in \{L, R\}$,
 and the machine halts in a superposition of
 configurations
 $\sum_i \alpha_i \ket{c_i}$ such that
 \begin{itemize}
  \item for all strings $x \in L$: the probability
        $p=\sum_i\alpha_i\abs{\braket{c(x)}{c_i}}^2 > c$,
  \item for all strings $x \notin L$: the probability
        $p=\sum_i\alpha_i\abs{\braket{c(x)}{c_i}}^2 \leq s$.
 \end{itemize}
}

{\definition[Exact or error-free quantum polynomial time,
 EQP] EQP is the class of decision problems efficiently
 solvable on a quantum Turing machine QTM with zero-error
 probability. A language $L$ is in the class EQP if there
 exists a QTM such that
 \begin{itemize}
  \item for all $x \in L$:
        the machine accepts $x$ with probability $p=1$,
  \item for all $x \notin L$:
        the machine accepts $x$ with probability $p=0$,
 \end{itemize}
 and the time complexity is upper-bounded by some
 polynomial $poly(x)$.
}\sq

The EQP class was defined in \cite{Bernstein-Vazirani97}
and, unfortunately,
there is no theory of universal quantum Turing machines
for exact quantum computing. This is due to the formulation
"if there exists a QTM such that \ldots" in the definition
of the class. Therefore showing that a problem is in the
EQP is of limited theoretical value. In the circuit model
for quantum computing (the model is described in 
Section~\ref{sec:circuit-model}), Deutsch-Jozsa and Simon
promise problems are known to have error-free efficient
circuits while using only gates from a universal set of
gates.

{\definition[Bounded-error quantum polynomial time, BQP]
 BQP is the class of decision problems solvable in a
 polynomial time on a quantum Turing machine with bounded
 error probability. A language $L$ is in the class BQP if
 \begin{itemize}
  \item for all $x \in L$:
        the machine accepts $x$ with probability $p > 2/3$,
  \item for all $x \notin L$:
        the machine accepts $x$ with probability $p \leq 1/3$,
 \end{itemize}
 and the time complexity is upper-bounded by some
 polynomial $poly(x)$.
}\sq

The class BQP is considered to be the class of feasible
problems for a quantum computer. It is a quantum analogue
of
the class BPP. It contains problems such as factoring and
the discrete logarithm problem \cite{Shor-factoring97},
Pell's equation \cite{Hallgren97} and simulation of
quantum systems having a local Hamiltonian \cite{Lloyd96}.
The authors of \cite{Bernstein-Vazirani97, Adleman97}
proved that BQP contains BPP and is contained in PP;
$BPP \subseteq BQP \subseteq PP$. The relation of BQP
and NP is unknown.

\subsection*{Concluding remarks for quantum Turing
             machines}
{\lemma
 Any quantum Turing machine $M_1$ can be simulated with
 constant slowdown by a quantum Turing machine $M_2$ which
 uses only real amplitudes.
 \label{qtm-with-real-amplitudes}
}

{\bf Proof.} Let us have a state
$\ket{\psi}=\sum_k \alpha_k \ket{k}$. A complex number
$\alpha= a + ib$
can be written as a two dimensional real vector:
$$
 \alpha = \left(\begin{array}{c}a\\b\end{array}\right)
 =a\ket{0}+b\ket{1},
 \qquad a, b \in \mathbb{R}
 \; .
$$
Then for a system composed out of qubits and amplitudes
$r_k \in \mathbb{R}$, we have
\begin{align*}
 M_1: \sum_{k \in \{0, 1\}^n} \alpha_k \ket{k}
 \longrightarrow
 M_2:
 &\sum_{k \in \{0, 1\}^n}
 \left(\begin{array}{c}a_k\\b_k\end{array}\right)\ket{k}
 \\
 &=
 \sum_{k \in \{0, 1\}^n} a_k\ket{0,k} + b_k\ket{1,k}\\
 &=
 \sum_{k \in \{0, 1\}^{n+1}} r_k \ket{k}\,.
\end{align*}

The transition unitary matrix is changed accordingly:
$$
 U_{M_1}:
 \left(\begin{array}{cc}\framebox{$a+ib$}
 & \cdots\\ \vdots & \ddots\end{array}\right)
 \quad \longrightarrow \quad
 U_{M_2}:
 \left(\begin{array}{cc}\framebox{$\begin{array}{cr}
 a & -b\\
 b & a
 \end{array}$} & \cdots \\
 \vdots & \ddots \end{array}\right)\, .
$$
Mutual simulation of quantum Turing machines with only
real numbers is in greater details discussed in
\cite{Bernstein-Vazirani97, Adleman97}.
Additionally, Adleman, DeMarrais
and Huang in \cite{Adleman97} proved that problems contained
in BQP can
be efficiently solved using only amplitudes from the set
$\{0, \pm \frac{3}{5}, \pm \frac{4}{5}, \pm 1\}$.
Kitaev \cite{KitaevBook97} proved even a smaller set
$\{0, \pm \frac{1}{\sqrt{2}}, \pm 1\}$ to be sufficient.
This is to be compared with a sufficient set for a
probabilistic Turing machine $\{0, \frac{1}{2}, 1\}$.
In \cite{Adleman97}, the authors raised
an interesting question whether BQP = BPP$_{\mathbb{R}}$,
where BPP$_{\mathbb{R}}$ is appropriately defined BPP class
for a probabilistic Turing machine with restricted set of
real amplitudes (probabilities). An affirmative
answer would imply that the assumed power of quantum
Turing machine comes only from possibly negative amplitudes
(destructive interference) and it is independent of the use
of the $2$-norm.

{\proposition
A quantum Turing machine can be simulated by a
deterministic Turing machine.
}

{\bf Proof.} The evolution of a quantum system is described
by an equation $\ket{\psi_2} = U \ket{\psi_1}$, where
$\ket{\psi_1}, \ket{\psi_2}$ are complex column vectors and
U is a unitary matrix. Thus a deterministic Turing machine
can simulate the evolution by a matrix-vector
multiplication.

{\proposition
Naive simulation of a quantum Turing machine on a
deterministic Turing machine by a matrix-vector
multiplication is not efficient.
}

{\bf Proof.} An $n$-qubit quantum system is described by
an $2^n$-dimensional complex vector. Corresponding
unitary evolution operator is a $2^n\text{-by-}2^n$ matrix.
Thus the matrix-vector multiplication takes $O(2^{2n})$
steps. Simulation of a quantum physical system,
described by a Hamiltonian $H$ during time $\Delta t$,
is even more complicated since it involves large matrix
exponentiation
\begin{equation}
 U(\Delta t) = e^{-iH\Delta t} = \sum_{k=0}^\infty
 \frac{(-iH\Delta t)^k}{k!}\;.
\end{equation}

{\lemma
 A quantum Turing machine with transitions
 restricted only to product states (no entanglement)
 can be simulated by a deterministic Turing machine
 in linear time.
}

{\bf Proof.} Since there is no entanglement, the qubits
evolve independently of each other. A simulation of
a single qubit is efficient - it involves only
multiplication of two-by-two matrix with
a two-dimensional vector. An $n$-qubit system is thus
simulated in time $O(n)$.

{\lemma
 A deterministic Turing machine can efficiently simulate
 a quantum Turing machine with transitions restricted
 only to states such that entanglement is localized in
 chunks. A chunk consists of at most $\log (poly(n))$
 qubits and there are at most polynomially many chunks.
}

{\bf Proof.} Chunks can be simulated independently of
each other. A simulation of one chunk is efficient.
Chunks are at most polynomially many.

\section{Quantum circuit model}
\label{sec:circuit-model}
The quantum circuit model is a computational model analogous
to the standard Boolean circuit model. The model was
introduced by D. Deutsch \cite{Deutsch89} in 1989 and called
quantum computational networks. Deutsch in his work also
identified a three-qubit gate, the Deutsch gate, which is
universal for quantum computing. The Deutsch gate is a quantum
analogue of the Toffoli gate.

\begin{figure}[h!]
\begin{center}
\begin{tabular}{l|l} \hline \hline
 Wire carrying a qubit                 
 &\includegraphics{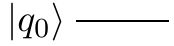}\\ \hline
 Wire carrying a bit
 &\includegraphics{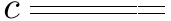}\\ \hline
 Projection onto $\ket{0}$ and $\ket{1}$
 &\includegraphics{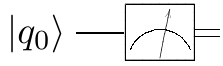}\\ \hline
 Unitary operation U
 &\includegraphics{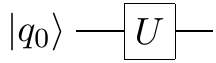}\\ \hline
 Controlled-U operation
 &\includegraphics{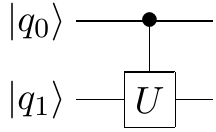}\\ \hline
 Controlled-NOT operation
 &\includegraphics{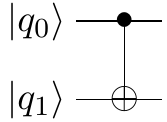}\\ \hline
 Swap
 &\includegraphics{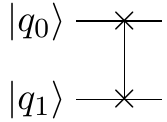}\\ \hline \hline
\end{tabular}
\caption[The circuit model notation]
        {The circuit model notation.}
\label{circuit-notation}
\end{center}
\end{figure}

A quantum circuit
presents quantum gates in a sequence as they are applied
to a quantum register. A quantum register
comprises of individually addressable qubits which are
depicted by horizontal lines (quantum wires) in a circuit
diagram. The quantum register is initially prepared in the
computational basis, typically in the state
$\ket{0,0, \ldots ,0}$. 
Vertical lines denote joint operations over two
or more qubits. This model allows us to express
computational complexity of a given algorithm in the number
of required elementary gates and/or a circuit depth.
The notation for quantum circuitry is summarized in the
Figure~\ref{circuit-notation}.

In 1993, Andrew Yao \cite{Yao93} elaborated the model and
showed that any function computable in polynomial
time on a quantum Turing machine has a polynomial-size
quantum circuit. Two years later, several papers
\cite{DiVincenzo95, Lloyd95, Barenco95, Deutsch-Barenco-Ekert95,Barenco-elem-gates95}
addressed questions related to gates construction and
decomposition, proofs of universality of almost all
two-qubit gates, and in particular, universality of
a two-qubit gate called Controlled-NOT accompanied with
single qubit gates. 

Since single qubit gates can be seen
as continuous operators, they can be at best only
approximated to some precision $\varepsilon$ by a
circuit consisting of gates from a fixed set of
single qubit gates. The Solovay-Kitaev theorem 
\cite{KitaevBook97,Solovay95} proves
that this approximation is indeed possible and induces only
a polylogarithmic overhead. Importantly, Bernstein and
Vazirani \cite{Bernstein-Vazirani97} showed that the total
error caused by a
sequence of imperfect (approximated) gates is at most
the sum of errors of individual gates.

\subsection{Single qubit gates}

Operations on a single qubit are described by 2-by-2
unitary matrices. Some of the most important unitary
matrices are Pauli matrices $X, Y, Z$ and identity matrix
$I$.
\begin{equation}
 X \equiv \left(\begin{array}{cc}
  0 & 1\\
  1 & 0 \end{array}\right); \quad
 Y \equiv \left(\begin{array}{cc}
  0 & -i\\
  i & 0 \end{array}\right); \quad
 Z \equiv \left(\begin{array}{cc}
  1 & 0\\
  0 & -1 \end{array}\right); \quad
 I \equiv \left(\begin{array}{cc}
  1 & 0\\
  0 & 1 \end{array}\right)\,.
\end{equation}

Pauli matrices give rise to three useful classes
of unitary matrices - the rotation operators
\begin{eqnarray}
  R_{x}(\phi) & \equiv & e^{-i\phi X/2} 
  = \cos\frac{\phi}{2}I - i\sin\frac{\phi}{2}X
  = \left(\begin{array}{rr}
      \cos\frac{\phi}{2}& -i\sin\frac{\phi}{2}\\[4pt]
    -i\sin\frac{\phi}{2}& \cos\frac{\phi}{2}
    \end{array} \right), \\[3pt]
  R_{y}(\phi) & \equiv & e^{-i\phi Y/2}
  = \cos\frac{\phi}{2}I - i\sin\frac{\phi}{2}Y
    \hspace{0.92mm}
  = \left( \begin{array}{rr}
      \cos\frac{\phi}{2}& - \sin\frac{\phi}{2}\\[4pt]
      \sin\frac{\phi}{2}& \cos\frac{\phi}{2}
    \end{array} \right), \\[3pt]
  R_{z}(\phi) & \equiv & e^{-i\phi Z/2} 
  = \cos\frac{\phi}{2}I - i\sin\frac{\phi}{2}Z
    \hspace{1.13mm}
  = \left( \begin{array}{cc}
      e^{- i\phi/2}& 0\\[4pt]
      0& e^{i\phi/2}
    \end{array} \right).
\end{eqnarray}

One reason why $R_{\{x, y, z\}}(\phi)$ matrices are referred
to as rotation operators is the Bloch sphere interpretation
of their actions.

\begin{figure}[h!]
 \begin{center}
  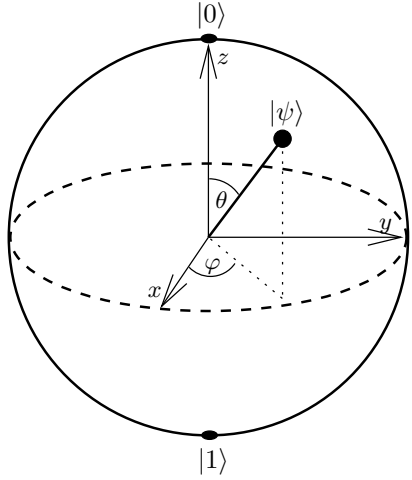
  \caption[The Bloch sphere]
          {The Bloch sphere.}
  \label{bloch-sphere}
 \end{center}
\end{figure}

\paragraph{Bloch sphere.}
A Bloch sphere is a unit sphere in Euclidean space
$\mathbb{R}^3$. A qubit state $\ket{\psi}=\alpha\ket{0}+
\beta\ket{1}$ parameterized by two complex numbers $\alpha,
\beta : \abs{\alpha}^2 + \abs{\beta}^2 =1$, can be rewritten
as
\begin{equation}
 \ket{\psi}
 = e^{i\delta}\left( \cos\frac{\theta}{2}\ket{0}
 + e^{i\varphi}\sin\frac{\theta}{2}\ket{1} \right),
\end{equation}
where $\delta, \theta,\varphi \in \mathbb{R}$. The global
phase factor $e^{i\delta}$ can be ignored because it has
no observable effect. The numbers $\theta, \varphi$,
interpreted as polar coordinates, define a point on a
Bloch sphere, see Figure~\ref{bloch-sphere}.
We write the $(x,y,z)$-coordinates as an unit Bloch vector
\begin{equation}
 \hat{r}
 =\left( \sin\theta\cos\varphi,\;
         \sin\theta\sin\varphi,\;
         \cos\theta \right).
\end{equation}

In the Bloch sphere picture, rotation operators
$R_{\{x,y,z\}}(\phi)$ corresponds to rotating a qubit state
about $\hat{x}, \hat{y}$ and $\hat{z}$ axes by $\phi$
radians. Moreover, since rows and columns of an
unitary matrix are orthonormal, i.e.
$UU^\dagger=U^\dagger U$,
arbitrary single qubit $U$ can be written as
\begin{equation}
 U =
 e^{i\delta}\left(\begin{array}{rr}
   e^{i (-\alpha/2-\gamma/2)}\cos\frac{\beta}{2} &
  -e^{i (-\alpha/2+\gamma/2)}\sin\frac{\beta}{2} \\[5pt]
   e^{i ( \alpha/2-\gamma/2)}\sin\frac{\beta}{2} &
   e^{i ( \alpha/2+\gamma/2)}\cos\frac{\beta}{2}
  \end{array}\right)\,,
\end{equation}
which actually corresponds to rotations about only two axes
\begin{equation}
 U=e^{i\delta}R_z(\alpha)\,R_y(\beta)\,R_z(\gamma)\,.
 \label{zy-decomposition}
\end{equation}
Equation \eqref{zy-decomposition} is called the Z-Y
decomposition of a single qubit gate. There is also
a Z-X decomposition following from the substitution of $R_y$
rotation by $R_x$ rotation: 
$R_y(\beta)=R_z(\pi/2)\, R_x(\beta)\, R_z(-\pi/2)$.
In fact, arbitrary single qubit gate can be written as
\begin{equation}
 U = e^{i\delta} R_{\hat{n}}(\alpha)\,R_{\hat{m}}(\beta)\,
 R_{\hat{n}}(\gamma)\,,
 \label{mn-decomposition}
\end{equation}
where $\hat{m}$ and $\hat{n}$ are non-parallel real unit
vectors in three dimensions. Therefore all single qubit
gates can be seen as operators performing rotations of
a qubit state on a Bloch sphere. 

Important single qubit gates which appear very often in
quantum circuits are the Hadamard gate (denoted $H$),
$\pi/8$ gate (denoted $T$), and phase gate (denoted $S$):
\begin{equation}
 H=\frac{1}{\sqrt{2}}\left(\begin{array}{rr}
  1 &  1 \\
  1 & -1
 \end{array}\right);\quad
 T=\left(\begin{array}{cc}
  1 & 0 \\
  0 & e^{i\pi/4}
 \end{array}\right);\quad
 S=\left(\begin{array}{cc}
  1 & 0 \\
  0 & i
 \end{array}\right)\,.
\end{equation}

It follows that
$H=e^{i \pi/2}\,R_z(\pi/2)\,R_x(\pi/2)\,R_z(\pi/2),\quad
T=e^{i \pi/8}\,R_z(\pi/4),\;\text{and}\quad S=T^2$.

Several useful gate identities are:
\begin{itemize}
 \item $X^2 =  Y^2 = Z^2 = I$,
 \item $HZH =  X, \quad
        HXH =  Z, \quad
        HYH = -Y, \quad
        HTH =  R_x(\pi/4)$,
 \item $XR_z(\theta)X = R_z(-\theta), \quad
        XR_y(\theta)X = R_y(-\theta)$.
\end{itemize}

\subsection{Two-qubit gates}
The set of all gates over two qubits consists of separable
and unseparable gates. Separable gates can be decomposed to
single qubit gates and in this sense they are not
considered to be two-qubit gates. Separable gates can be
always simulated by single qubit gates. On the other hand,
results of unseparable gates related to entanglement cannot
be simulated by single qubit gates. For this reason,
unseparable gates are referred to as entangling gates.

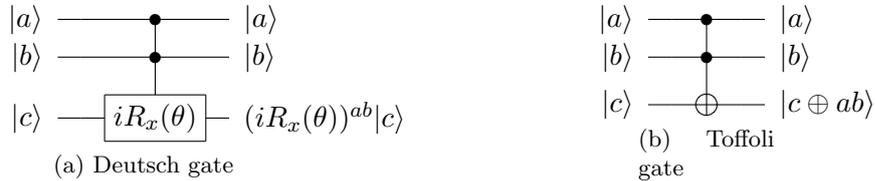
\begin{figure}[h!]
 \begin{center}
 \mbox{
 \subfloat[Deutsch gate]{ 
 $\Qcircuit  @R=1.3em @C=0.8em {
	 \lstick{\ket{a}} & \qw & \ctrl{1} &
	 \rstick{\ket{a}} \qw \\
	 \lstick{\ket{b}} & \qw & \ctrl{1} &
	 \rstick{\ket{b}} \qw \\
	 \lstick{\ket{c}} & \qw & \gate{iR_x(\theta)} &
	 \rstick{(iR_{x}(\theta))^{ab}\ket{c}} \qw
 }$
 \label{deutsch-gate}
 }
 \hspace{5cm}
 \subfloat[Toffoli gate]{
 $\Qcircuit @R=1.3em @C=0.8em {
	 \lstick{\ket{a}} & \qw & \ctrl{1} & \qw &
	 \rstick{\ket{a}} \qw  \\
	 \lstick{\ket{b}} & \qw & \ctrl{1} & \qw &
	 \rstick{\ket{b}} \qw  \\
	 \lstick{\ket{c}} & \qw & \targ    & \qw &
	 \rstick{\ket{c \oplus ab}} \qw
 }$
 \label{toffoli-gate}
 }
 }
\caption[Universal gates for quantum and classical 
         reversible computing]
        {Universal gates for quantum and classical
         reversible computing.}
\end{center}
\end{figure}

DiVincenzo \cite{DiVincenzo95} has shown that certain
two-qubit gates are already adequate in quantum computing.
Barenco in \cite{Barenco95}
improved on results of DiVincenzo and original work of
Deutsch \cite{Deutsch89} by identifying a large class of
two-qubit gates of form
\begin{equation}
 A(\phi,\alpha,\theta) =
 \left(\begin{array}{cccc}
  1 & 0 & 0 & 0 \\
  0 & 1 & 0 & 0 \\
  0 & 0 &
  e^{i\alpha}\cos\theta & -i e^{i(\alpha - \phi)}\sin\theta
  \\
  0 & 0 &
  -i e^{i(\alpha+\phi)}\sin\theta & e^{i\alpha}\cos\theta
 \end{array}\right)\,,
\end{equation}
which is sufficient for construction of the Deutsch gate,
$D(\theta)$, see Figure~\ref{deutsch-gate}. Parameters 
$\phi, \alpha$, and $\theta$ in $A(\phi,\alpha,\theta)$
and $D(\theta)$ are fixed irrational multiples of
$\pi$ and of each other. 

Without depending on an exact proof
\cite{Deutsch89} of quantum universality of the Deutsch gate,
we can
say the following. Due to irrationality properties that are
required, a sequence of $D(\theta_0)$ gates with a chosen
parameter $\theta_0$ can generate
the whole family of gates, $D(\theta)$, asymptotically.
The same applies to the $A(\phi,\alpha,\theta)$ gate, where,
in particular
$$
 A^n(\phi_0,\alpha_0,\theta_0) = A(\phi_0,\;
 n\alpha_0 \hspace{-2mm}\pmod {2\pi},\;
 n\theta_0 \hspace{-2mm}\pmod {2\pi})\,.
$$

The family of gates $D(\theta)$ comprises the Toffoli gate,
Toffoli $=D(\pi)$, and since $A(\phi_0,\alpha_0,\theta_0)$
gate is sufficient for construction of the $D(\theta_0)$ gate,
this implies that the class of gates $A(\phi,\alpha,\theta)$
is capable of classical universal reversible computation, at
least. The Toffoli gate is depicted in 
Figure~\ref{toffoli-gate}.

To achieve the power of a quantum computer as assumed by
the BQP class, the Toffoli gate must be accompanied with a
gate which can manipulate the interference phenomena. The
Hadamard gate or $A(\phi,\alpha,\theta)$ gates are examples
of such gates. Therefore sets $\{\text{Toffoli}, H\}$ and 
$\{A(\phi,\alpha,\theta)\}$ constitute quantum universal
sets in computational sense. 

{\definition[Computationally universal set]
 A finite set of unitary gates $G$ is called quantum
 computationally universal if every quantum Turing machine
 can be efficiently simulated to arbitrary fixed accuracy
 by a circuit composed out of gates from the set $G$.\sq
}

Quantum computational universality of the set
$\{\text{Toffoli}, H\}$ was proven by Y. Shi \cite{Shi03}
in 2002.
This result is in line with the minimal set of real
amplitudes sufficient for universal quantum Turing machine; 
see  Lemma~\ref{qtm-with-real-amplitudes}
on page~\pageref{qtm-with-real-amplitudes}. Moreover, since
the Hadamard transform is the Fourier transform over the
group $\mathbb{Z}_2$, quantum computing can be roughly viewed
as the Fourier transform plus a classical reversible
computing.
Indeed, the quantum part of the Shor factoring algorithm
consists exactly of a reversible circuit calculating the
function $f_{k,N}(x)=k^x \pmod N$ for given coprimes $k$ and
$N$, and the Fourier transform over the Abelian group
$\mathbb{Z}_{2^n}$.

Universality of the $A(\phi,\alpha,\theta)$ gate as proven
by Barenco \cite{Barenco95} is also remarkable. No classical
reversible universal gate with only two input and output
bits is known. The Toffoli gate has three input/output wires
and so has the primordial Deutsch gate.
Furthermore, Deutsch, Barenco and Ekert in
\cite{Deutsch-Barenco-Ekert95} and
Lloyd \cite{Lloyd95} showed than almost any two-qubit gate is
quantum universal. Thus the $A(\phi,\alpha,\theta)$ gate is
not a special gate in this sense. It seems that universal
computation independently of the classical or quantum
framework is deeply embodied in nature.

\begin{figure}[h!]
\[
\Qcircuit @R=1em @C=0.7em {
    \lstick{\ket{c}}
  & \qw 
  & \ctrl{1} 
  & \qw
  & \rstick{\ket{c}} \qw 
  \\
    \lstick{\ket{t}}
  & \qw 
  & \gate{U} 
  &\qw
  & \rstick{U^c\ket{t}} \qw
}
\]
\caption[Controlled-U operation]
        {Controlled-U operation.}
\label{controlled-u}
\end{figure}
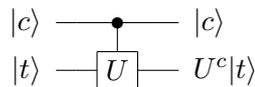

However, since  $A(\phi,\alpha,\theta)$ already contains
the essence of quantum universal computing, it is useful
to study its structure. A matrix representation
of the gate shows a layout of  form
$$
 \left(\begin{array}{cc}
  I & 0 \\
  0 & U
 \end{array}\right)\,,
$$
where $U$ is a single qubit gate.
Such a two-qubit gate applied to two qubits effectively
acts like a Controlled-U gate, see 
Figure~\ref{controlled-u}. The two input qubits are
known as the control qubit and the target qubit.
If the control qubit is set, i.e.
$\ket{c}=\ket{1}$, then $U$ is
applied to the target qubit, otherwise identity gate is
applied, $U^0=I$, and thus both qubits are kept untouched.

\begin{figure}[h!]
\[
\Qcircuit @R=1em @C=0.7em {
 & \qw 
 & \gate{H} 
 & \ctrl{1} 
 & \gate{H} 
 & \qw 
 & \qw 
 & \push{\rule{3em}{0em}=\rule{3em}{0em}} 
 & \qw 
 & \targ 
 & \qw 
 & \qw 
 \\
 & \qw 
 & \gate{H} 
 & \targ 
 & \gate{H}
 & \qw
 & \qw
 & \push{\rule{3em}{0em}=\rule{3em}{0em}} 
 & \qw 
 & \ctrl{-1} 
 & \qw 
 & \qw
}
\]
\caption[Interpreting who is controlling who]
        {Difficulties with interpreting who is controlling
         who.}
\label{controlling-who-is-who}
\end{figure}
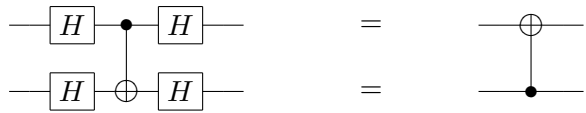

Due to the 'controlling' interpretation of some entangling
gates they are also referred to as controlling gates. But,
this term might be misleading, since it is not
always clear who is controlling who. See 
Figure~\ref{controlling-who-is-who}, where Hadamard gates
lead to the exchange of the control.

Out of controlling gates, the Controlled-NOT (CNOT) is a
prototypical operation. Its action is described as
\begin{equation}
 {CNOT:}\quad 
 \ket{c}\ket{t} \longrightarrow \ket{c}\ket{t \oplus c},
\end{equation}
and the matrix representation is
\begin{equation}
 CNOT =
 \left(\begin{array}{cccc}
  1 & 0 & 0 & 0 \\
  0 & 1 & 0 & 0 \\
  0 & 0 & 0 & 1 \\
  0 & 0 & 1 & 0
 \end{array}\right)\,.
\end{equation}

The importance of the CNOT gate comes from the fact that it
is relatively easy to understand its action, and moreover,
there is a constructive proof that any controlled single
qubit gate U, controlled-U, can be constructed using at most
two CNOTs and three single
qubit gates. This proof is due to Barenco {\it et. al.}
\cite{Barenco-elem-gates95}.
Since the $A(\phi,\alpha,\theta)$ gate is sufficient for
universal quantum computing this implies that CNOT and all
single qubit gates are sufficient too.

{\theorem[Controlled-U simulation]
\label{controlled-u-simulation}
Let $U$ be a single qubit gate. Then there exist single
qubit gates $A, B, C$ such that $ABC=I$ and $U=AXBXC$.
Controlled-U is then implemented by the following circuit.
\[
\Qcircuit @R=1em @C=0.7em {
 & \qw 
 & \ctrl{1} 
 & \qw  
 & \push{\rule{3em}{0em}=\rule{3em}{0em}} 
 & \qw      
 & \ctrl{1} 
 & \qw      
 & \ctrl{1} 
 & \qw 
 & \qw
 \\
 & \qw 
 & \gate{U} 
 & \qw  
 & \push{\rule{3em}{0em}=\rule{3em}{0em}} 
 & \gate{C} 
 & \targ   
 & \gate{B} 
 & \targ 
 & \gate{A} 
 & \qw
}
\]
The control qubit decides whether operation $ABC$ or
$AXBXC$ will be applied to the target qubit.
}

\textbf{Proof.}
By Z-Y decomposition,
$U=R_z(\alpha)\,R_y(\beta)\,R_z(\gamma)$ for some
$\alpha, \beta, \gamma \in \mathbb{R}$.
Set $A~=~R_z(\alpha)\,R_y(\beta/2)$,
$\;B=R_y(-\beta/2)\,R_z(-(\alpha+\gamma)/2)\;$ and
$C=R_z((-\alpha+\gamma)/2)$. Then
$$
 ABC 
 =R_z(\alpha)\,
  \underset{I}{\underbrace{R_y(\beta/2)\,R_y(-\beta/2)}}
  \,
  \underset{R_z(-\alpha)}
           {\underbrace{
         R_z(-(\alpha+\gamma)/2)\,R_z((-\alpha+\gamma)/2)}}
 =I
$$
and using the identity $I=XX$, we have
\begin{equation}
\begin{split}
 AXBXC
 &=R_z(\alpha)\,R_y(\beta/2)
  \, X \,
  R_y(-\beta/2)\,R_z(-(\alpha+\gamma)/2)
  \, X \,
  R_z((-\alpha+\gamma)/2)
 \\[2mm]
 &=R_z(\alpha)\,R_y(\beta/2)\,
  \underset{R_y(\beta/2)}{\underbrace{X\,R_y(-\beta/2)\,(X}}
  \,
  \underset{R_z((\alpha+\gamma)/2)}
  {\underbrace{
   X)\,R_z(-(\alpha+\gamma)/2)\,X}}\,R_z((-\alpha+\gamma)/2)
 \\
 &=R_z(\alpha)\,R_y(\beta)\,R_z(\gamma). \nonumber
\end{split}
\end{equation}
\sq

Barenco {\it et. al.} \cite{Barenco-elem-gates95}
in fact showed that
arbitrary gate over $n$ qubits can be constructed using
only CNOTs and single qubit gates. With an assumption that
arbitrary single qubit gates are available or can be
efficiently approximated, the number of CNOTs basically
describes the length of a circuit since at most one single
qubit gate makes sense between two successive CNOTs acting
on the same qubit.

The authors of \cite{Barenco-elem-gates95}
gave the upper bound
$O(n^3 4^n)$ and the lower bound $\Omega (4^n)$ on the
number of needed CNOTs for $n$-qubit entangling gates.
Since the vast
majority of all gates over $n$ qubits are indeed entangling
gates this implies that computationally interesting unitary
transforms requiring only polynomially many CNOTs are rare.
Clearly, those rare ones can be hardly found by automatized
brute-force decomposition. The upper bound $O(n^3 4^n)$ was
derived using the QR-decomposition \cite{QR-decomposition}.
The lower
bound $\Omega (4^n)$ is a consequence of $2^n \times 2^n$
degrees of freedom of an $n$-qubit gate.

\begin{figure}[h!]
\[
\Qcircuit @R=1em @C=0.7em {
   \lstick{\ket{\psi}}     
 & \ctrl{1} 
 & \targ     
 & \ctrl{1} 
 & \rstick{\ket{\upsilon}} \qw
 \\
   \lstick{\ket{\upsilon}} 
 & \targ    
 & \ctrl{-1} 
 & \targ    
 & \rstick{\ket{\psi}} \qw
}
\]
\caption[Implementation of the SWAP gate]
        {The SWAP gate implementation using three CNOT
	 gates.}
\label{swap-implementation}
\end{figure}
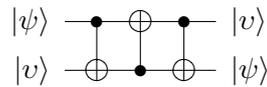

In 2003, Vartiainen {\it et. al.} \cite{Vartiainen04}
improved the
upper bound to $O(4^n)$. Vidal and Dawson \cite{Vidal04}
studied
the lower bound for the limit case, $n=2$, and showed that
arbitrary unitary transformation over two qubits can be
decomposed to a circuit containing at most three CNOTs and
eight single qubit gates. An example of a gate requiring
three CNOTs is the SWAP gate, see 
Figure~\ref{swap-implementation}.

Additional reason for choosing CNOT as an exemplary 
two-qubit gate is an easy entangled pair creation from a
trivial product state  $\ket{0,0}$.

{\example[Entangled pair creation]$\,$\\
\begin{eqnarray}
\ket{0,0} & \overset{H\otimes I}{\longrightarrow} 
          & \frac{1}{\sqrt{2}}(\ket{0,0}+\ket{1,0})
	  \nonumber
	  \\
	  \nonumber
          & \overset{CNOT}{\longrightarrow} 
	  & \frac{1}{\sqrt{2}}(\ket{0,0}+\ket{1,1})
\end{eqnarray}
\[
\Qcircuit @R=1em @C=0.7em {
 \lstick{\ket{0}} &\qw & \gate{H} & \ctrl{1} & \qw & \qw \\
 \lstick{\ket{0}} &\qw & \qw      & \targ    & \qw & \qw
}
\]
}

\subsection{Evolutionary universal set of gates}

In order to capture most of the abstraction offered by
a circuit model, it is useful to step from quantum 
computational universality to more strict evolutionary
universality where every unitary evolution may be
arbitrarily well approximated by a circuit composed out
of gates from a fixed finite set. For example, the set
\{Toffoli, H\} is  evolutionary universal for gates with
real entries but it clearly cannot be evolutionary
universal for gates with complex entries.

To formally define evolutionary universality we need a notion
of approximation. For that purpose we define a distance
function between two unitary operators $U$ and $U'$ as
\begin{equation}
 E(U,U') = \norm{U-U'}
         = \max_{\norm{\psi}=1}\norm{(U-U')\ket{\psi}},
 \label{distance-function}
\end{equation}
where the maximum is
over all normalized quantum states $\ket{\psi}$, and
we say that operator $U'$ approximates operator $U$ to
within $\varepsilon$ if 
$$
 E(U,U') \leq \varepsilon\,.
$$

The justification of this distance function  comes from
the fact that
measurement outcomes are the only things that matter
regardless whether $U$ or $U'$ was actually used. Let
$M~=~\sum_m r_m P_m,$ where $\sum_m P_m=I$, be an observable with
orthogonal projectors $P_m$.
Then $p_U(r_m)$ and $p_{U'}(r_m)$ are
probabilities of obtaining the outcome $r_m$ if $U$ and
$U'$ were performed, respectively. The distance function
$E(U,U')$ as defined by \eqref{distance-function} obeys
the rule that if $E(U,U')$ is small then the difference
between outcome probabilities is small.
In particular,
\begin{equation}
 \abs{p_U(r_m) - p_{U'}(r_m)} \leq 2 E(U,U').
\end{equation}

\textbf{Proof.} Let $\ket{\psi}$ be an initial state. 
\begin{align}
 \abs{p_U(r_m) - p_{U'}(r_m)} 
 &= \abs{ \bra{\psi}U^\dagger  P_m U  \ket{\psi} 
         -\bra{\psi}U'^\dagger P_m U' \ket{\psi}}
	  \qquad (\text{projective measurement})
	  \nonumber
	  \\
 &= \abs{ \bra{\psi}U^\dagger  P_m U  \ket{\psi}
         -\bra{\psi}U^\dagger  P_m U'  \ket{\psi}
	 +\bra{\psi}U^\dagger  P_m U'  \ket{\psi}
	 -\bra{\psi}U'^\dagger  P_m U'  \ket{\psi}}
	 \nonumber
	 \\
 &= \abs{ \bra{\psi}U^\dagger  P_m (U-U') \ket{\psi}
         +\bra{\psi}
	  (U^\dagger - U'^\dagger) P_m U'
	  \ket{\psi}}
	 \nonumber
	 \\
 &= \abs{ \bra{\psi} U^\dagger P_m \ket{\Delta}
         +\bra{\Delta} P_m U' \ket{\psi}}
	 \qquad (\text{substitution}\;\;
	 \ket{\Delta} = (U-U')\ket{\psi}\, )
	 \nonumber
	 \\
 &\leq  \abs{ \bra{\psi} U^\dagger P_m \ket{\Delta} }
       +\abs{ \bra{\Delta} P_m U' \ket{\psi}}
        \qquad (\text{triangle inequality})
        \nonumber
	\\
 &\leq \sqrt{\bra{\psi}U^\dagger U \ket{\psi}\;
             \bra{\Delta}P_m\ket{\Delta}}
      +\sqrt{\bra{\psi}U'^\dagger U' \ket{\psi}\;
             \bra{\Delta}P_m\ket{\Delta}}
       \quad (\text{Schwarz ineql.})
       \nonumber
       \\
 &=    2 \Bigl( {\bra{\Delta}
                 P_m
		 \ket{\Delta}}
	 \Bigr)^{1/2}
	 \qquad (\text{\ket{\psi} is normalized,
	 and U, U' preserve the norm})\nonumber \\
 &\leq 2 \Bigl( {\sum_m \bra{\Delta}
                        P_m
			\ket{\Delta}}
	 \Bigr)^{1/2}
  = \; 2 \Bigl( {\bra{\Delta}
                 \Bigl(\sum_m P_m\Bigr)
		 \ket{\Delta}}
	 \Bigr)^{1/2}
  = \; 2 \bigl( 
                {\braket{\Delta}{\Delta}}
	 \bigr)^{1/2}
	 \nonumber \\
 &=    2\norm{\Delta} = 2\norm{(U-U')\ket{\psi}}\nonumber \\
 &\leq 2E(U,U')\,. \nonumber
\end{align}

{\definition[Evolutionary universal set]
 A finite set of unitary gates $G$ is called
 evolutionary universal if for every unitary gate U
 and $\varepsilon>0$ there exists a
 sequence of gates $u'_1, u'_2,\ldots,u'_k \in G$ and
 ancilla state $\ket{a}$ such that
 \begin{equation}
  \max_{\norm{\psi}=1}
  \norm{U(\ket{\psi})\otimes\ket{a}
        - U'_1 U'_2 \cdots U'_k (\ket{\psi}\otimes\ket{a})}
  \leq \varepsilon,
 \end{equation}
 where $U'_i$ gate corresponds to the gate $u'_i$ surrounded
 with identity gates for an appropriate number of qubits
 and ancilla \ket{a} is an efficiently preparable state by
 means of gates from $G$. A set $G$ typically consists of
 gates acting on $1, 2$ and/or $3$ qubits.\sq
}

The definition of evolutionary universal set utilizes a special
form of the distance function where part of an input state
is regarded as an ancilla state. The ancilla state allows to sets
which are missing gates of small dimension still to be said
to be evolutionary universal in all dimensions.

Well-known evolutionary universal sets include:
\begin{itemize}
  \item \{$D(\theta)$\},\quad Deutsch \cite{Deutsch89},
        \vspace{-2mm}
  \item \{$A(\phi,\alpha,\theta)$\},\quad Barenco
        \cite{Barenco-elem-gates95},
        \vspace{-2mm}
  \item \{Toffoli, S, H\},\quad  Shor
        \cite{Shor-fault-tolerant96},
        \vspace{-2mm}
  \item \{Controlled-S, H\},\quad Kitaev
        \cite{KitaevBook97},
        \vspace{-2mm}
  \item \{Controlled-NOT, H, T\},\quad Boykin {\it et. al.}
        \cite{Boykin00}.
\end{itemize}

Gates in these sets create only a tiny fraction of gates
which can be implemented in a more or less straightforward way
using current technologies. In light of evolutional
universality of almost any two-qubit gate 
\cite{Deutsch-Barenco-Ekert95,Lloyd95}
it should be easy to find different evolutionary universal
sets using only the most feasible gates for a given physical
system. Surprisingly, a borderline between evolutionary
universal sets and sets which are not even quantum
computationally universal can be very thin. 

D. Gottesman and E. Knill 
\cite{Gottesman-Knill-theorem-a,Gottesman-Knill-theorem-b}
identified an
important class of quantum circuits which can be efficiently
simulated classically. This is the class of stabilizer
circuits introduced in study of quantum error-correcting
codes \cite{Gottesman-thesis97}. Consequently, the set of
gates which generates
this class of quantum circuits cannot be quantum 
computationally universal.

{\theorem[Gottesman-Knill]\quad\\
 Every circuit involving only the following elements:
 \vspace{-2mm}
 \begin{itemize}
  \item state preparation in the computational basis,
        \vspace{-2mm}
  \item gates Controlled-NOT, H, S, and the Pauli gates,
        \vspace{-2mm}
  \item and measurements of observables in the Pauli group,
        \vspace{-2mm}
 \end{itemize}
 can be efficiently simulated on a classical computer.
}

Stabilizer circuits embrace not only all linear quantum
error-correcting codes but schemes for entanglement
purification \cite{Entanglement-purf-a,Entanglement-purf-b},
quantum state teleportation \cite{Bennett-teleportation93}, 
superdense coding \cite{Bennett-dense-coding92}, and
Greenberger-Horne-Zeilinger paradox \cite{GHZ-paradox89} too. 
Anders and Briegel \cite{Anders06} presented recently an
algorithm
based on graph-state formalism which simulates stabilizer
circuits consisting of $n$ qubits in time and space 
$O(n \log(n))$. This result makes a numerical study of the above
mentioned problems easily handled and shows how subtle is
the power of quantum computation.

An important figure of merit of a particular evolutionary
universal set is whether its gates can operate in a 
fault-tolerant manner. Foundations of fault-tolerant quantum
circuits were laid by P. W. Shor \cite{Shor-fault-tolerant96}
in 1996. He 
suggested
to perform the computation and error-correction on encoded
logical qubits without a need to ever decode them, and in
such a way that an error on a single physical qubit cannot
propagate to other physical qubits in the same logical qubit.
This imposes constraints on both the type of unitary
operations that can be performed and error-correcting codes
used to encode logical
qubits. For example, gates that perform rotation by an
irrational multiple of $\pi$ cannot be realized in a 
fault-tolerant way \cite{Boykin00}. Shor showed that the set
\{Toffoli, H, S\} and CSS codes \cite{CSS-a,CSS-b}
conform to the
restrictions and  that a quantum circuit with $t$ gates can
tolerate $O(1/\log^c t)$ amounts of inaccuracy and
decoherence per gate for some small constant $c$ at a cost
of polylogarithmic increase of the circuit. This result was
later improved to hold for all codes encompassed by the 
stabilizer formalism \cite{DiVincenzo-fault-tolerant96}. Further
utilization of concatenation of
codes led to the threshold theorem 
\cite{Gottesman-thesis97, Seane-threshold03,Knill-threshold},
which says that if the basic error rate is below some 
threshold,
we can do arbitrary long computations. Currently the
threshold is above $10^{-3}$.  Next to the set 
\{Toffoli, H, S\}, the Kitaev's set \{Controlled-S,
H\} and the set \{Controlled-NOT, H, T\} are fault-tolerant
too. 

Note that general methods for classical 
fault-tolerant computation with noisy gates also require
logarithmic increase in the size of circuits 
\cite{Pippenger91}.

The set \{Controlled-NOT, H, T\} is the most commonly used
evolutionary universal set. Hadamard and $\pi/8$ gates can
efficiently approximate arbitrary single qubit gate U and
hereby together with the Controlled-NOT are sufficient to
approximate the set \{Controlled-NOT, all single qubit
gates\}. The possibility of approximation follows from the
fact that a sequence
\begin{equation}
 HTHT=R_x(\pi/4)\,R_z(\pi/4)
 =e^{-i(\pi/8)X}\,e^{-i(\pi/8)Z}=R_{\hat{n}}(\theta_c)
\end{equation}
produces a rotation along the axis 
$$
 \hat{n}=(\cos \pi/8,\; \sin \pi/8,\; \cos \pi/8)
$$
by a constant angle $\theta_c$ which is incommensurable
with $\pi$ \cite{Boykin00}; accordingly, 
\begin{equation}
 H R_{\hat{n}}(\theta_c) H = THTH = R_{\hat{m}}(\theta_c)
\end{equation}
with
$$
 \hat{m}=(\cos \pi/8,\; -\sin \pi/8,\; \cos \pi/8)\,.
$$

Using Equation~\eqref{mn-decomposition} we know that
a single qubit gate $U$ can be written as
$$
 U
 =R_{\hat{n}}(\alpha)\,
  R_{\hat{m}}(\beta)\,
  R_{\hat{n}}(\gamma)\,,
$$
and, additionally, thanks to $\theta_c/\pi$ being irrational,
a sequence of gates
$R_{\{\hat{m},\hat{n}\}}(\theta_c)$ can approximate any
$R_{\{\hat{m},\hat{n}\}}(\theta)$ gate:
\begin{equation}
   R_{\{\hat{m},\hat{n}\}}^k(\theta_c)
 = R_{\{\hat{m},\hat{n}\}}
     (k \theta_c \hspace{-1mm}\pmod{2\pi}).
\end{equation}
Thus for given $\xi > 0$ there exists a positive
integer $k$ such that
\begin{equation}
 E(R_{\{\hat{m},\hat{n}\}}(\theta),\;
   R_{\{\hat{m},\hat{n}\}}^k(\theta_c))
 \leq \xi\,.
\end{equation}

The question now is how the error accumulates while using
approximate gates. Suppose that gates in a sequence
$U_1 U_2$ are replaced with their approximate version
$U'_1$ and $U'_2$. Then
\begin{align*}
E(U_1 U_2, U'_1 U'_2) 
 &=    \norm{U_1 U_2 - U'_1 U'_2}\\
 &=    \norm{U_1 U_2 - U_1 U'_2 + U_1 U'_2 - U'_1 U'_2}\\
 &\leq \norm{U_1 (U_2 - U'_2)} + \norm{(U_1 - U'_1) U'_2}
       \quad (\text{triangle inequality})\\
 &\leq   \norm{U_1}\norm{U_2 - U'_2} 
       + \norm{U_1 - U'_1}\norm{U'_2}
       \quad\!\! (\text{submultiplicative prop. of Eq.
       \eqref{distance-function}})\\
 &=    E(U_2,U'_2) + E(U_1,U'_1),
       \quad ( \norm{U_1}=\norm{U'_2}=1)
\end{align*}
and, by induction, we can conclude that the error inccured
by approximate gates adds at most linearly:
\begin{equation}
 E(U_1 U_2 \ldots U_m, U'_1 U'_2 \ldots U'_m)
 \leq
 \sum_{j=1}^m E(U_j, U'_j)\,.
 \label{chaining-inequality}
\end{equation}

Thus if each gate in a sequence with $t$ gates is accurate
to $\xi=\varepsilon/t$ then the sequence is
accurate to $\varepsilon$. In conclusion, there exist
positive integers $a, b, c$ such that
\begin{equation}
 E(U,U')
 =E(R_{\hat{n}}(\alpha)\,
    R_{\hat{m}}(\beta)\,
    R_{\hat{n}}(\gamma)
    ,\;
    R_{\hat{n}}^a(\theta_c)\,
    R_{\hat{m}}^b(\theta_c)\,
    R_{\hat{n}}^c(\theta_c)
   )
 \leq \varepsilon\,.
 \label{h-t-is-dense}
\end{equation}

Regarding the rate of convergence, a sequence of angles
$\theta_c$ fills in the interval $[0,2\pi)$ roughly
uniformly and therefore it takes at most $O(1/\varepsilon)$
Hadamard
and $\pi/8$ gates to approximate a single qubit gate to
accuracy $\varepsilon$. Surprisingly, the Solovay-Kitaev
theorem discovered by Solovay \cite{Solovay95} and Kitaev
\cite{KitaevBook97} proves much faster convergence.

{\theorem[Solovay-Kitaev]
 Let $SU(d)$ denote the group of $d$-dimensional unitary
 matrices with unit determinant. Let $G$ be a finite set
 of gates in $SU(d)$ containing its own inverse such
 that the group generated by $G$ is dense in $SU(d)$. Let
 $\varepsilon > 0$ be given. There is a constant $c\approx
 4$ such that for any $U \in SU(d)$ there exists
 a finite sequence $U'=U'_1U'_2\ldots U'_l$ of gates from
 $G$ of length $l=O(\log^c(1/\varepsilon))$ such that 
 $E(U,U')\leq \varepsilon$.
}
 
The Solovay-Kitaev theorem for $SU(2)$, the set of all
single qubit gates up to the overall phase, manifestly
applies to Hadamard and $\pi/8$ gates: 
$\{H,\, H^\dagger = H,\, T,\, T^\dagger = T^7\}=\{H, T\} \in
SU(2)$, up to the overall phase, and according to
Equation~\eqref{h-t-is-dense} the set \{H, T\} generates
a dense subgroup in $SU(2)$. 

The constant $c$ is different
for different approaches to the proof of the theorem.
The algorithm described in \cite{KitaevBook02} produces a
sequence of
length $O(\log^{3+\delta}(1/\varepsilon))$, where $\delta$
is a chosen positive real number. With additional
restrictions imposed on $G$ and usage of ancilla qubits
there is an algorithm described in \cite{Kitaev97} which
produces a sequence of
length $O(\log^2(1/\varepsilon)\log(\log(1/\varepsilon)))$.
The running time of both algorithms is polylogarithmic
since they effectively produce a gate of the output sequence
per step. Moreover, the authors of \cite{Harrow02} gave a
non-constructive proof that for a suitable $G$ a sequence
of length $O(\log(1/\varepsilon))$ can be achieved. They
also showed this result to be the best possible. However,
it is not known which sets $G$ are suitable and whether the
compiling algorithm will be efficient.

In conclusion, the Solovay-Kitaev theorem implies that an
approximate
circuit simulating a circuit with $n$ qubits, $f(n)$ CNOTs
and consequently $O(f(n))$ single qubit gates to an accuracy
$\varepsilon$ requires $O(f(n)\log^c(f(n)/\varepsilon))$
gates from a universal set.

 \chapter{Phase estimation and applications}
 \label{chap:phase-estimation}
\begin{center}\end{center}

 \section{Fast quantum algorithms}
 \label{sec:fast-quantum-algorithms}
\begin{center}\end{center}

The discovery of the fast quantum factoring algorithm
\cite{Shor-factoring97} by Shor in 1994 was immediately
followed by a more
general work of A. Yu. Kitaev \cite{Kitaev96}, who using
group-theoretic approach gave an efficient algorithm for
the abelian stabilizer problem. 
This problem subsumes the Deutsch-Jozsa problem
\cite{Deutsch-Jozsa92},
the Simon problem \cite{Simon97}, factoring \& discrete
logarithms and many other problems as particular instances.
A study of techniques used by Shor and Kitaev showed that
the discovery of an efficient circuit for quantum Fourier
transform over finite abelian groups enables efficient
solutions to the abelian Hidden Subgroup Problem (HSP) and
the approach can be described as quantum Fourier sampling.
For example, an efficient quantum algorithm for solving
Pell's equation \cite{Hallgren97} has been found using this
framework.

{\definition[Abelian hidden subgroup problem]
 Let $G$ be a finite abelian group, $X$ a finite set, and
 $f: G \rightarrow X$ a function such that there exists a
 subgroup $H \subset G$ for which $f$ separates cosets of
 $H$, i.e. $f(g_1)=f(g_2)$ if and only if $g_1$ and $g_2$
 are members of the same coset $gH$. By querying a quantum
 oracle for performing the unitary transform 
 $U_f \ket{g}\ket{x}=\ket{g}\ket{x \oplus f(g)}$, for
 $g \in G, x \in X$, and $\oplus$ an appropriately chosen
 binary operation on $X$, determine a generating set for
 $H$.
}

A central challenge is to determine the hidden subgroup
in time $O(poly(\log{\abs{G}}))$ including encoding of
elements of $G$ and $X$ in terms of bitstrings, oracle
calls, quantum circuit for (optimal) measurement and any
needed classical postprocessing time. In the case of abelian
groups, the key point is that an efficient circuit for the 
Fourier transform is known and the transform performs 
mapping from the group to its dual while respecting
subgroups and cosets. 

Let $G$ be a finite abelian group. The characters of $G$
are homomorphisms 
$\varkappa : G \rightarrow \mathbb{C}\backslash\{0\}$,
that is mapping such that 
$\varkappa(g_1 + g_2) = \varkappa(g_1)\varkappa(g_2)$
assuming additive structure of $G$ and multiplicative
structure of $\mathbb{C}\backslash\{0\}$. The characters
form a group, called the dual group 
$\hat{G}$, $\vert\hat{G}\vert=\abs{G}$, and using the
characters the Fourier transform over the group $G$ is
given by:
\begin{equation}
 FT_G =
 \frac{1}{\sqrt{\abs{G}}}
 \sum_{i,j}
 \varkappa_j(i)\ket{j}\bra{i},\qquad \text{i.e.} \qquad
 \ket{g} \;\overset{FT_G}{\longrightarrow}\;
 \frac{1}{\sqrt{\abs{G}}}
 \sum_j \varkappa_j (g) \ket{j}.
\end{equation}
For the case $G=\mathbb{Z}_N$, the characters are described
as $\varkappa_j(k)=e^{-2\pi i j k / N}=\omega^{jk}$.

Corresponding to any subgroup $H \subseteq G$, there is a
set of elements in $G$ orthogonal to $H$: 
\begin{equation}
 H^\perp = 
 \{g \in G \mid \varkappa_g(h)=1 
 \quad \text{for all} \quad h \in H\}.
\end{equation}
The set of elements $H^\perp$ forms the orthogonal subgroup 
$H^\perp \subseteq \hat{G}$ to the subgroup $H \subseteq G$,
where $\vert H^\perp \vert =\abs{G}/\abs{H}$. The Fourier
transform over $G$, $FT_G$, maps a subgroup $H$ to its
orthogonal subgroup $H^\perp$. In particular, an equally
weighted superposition on $H$ is mapped to equally weighted
superposition over
$H^\perp$:
\begin{equation}
 \frac{1}{\sqrt{\abs{H}}} 
 \sum_{h \in H} \ket{h}
 \;\overset{FT_G}{\longrightarrow}\;
 \frac{\sqrt{\abs{H}}}{\sqrt{\abs{G}}}
 \sum_{k \in H^\perp} \ket{k}
 =
 \frac{1}{\sqrt{\vert H^\perp \vert}}
 \sum_{k \in H^\perp} \ket{k},
\end{equation}
and for a particular coset of $H$:
\begin{equation}
 \ket{gH}=\frac{1}{\sqrt{\abs{H}}}
 \sum_{h \in H} \ket{gh}
 \;\overset{FT_G}{\longrightarrow}\;
 \frac{1}{\sqrt{\vert H^\perp \vert}}
 \sum_{k \in H^\perp} \varkappa_g(k) \ket{k}.
\end{equation}
The standard approach for solving the abelian HSP is to
create a random coset state, 
perform a quantum
Fourier transform and sample the resulting state from
$H^\perp$. Since $(H^\perp)^\perp=H$, determining a
generating set for $H^\perp$ determines $H$ uniquely.
For example, in the factoring algorithm we have 
$f_{k,N}: x \rightarrow k^x \pmod N$, underlying group is
$\mathbb{Z}_{M}$, where $M \gg N$, the hidden subgroup 
is $H=\langle r \rangle = \{0, r, 2r, \ldots\}$, and 
$H^\perp=\langle M/r \rangle = \{0, M/r, 2M/r,\ldots \}$.
\subsection[Algorithm solving the abelian hidden subgroup
            problem]
	    {Algorithm for solving the abelian HSP}
\begin{description}
 \item [1. Create a random coset state]$\,$\\
 Create a superposition over all elements of $G$ in the
 first register and call the oracle:
 \begin{equation*}
 \begin{split}
  \ket{0,0} 
  & \;\overset{FT_G}{\longrightarrow}\;
    \frac{1}{\sqrt{\abs{G}}} \sum_{g \in G} \ket{g,0} \\
  & \;\overset{U_f}{\longrightarrow}\;
    \frac{1}{\sqrt{\abs{G}}} \sum_{g \in G} \ket{g,f(g)}.
 \end{split}
 \end{equation*}
 Next, measure the second register. This results in a
 uniformly chosen coset $g_iH$ of $H$ in the first
 register:
 \begin{equation*}
    \frac{1}{\sqrt{\abs{H}}}
    \sum_{h \in H}
    \ket{g_i h, f(g_i)}.
 \end{equation*}
 The second register is no longer in use now.

 \item [2. Fourier sampling]$\,$\\
 Perform the Fourier transform on the first register again:
 \begin{equation*}
  \frac{1}{\sqrt{\abs{H}}}
  \sum_{h \in H} \ket{g_i h}
  \;\overset{FT_G}{\longrightarrow}\;
  \frac{1}{\sqrt{\vert H^\perp \vert}}
  \sum_{k \in H^\perp} \varkappa_{g_i}(k)\ket{k}.
 \end{equation*}
 The phase $\varkappa_{g_i}(k)$ has no effect on a
 subsequent measurement and uniformly random $k \in H^\perp$
 is obtained.

 \item [3. $H$ reconstruction]$\,$\\
 Reconstruct $H$ from polynomially many samples.
\end{description}

\nicebox{
{\note
Since the measured value $f(g_i)$ from the second
register is actually never used, this measurement might 
in fact be skipped. In the first step of the algorithm, we
prepare the state
\begin{equation*}
  \ket{0,0} 
   \;\overset{FT_G}{\longrightarrow}\;
    \frac{1}{\sqrt{\abs{G}}} \sum_{g \in G} \ket{g,0}
   \;\overset{U_f}{\longrightarrow}\;
    \frac{1}{\sqrt{\abs{G}}} \sum_{g \in G} \ket{g,f(g)},
\end{equation*}
and after discarding (tracing out) the second register, 
due to the promise that $f$ separates cosets, the first
register is then in a mixed state
\begin{equation*}
 \rho_H =
 \frac{\abs{H}}{\abs{G}} 
 \sum_{\tilde{g}} \ket{\tilde{g}H}\bra{\tilde{g}H}
 \quad \text{with} \quad
 \ket{\tilde{g}H} =
 \frac{1}{\sqrt{\abs{H}}}
 \sum_{h \in H} \ket{\tilde{g}h},
\end{equation*}
where $\tilde{g}$ is a coset representative. The state
$\rho_H$ is called the hidden subgroup state. Implicit 
measurement, however, simplifies the analysis.
}} 


\subsection{New quantum algorithms}
Attempts to find new fast quantum algorithms which might
make some of known hard problems tractable naturally focus
on various generalizations of the abelian HSP and underlying
techniques. Efficient algorithms for non-abelian HSP are of
particular importance because the HSP over the symmetric
group encompasses the graph isomorphism problem and the HSP
over the dihedral group subsumes the shortest vector in a
lattice problem. 

Efficient algorithms for the HSP over non-abelian groups
such as 'almost abelian' groups \cite{Almost-abelian04} or
'near Hamiltonian' groups \cite{Near-abelian04} have been
successfully derived from the
abelian approach, but in general straightforward quantum 
Fourier sampling fails since there is no dual group in
the non-abelian case. An important result was the result of
Ettinger {\it et. al.} \cite{Ettinger99} who showed that the
query complexity
of non-abelian HSP is polynomial as well as for abelian HSP.
This means that there exists a measurement (a unitary
transform followed by a projective measurement) over
polynomially many samples of the hidden subgroup state
$\rho_H$ which reveals information sufficient to reconstruct
$H$. It was shown that for certain groups (including the
dihedral group \cite{Ettinger-dihedral00}) the measurement
may be performed by
repeated measurement on a single register containing the 
hidden subgroup state $\rho_H$ while other groups 
(including the symmetric group \cite{Moore05,Hallgren06})
require
measurement across multiple copies of $\rho_H$. The problem
is that the measurement even if correctly identified does not
always have to have an efficient implementation. Moreover,
classical postprocessing is not guaranteed to be efficient as
well. 

Measurement across multiple copies of the hidden
subgroup state was for the first time used by Kuperberg
\cite{Kuperberg05} who gave a subexponential but not polynomial
algorithm for the dihedral HSP. Utilizing measurement across
multiple registers likewise, Bacon {\it et. al.} 
\cite{HSP-heisenberg-grp05} gave
recently an efficient algorithm for solving the HSP over
certain semidirect product groups of the form 
$\mathbb{Z}_p^r \rtimes \mathbb{Z}$, for a fixed $r$ 
(including the Heisenberg group, $r=2$) and prime $p$. Bacon
{\it et. al.} achieved this result in the framework of 
so-called 'pretty good measurement' (PGM) \cite{Bacon-PGM06}
and later reinterpreted the result using the Clebsch-Gordan
transform 
\cite{Bacon-Clebsch-Gordan-approach06}.
PGM turned out to be an optimal measurement for the
dihedral HSP, however, no efficient implementation is known
in this case.
The Clebsch-Gordan transform is the key component in the
recently discovered efficient circuit for the Schur
transform \cite{Bacon-Schur-transform06}. The Schur transform
is known to be useful
in tasks such as an estimate of the spectrum of a density
operator \cite{Keyl-density-spectrum01} or communication
without a shared reference
frame \cite{Bartlett-without-srf03}. Whether the Schur
transform opens up for
advances in non-abelian HSP is an open question.

An alternative effort in finding new quantum algorithms focuses
on what kind of other structures may be efficiently revealed
on a quantum computer. Partial results along this lines are
discussed for the hidden shift problem
\cite{Dam-some-hidden-shifts06,Childs-hidden-shifts07} and
hidden nonlinear structures over finite fields
\cite{Childs-hidden-nonlinear07}.

 \section{Quantum Fourier transform based phase estimation}
 \label{sec:qft-based-pea}
 \begin{center}\end{center}

While theory based on a quantum Fourier sampling provides a
rather unifying view of the core of many quantum algorithms,
the actual problem-solution approach is often quite
different. One of the most fruitful approaches up to date is
to cast the problem as an estimation of an eigenvalue of a
certain unitary operator. Additional insight is that since
unitary operators preserve the $2$-norm the eigenvalues
are of the form $\lambda=e^{2\pi i \phi}$ for some phase
$\phi \in [0,1)$. Hence the goal is to estimate the phase.

In general, designing efficient and numerically stable
algorithms for finding the eigenvalues of an operator is one
of the most important problems in linear algebra.
Eigenvalues of small
operators can be found by finding roots of characteristic
polynomial, but this method cannot be used for large
operators, see Abel-Ruffini theorem \cite{Abel-Ruffini}.
One implication of the theorem is an expected iterative
nature for any general eigenvalue algorithm. A  stable
classical method for finding eigenvalues is the
QR-algorithm \cite{QR-algorithm} 
which iteratively executes the QR-decomposition 
\cite{QR-decomposition}.
For an $N \times N$ operator $U$, each QR iteration requires
$O(N^3)$ operations. Other methods utilize calculation of
$U, U^2, U^3,\ldots$ (power iteration) and the overall
complexity of these algorithms scale as $O(N^3)$ too. Thus
eigenvalues (or even a single eigenvalue) of an operator of
size $2^n \times 2^n$ are found inefficiently with respect
to $n$. A quantum computer can often perform better given
certain constraints on $U$. These constraints relate to the
existence of efficient quantum circuits (efficient with
respect to the size of the input $n$) for performing 
controlled powers $U^{2^k}$. Such constraints are not known
to be helpful for classical algorithms.

\subsection{Quantum phase estimation}
The main idea in quantum phase estimation is to consider a
register in a basis containing eigenvectors of the unitary
operator related to the problem being solved.
From the definition of eigenvalues and eigenvectors, when
an operator $U$ acts on one of its eigenvectors the result
is a scaling of this vector by a corresponding eigenvalue:
\begin{equation*}
 U\ket{\psi} = \lambda \ket{\psi}.
\end{equation*}
In another words, the eigenvalue is kicked in front of the
register. The trick behind quantum phase estimation is to
exploit the eigenvalue kick-back in combination with the
ability to efficiently perform controlled powers $U^{2^k}$
and transfer the phase $2 \pi (2^k \phi)$ into the relative
phase of qubit in another register.
Since $U$ is a linear operator, the same concept applies
to an arbitrary input superposition of eigenvectors of $U$.

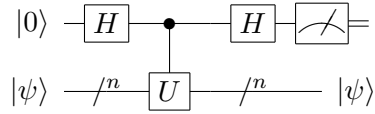
\begin{figure}[h!]
 \[
 \Qcircuit @R=1em @C=0.7em {
  \lstick{\ket{0}}    & \gate{H} & \ctrl{1} & \qw & \gate{H}
    & \meter &  \cw \\
  \lstick{\ket{\psi}} &  {/^n} \qw      & \gate{U} & \qw
    &  {/^n} \qw    & \rstick{\ket{\psi}} \qw }
 \]
 \caption[Eigenvalue kick-back circuit]
         {Eigenvalue kick-back circuit.}
 \label{naive-pea}
\end{figure}

The very basic scheme which utilizes only a single 
controlled-$U$
gate is shown in the Figure~\ref{naive-pea}. The upper line
is the ancillary qubit which is measured, and the lower line
represents space of $n$ qubits on which $U$ operates.
Initially the ancillary qubit is set to $\ket{0}$ and the
lower line register to an eigenstate $\ket{\psi}$ of the
operator $U$ with eigenvalue $e^{2\pi i \phi}$.

\begin{figure}[h!]
\begin{center}
 \includegraphics{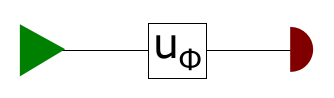}
\end{center}
\caption[Schematic diagram of the eigenvalue kick-back
         circuit]
        {Schematic diagram of the eigenvalue kick-back
         circuit.}
\label{phase-model-box}
\end{figure}

The process of estimation can be divided into two stages.
At the first stage, the eigenvalue $e^{2 \pi i \phi}$ is
kicked in front of the $\ket{1}$ component of the
first qubit:
\begin{align*}
 \ket{0}\ket{\psi}
 & \; \overset{H \otimes I}{\longrightarrow} \;
   \frac{1}{\sqrt{2}}(\ket{0}+\ket{1})\ket{\psi}
   = \frac{1}{\sqrt{2}}
   (\ket{0}\ket{\psi} + \ket{1}\ket{\psi})
   \\
 & \; \overset{c-U}{\longrightarrow} 
   \frac{1}{\sqrt{2}}
   (\ket{0}\ket{\psi} + \ket{1}U\ket{\psi})
   = \frac{1}{\sqrt{2}}
   (\ket{0}\ket{\psi} + e^{2\pi i \phi}\ket{1}{\ket{\psi}})
   \\ 
 & \hspace{1cm} = \frac{1}{\sqrt{2}}
   (\ket{0} + e^{2\pi i \phi}\ket{1})\ket{\psi}.
\end{align*}
The second stage is to apply the Hadamard transform to the
first qubit in order to make the phase an observable quantity
in the standard basis:
\begin{equation*}
 \frac{1}{\sqrt{2}}
 (\ket{0} + e^{2\pi i \phi}\ket{1})\ket{\psi}
 \; \overset{H \otimes I}{\longrightarrow} \;
 \frac{1}{2} 
 \left( (1+e^{2\pi i \phi})\ket{0}
      + (1-e^{2\pi i \phi})\ket{1}\right) \ket{\psi}.
\end{equation*}
The outcome '$0$' is observed with probability
\begin{equation}
 P_0 = \abs{\frac{1+e^{2\pi i \phi}}{2}}^2
     = \frac{1}{4}\abs{(1+\cos{2\pi\phi}+i\sin{2\pi\phi})}^2
     = \cos^2{\pi \phi}.
\end{equation}
Accordingly, the outcome '$1$' is observed with probability 
$P_1=1-P_0=\sin^2{\pi\phi}$. The whole procedure is often
modelled by a diagram shown in the 
Figure~\ref{phase-model-box}.
The gate $u_\phi$ represents a phase shift caused by some
physical process and maps a prescribed single qubit state
$\ket{0}+\ket{1}$ to $\ket{0}+e^{2\pi i \phi}\ket{1}$.

\begin{figure}[h!]
 \begin{center}
  \includegraphics{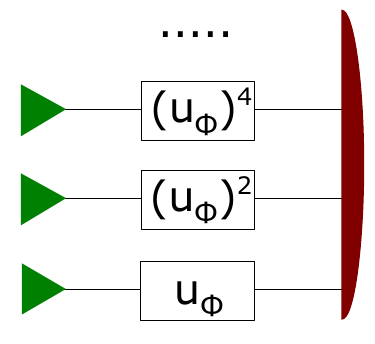}
 \end{center}
 \caption[Advanced setup for the phase estimation]
         {Advanced setup for the phase estimation.}
 \label{advanced-setup-for-pea}
\end{figure}

The question is now how to determine the phase in the most
efficient way. By repeating the simple eigenvalue
kick-back procedure $M$-times, $P_0$ and consequently
$\phi$ can be determined to an accuracy of 
$1/\sqrt{M}$. This result follows from the central
limit theorem: for large $M$ the error on average
decreases as $\Delta/\sqrt{M}$, where $\Delta^2$ is the 
variance of the measurement results associated to the 
physical apparatus. Thus one needs go to through at least
$M \sim 2^{2m}$ independent rounds to obtain $m$ accurate
binary digits of $\phi$. The scaling $1/\sqrt{M}$ is known
as the standard quantum limit (SQL). However, using an
advanced setup, the scaling may be improved up to $1/M$
(the Heisenberg limit).
Giovannetti {\it et. al.} \cite{Quantum-metrology06} showed
that after employing
the phase shift $u_\phi$ at total $M$-times the scaling
$1/M$ is the general lower bound to the estimation error.
A setup which achieves the Heisenberg limit is shown in the
Figure~\ref{advanced-setup-for-pea}. This setup with a 
register of ancillary qubits exploits namely a sequential
applying of the phase shift $u_\phi$ to the same qubit.
Dam {\it et. al.} \cite{Dam-optimal-pea07} proved that
this scheme is indeed optimal.

\subsection{Advanced scheme for phase estimation}
The algorithm based on the powers of the phase shift is
called the phase estimation algorithm (PEA). The PEA was
for the first time used by Kitaev \cite{Kitaev96} in his
work on the abelian
stabilizer problem and the modern formulation of the PEA
based on the quantum Fourier transform is due to Cleve
{\it et. al.} \cite{Cleve98}. The phase estimation problem
itself is defined as follows.

{\definition [Phase estimation problem]
 Let $U$ be a $2^n \times 2^n$ unitary transform. 
 Let $\ket{\psi_1}, \ket{\psi_2},\ldots,\ket{\psi_{2^n}}$
 denote eigenvectors of $U$ with corresponding eigenvalues
 $\lambda_1, \lambda_2, \ldots, \lambda_{2^n}$, 
 where $\lambda_j$ is of the form $e^{2\pi i \phi_j}$, i.e.,
 \begin{equation*}
  U = \sum_j
      e^{2 \pi i \phi_j} \ket{\psi_j}\bra{\psi_j}.
 \end{equation*}
 Given a set of black-boxes (oracles) capable of preparing
 a particular eigenstate $\ket{\psi}$ and performing the 
 controlled-$U^{2^k}$ operation for $k=0,1,2,\ldots,m-1$,
 efficiently determine phase $\phi \in [0,1)$ to the
 precision of order $1/2^m$ with reasonable high probability.
 \sq
}

A proof that the advanced scheme
(Figure~\ref{advanced-setup-for-pea}) for phase estimation
reaches the Heisenberg limit follows. Right before the
measurement the state of $m$ ancillary qubits can be
described as:
\begin{equation}
\begin{split}
 &
 \frac{1}{\sqrt{2^m}}
 \bigotimes_{l=m-1}^2
         \Bigl(\ket{0}+e^{2 \pi i (2^l \phi)}\ket{1}\Bigr)
 \otimes \bigl(\ket{0}+e^{2 \pi i (2^1 \phi)}\ket{1}\bigr)
 \otimes \bigl(\ket{0}+e^{2 \pi i (2^0 \phi)}\ket{1}\bigr) 
 =\\
 &
 \frac{1}{\sqrt{2^m}}
 \bigotimes_{l=m-1}^2
         \Bigl(\ket{0}+e^{2 \pi i (2^l \phi)}\ket{1}\Bigr)
 \otimes \bigl(                 \ket{00} + 
               e^{2\pi i \phi}  \ket{01} +
	       e^{2 \pi i 2\phi}\ket{10} +
	       e^{2 \pi i 3\phi}\ket{11} \bigr)
 =\\
 &
 \frac{1}{\sqrt{2^m}}
 \bigotimes_{l=m-1}^2
         \Bigl(\ket{0}+e^{2 \pi i (2^l \phi)}\ket{1}\Bigr)
 \otimes \sum_{k=0}^3 e^{2\pi i k \phi} \ket{k} 
 =\\
 &
 \frac{1}{\sqrt{2^m}}
 \sum_{k=0}^{2^m-1} e^{2\pi i k \phi}\ket{k} \,.
 \label{tensor-product-to-sum-for-qft}
\end{split}
\end{equation}

In order to prepare this superposition, the phase shift
$u_\phi$ has been employed 
$\sum_{l=0}^{m-1}2^l=2^m-1\cong 2^m$-times. Now, the
essential role of the (inverse) quantum Fourier transform
is the ability to perform the transformation
\begin{equation}
  \frac{1}{\sqrt{2^m}}
  \sum_{k=0}^{2^m-1} e^{2\pi i k \phi}\ket{k}
  \;\longrightarrow\;
  \ket{j},
\end{equation}
where $j$ is a non-negative integer and  $j/2^m$ is a
good estimator for $\phi$ with high probability.

\begin{figure}[h!]
 \begin{center}
  \includegraphics[scale=0.45]
                  {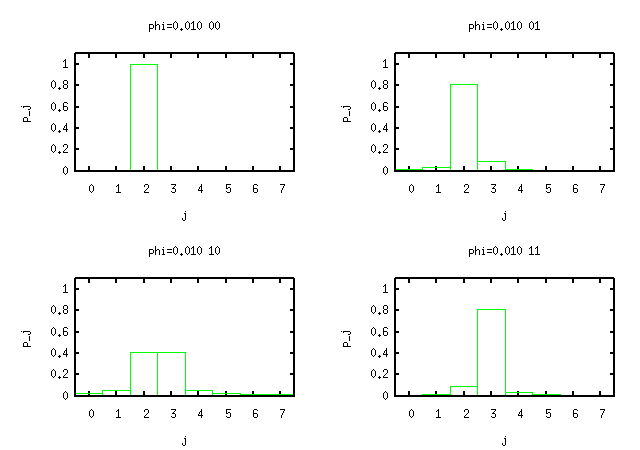}
 \end{center}
 \caption[The PEA output probability distribution.]
         {Output probability distribution for $m=3$ and
          various $\phi$.}
 \label{pea-probability-bars}
\end{figure}

Let us use the following notation:
\begin{align}
  & \ket{\nu} =
    \frac{1}{\sqrt{2^m}}
    \sum_{k=0}^{2^m-1} e^{2\pi i k \phi}\ket{k},
    \quad \text{and} \\
  &
    QFT^\dagger: \ket{k}
    \;\longrightarrow\;
    \frac{1}{\sqrt{2^m}}
    \sum_{j=0}^{2^m-1} e^{-2\pi i k j/2^m}\ket{j}\,.
\end{align}
Then,
\begin{align}
 QFT^\dagger\ket{\nu}
 &=
  \frac{1}{\sqrt{2^m}}
  \sum_{k=0}^{2^m-1}
  e^{2\pi i k \phi} \Bigl( \frac{1}{\sqrt{2^m}}
                           \sum_{j=0}^{2^m-1}
			   e^{-2\pi i k j/2^m} \ket{j}\Bigr)
  \\
 &=
  \frac{1}{2^m}
  \sum_{k=0}^{2^m-1}\sum_{j=0}^{2^m-1}
  e^{2\pi i k (\phi - j/2^m)} \ket{j}
  \quad \text{(change the order of sums)} \\
 &=
  \sum_{j=0}^{2^m-1}\Bigl(\frac{1}{2^m}\sum_{k=0}^{2^m-1}
  e^{2\pi i k (\phi - j/2^m)} \Bigr) \ket{j}\;.
\end{align}
Therefore the probability to measure a particular outcome
$j$ is
\begin{equation}
 p_j
 =
 \abs{ \frac{1}{2^m} 
       \sum_{k=0}^{2^m-1} 
       e^{2\pi i k (\phi - j/2^m)} }^2 \,.
 \label{p-j-prob-given-by-sum}
\end{equation}
Figure~\ref{pea-probability-bars} shows the output
probability distribution for $m=3$ and various $\phi$.

\begin{figure}[h!]
 \begin{center}
  \includegraphics[scale=0.8]{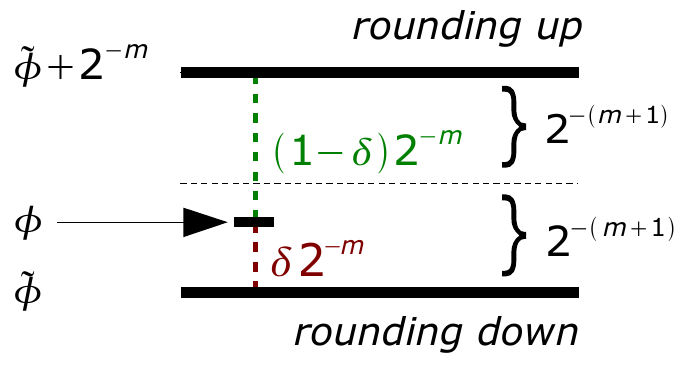}
 \end{center}
 \caption[Two closest estimates of the phase]
         {Two closest estimators of the phase $\phi$.}
 \label{rounding-details}
\end{figure}

In order to analyze (Eq.~\ref{p-j-prob-given-by-sum}) and 
to understand the Figure~\ref{pea-probability-bars} let
us write $\phi$ as $\phi~=~\tilde{\phi}~+~\delta~2^{-m}$,
where $\tilde{\phi}~=~0.\phi_1 \phi_2 \ldots \phi_m$ denotes
the first $m$ bits of the binary expansion and  
$0 \leq \delta < 1$ is a remainder. 
A special case where $\delta=0$ means that
$\phi$ has a binary expansion with no more then $m$ bits,
i.e.  $\phi=\tilde{\phi}=j/2^m$ for some 
$j \in \{0,1,\ldots,2^m-1\}$. Given $\delta=0$, the
probability of observing $j$ such that $j/2^m=\phi$ is
\begin{equation}
 p_j (\phi=j/2^m) =
 \abs{ \frac{1}{2^m}
       \sum_{k=0}^{2^m-1}
       e^{2\pi i k (\phi - j/2^m)}}^2
 = \abs{\frac{1}{2^m}\sum_{k=0}^{2^m-1} 1}^2 = 1.
\end{equation}
Thus the Fourier transform deterministically reveals the
exact phase in this case. In other cases, $0<\delta<1$, the
best we can hope for is to observe
$$
j\begin{cases}
 \text{such that }
   j/2^m=\tilde{\phi}=0.\phi_1\phi_2\ldots\phi_m 
   \quad \text{ (rounding down) }, \\
 \text{such that } 
   j/2^m=\tilde{\phi}+2^{-m} 
   \quad\hspace{1.5cm} \text{ (rounding up) },
\end{cases}
$$
with sufficiently large probability. Both estimates
$\tilde{\phi}$ and $\tilde{\phi}+2^{-m}$ guarantee error
smaller then $1/2^m$. Figure~\ref{rounding-details}
details the situation.

\begin{figure}[h!]
 \begin{center}
  \includegraphics[scale=0.45]{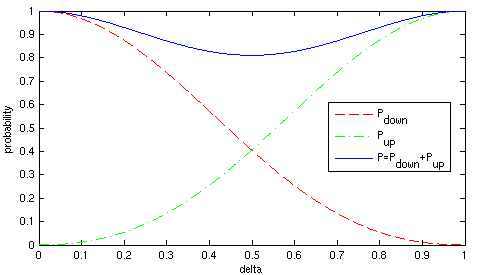}
 \end{center}
 \caption[Success probability of observing a good estimator]
         {Success probability of observing a good
          estimator.}
 \label{pea-lower-bound}
\end{figure}

To determine the lower bound of the probability
to observe either $\tilde{\phi}$ or $\tilde{\phi}+2^{-m}$,
we simplify (Eq.~\ref{p-j-prob-given-by-sum}) using the
formula for the sum of a geometric series and trigonometric
identities:
\begin{equation}
 p_j
 =
 \abs{ \frac{1}{2^m} 
       \sum_{k=0}^{2^m-1} 
       e^{2\pi i k (\phi - j/2^m)} }^2 
 =
 \frac{1}{2^{2m}}
 \abs{ \frac{1-e^{2 \pi i(2^m\phi - j)}}
            {1-e^{2\pi i (\phi - j/2^m)}}}^2
 =
 \frac{1}{2^{2m}}
 \frac{\sin^2({\pi(2^m\phi-j))}}
      {\sin^2({\pi(\phi-j/2^m)})} \, .
 \label{p-j-prob-given-by-sinus}
\end{equation}
Using the substitution $\phi= \tilde{\phi} + \delta 2^{-m}$,
we get
\begin{equation}
 \label{probability-p-down-qft-based}
 P_{down}
 = p_j(j/2^m=\tilde{\phi}) 
 = \frac{1}{2^{2m}}
   \frac{\sin^2{(\pi\delta)}}
        {\sin^2{(\pi\delta 2^{-m})}}\, ,
\end{equation}
and
\begin{equation}
 \label{probability-p-up-qft-based}
 P_{up}
 = p_j(j/2^m=\tilde{\phi}+2^{-m})
 = \frac{1}{2^{2m}}
   \frac{\sin^2{(\pi(1-\delta))}}
        {\sin^2{(\pi(1-\delta)2^{-m})}}\, .
\end{equation}
The total probability is then  $P=P_{down}+P_{up}$.
Corresponding function plots for $0<\delta<1$ and $m=10$ are
shown in the Figure~\ref{pea-lower-bound}.

The success probability $P$ decreases monotonically for
increasing $m$ and in the limit $m \rightarrow \infty$,
the lower bound is 
\begin{equation}
 P(\delta=1/2)
 = P_{down}(\delta=1/2)+P_{up}(\delta=1/2)
 = \frac{4}{\pi^2}+ \frac{4}{\pi^2} > 0.81.
\end{equation}

Here, I provide sketch of a simple proof which is
conceptually different from the conventional proof
based on the ratio of the minor arc length to the
chord length \cite{Nielsen-Chuang00,Cleve98}.

\begin{subequations}
\begin{align}
   \delta    
     &>
     \sin (\delta)\, , 
     \quad \text{ for $0<\delta$ },
     \qquad \text {(follows from the Taylor series of
                   $\sin \delta$)} \\
   \pi\delta
     &>
     \sin(\pi\delta)\, , \\
   \frac{\pi\delta}{2^m} 
     &>
     \sin{\Bigl(\frac{\pi\delta}{2^m}\Bigr)}\, ,
     \quad \text{ for $m=0,1,2,\ldots$} \\
   \pi\delta 
     &>
     2^m \sin{\Bigl(\frac{\pi\delta}{2^m}\Bigr)}\, ,
\end{align}
\end{subequations}
\begin{align}
   P_{down}(0<\delta\leq 1/2) 
   &= 
   \left(\frac{\sin(\pi\delta)}
              {2^m\sin{\Bigl(\frac{\pi\delta}{2^m}\Bigr)}}
   \right)^2
   \geq \frac{\sin^2{(\pi\delta)}}
             {(\pi\delta)^2}
   \geq \frac{\sin^2{(\frac{\pi}{2})}}
             {(\frac{\pi}{2})^2}=\frac{4}{\pi^2}\, ,\\
   P_{up}(1/2 \leq\delta < 1) 
   &=
   \left(\frac{\sin(\pi\delta)}
              {2^m\sin{\Bigl(\frac{\pi\delta}{2^m}\Bigr)}}
   \right)^2
   \geq \frac{\sin^2{(\pi\delta)}}
             {(\pi\delta)^2}
   \geq \frac{\sin^2{(\frac{\pi}{2})}}
             {(\frac{\pi}{2})^2}=\frac{4}{\pi^2}\, .
\end{align}\sq

In conclusion, the advanced scheme
(Figure~\ref{advanced-setup-for-pea}) can be used to 
determine the phase with precision of order $1/2^m$ after
the phase shift $u_\phi$ has been used at total $2^m$ times.
Thus the advanced scheme reaches the Heisenberg limit. In
particular, the accuracy of order $1/2^m$ is reached with a
success probability $ > 0.81$ independently of $m$.

In contexts where a higher probability of success is 
desirable, the success probability can be amplified to
$1-\varepsilon$ for any $\varepsilon>0$ by inflating $m$
to $m'=m+O(\log({\frac{1}{\varepsilon}}))$ and keeping
only $m$ most significant bits. For a proof see Cleve
{\it et. al.} \cite{Cleve98}.
A different approach, which avoids implementing $U^{2^k}$
for $k>m$, is to repeat the whole procedure a number of
times, choosing the most frequent result.

\subsection{Efficient circuit for the quantum Fourier
transform}
The next step in the search for efficient algorithms 
solving instances of the phase estimation problem is to
find an efficient circuit for the quantum Fourier
transform over $\mathbb{Z}_{2^m}$. An approach which is
essentially the inverse of the derivation used in
(Eq.~\ref{tensor-product-to-sum-for-qft}) leads us to
\begin{align}
 QFT: \ket{j} \rightarrow
 &
  \frac{1}{\sqrt{2^m}}
  \sum_{k=0}^{2^m-1}
  e^{2\pi i j k /2^m} \ket{k}
  \hspace{3cm} \left\vert\;
  k=\sum_{l=1}^m k_l\, 2^{m-l},\;
  k_l \in \{0,1\},\right.\\
 &=
  \frac{1}{\sqrt{2^m}}
  \sum_{k=0}^{2^m-1}
  e^{2\pi i j 
        \Bigl(\sum\limits_{l=1}^m k_l\, 2^{m-l}\Bigr)/2^m}
  \ket{k}
  \hspace{0.7cm} \left\vert\;
  M=2^m,\; 
  \ket{k}=\bigotimes_{l=1}^{m}\ket{k_l},\right.\\
 &=
  \frac{1}{\sqrt{M}}
  \sum_{k_1=0}^1\ldots\sum_{k_m=0}^1
  e^{2\pi i j \sum\limits_{l=1}^m k_l\, 2^{-l}}
  \ket{k_1,k_2,\ldots,k_m}\\
 &=
  \frac{1}{\sqrt{M}}
  \sum_{k_1=0}^1\ldots\sum_{k_m=0}^1
  e^{2\pi i j (k_1 2^{-1}
             + k_2 2^{-2}
	     + \cdots
	     + k_m 2^{-m})}
  \ket{k_1,k_2,\ldots,k_m}
\end{align}
\begin{align}
 &=
  \frac{1}{\sqrt{M}}
  \sum_{k_1=0}^1\ldots\sum_{k_m=0}^1
  e^{2\pi i j k_1 2^{-1}}
  e^{2\pi i j k_2 2^{-2}}
  \ldots
  e^{2\pi i j k_m 2^{-m}}
  \ket{k_1,k_2,\ldots,k_m}\\
 &=
  \frac{1}{\sqrt{M}}
  \sum_{k_1=0}^1\ldots\sum_{k_m=0}^1
  \bigotimes_{l=1}^m
  e^{2\pi i j k_l 2^{-l}} \ket{k_l}\\
 &=
  \frac{1}{\sqrt{M}}
  \sum_{k_1=0}^1\ldots\sum_{k_m=0}^1
  \bigotimes_{l=1}^{m-1}
  e^{2\pi i j k_l 2^{-l}} \ket{k_l}
  \otimes
  e^{2\pi i j k_m 2^{-m}} \ket{k_m}\\
 &=
  \frac{1}{\sqrt{M}}
  \sum_{k_1=0}^1\ldots\sum_{k_{m-1}=0}^1
  \bigotimes_{l=1}^{m-1}
  e^{2\pi i j k_l 2^{-l}} \ket{k_l}
  \otimes
  \sum_{k_m=0}^1 e^{2\pi i j k_m 2^{-m}}
  \ket{k_m}\\
 &=
  \frac{1}{\sqrt{M}}
  \bigotimes_{l=1}^m \sum_{k_l=0}^1 
  e^{2\pi i j k_l 2^{-l}} \ket{k_l}\\
 &=
  \frac{1}{\sqrt{M}}
  \bigotimes_{l=1}^m
  \Bigl( e^{2\pi i j (0) 2^{-l}} \ket{0} 
       + e^{2\pi i j (1) 2^{-l}} \ket{1} \Bigr)\\
 &=
  \frac{1}{\sqrt{M}}
  \bigotimes_{l=1}^m
  \Bigl( \ket{0} + e^{2\pi i j 2^{-l}} \ket{1} \Bigr)\\
 &=
  \frac{1}{\sqrt{M}}
  \Bigl(
   \ket{0} + e^{2\pi i 0.j_m}\ket{1} \Bigr)
  \Bigl(
   \ket{0} + e^{2\pi i 0.j_{m-1} j_m }\ket{1} \Bigr)
  \cdots
  \Bigl(
   \ket{0} + e^{2\pi i 0.j_1 j_2 \ldots j_m}\ket{1} \Bigr)
  \label{product-form}
\end{align}

\begin{figure}[h!]
 \begin{center}
 \scalebox{0.9}{
  \hspace{-3.3cm}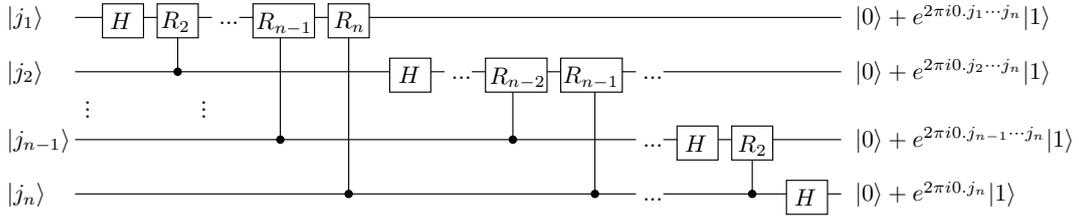}
 \caption[Efficient circuit for the quantum Fourier
          transform over $\mathbb{Z}_{2^m}$]
         {Efficient circuit for the quantum Fourier
          transform over $\mathbb{Z}_{2^m}$.}
 \label{qft-circuit}
 \end{center}
\end{figure}

Let us write the input state $\ket{j}$ in the tensor product
form 
$\ket{j}=\bigotimes_{l=1}^m\ket{j_l},\; j_l \in \{0,1\}$.
Then the first factor in (Eq.~\ref{product-form}) can be
prepared via the Hadamard transform applied to the state
$\ket{j_m}$.
\begin{equation}
 H\ket{j_m}
 = \frac{1}{\sqrt{2}}(\ket{0}+(-1)^{j_m}\ket{1})
 = \frac{1}{\sqrt{2}}(\ket{0}+e^{2\pi i 0.j_m}\ket{1})
\end{equation}
To prepare the second factor in (Eq.~\ref{product-form}) we
need a gate capable of adjusting the relative phase. A 
suitable gate is a z-rotation gate
\begin{equation}
 R_k = \left(\begin{array}{cc}
  1 & 0 \\
  0 & e^{2\pi i/2^k} \end{array}\right)\,.
\end{equation}
By applying the Hadamard transform and the gate $R_2$
conditioned on the bit $j_m$ to the state 
$\ket{j_{m-1}}$ we get
\begin{align}
 R_2^{j_m}H\ket{j_{m-1}}
 &=
  R_2^{j_m}
  \frac{1}{\sqrt{2}}
  (\ket{0}+e^{2\pi i 0.j_{m-1}}\ket{1})
  = 
  \frac{1}{\sqrt{2}}
  (\ket{0}
   + e^{2\pi i 0.j_{m-1}}\,
     e^{2\pi i j_m /2^2}
   \ket{1})
  \nonumber\\
 &=
  \frac{1}{\sqrt{2}}(\ket{0}+e^{2\pi i 0.j_{m-1}j_m}\ket{1})
\end{align}
The other factors can be prepared accordingly. The 
corresponding circuit, derived from the product
representation (Eq.~\ref{product-form}), for the quantum
Fourier transform over $\mathbb{Z}_{2^m}$ is shown in the
Figure~\ref{qft-circuit}. Note that the order of qubits
becomes reversed and swap gates should be used if needed.

The circuit utilizes $(m^2+m)/2=O(m^2)$ gates. An obvious
bottleneck is that some of these gate require an exponential
precision up to $\xi=1/2^m$. By the Solovay-Kitaev
theorem, each such gate will require
$O(\log^c(1/\xi))=O(m^c)$, where $c \approx 4$, gates from
a fixed universal set. This implies an efficient circuit, 
however, the corresponding polynomial is of unacceptable
degree. Barenco {\it et. al.} \cite{Barenco-aqft96}
studied approximated quantum Fourier transform (AQFT) where 
gates $R_k$ for $k \geq \log_2{m}+2$ (i.e. a logarithmic-depth
AQFT) are neglected and showed that the accuracy of the QFT
can be achieved after $O((m/\log{m})^3)$ iterations of the
AQFT. D. Cheung \cite{Cheung04} improved significantly on
this result and showed that the output  probability
distribution of a single logarithmic-depth AQFT iteration is
in fact very close to the QFT output distribution. In
particular, Cheung gave a lower bound 
$$
 P_{down, up}(\delta=1/2)
 \geq
 \frac{4}{\pi^2} - \frac{1}{4m},
$$
for $m \geq 4$. Therefore the logarithmic-depth AQFT can
be considered a direct replacement for the QFT. The
amount of $R_k$ gates is now $O(m\log{m})$ and the desired
gate  accuracy is $\xi~=~1/2^{\log{m}}$. Using the
Solovay-Kitaev theorem, only $O(m\log{m}\log^c{m})=O(m^2)$
gates from a fixed universal set are required in order to
implement that transform.

The history of efficient circuits for a quantum Fourier
transform goes back to D. Deutsch. In 1989, Deutsch
\cite{Deutsch89}
described an efficient circuit for QFT over 
$\mathbb{Z}_2^m$, that is the Hadamard transform. Shor 
\cite{Shor-factoring97}
inspired by the Simon promise problem \cite{Simon97}, where
the
Hadamard transform was used,  realized how to efficiently
implement QFT over the group $\mathbb{Z}_q$ for certain
special values of $q$. In 1994, D. Coppersmith improved
this result to hold for $\mathbb{Z}_{2^m}$ and one year
later Kitaev \cite{Kitaev96} gave an efficient circuit for
a quantum Fourier transform over an arbitrary finite abelian
group. Somewhat later, human-competitive circuits for
QFT (and other transforms) were also discovered using
genetic programming \cite{Spector04, Massey05}.

\begin{figure}[h!]
 \begin{center}
  \hspace{-6mm}\scalebox{0.88}
	                {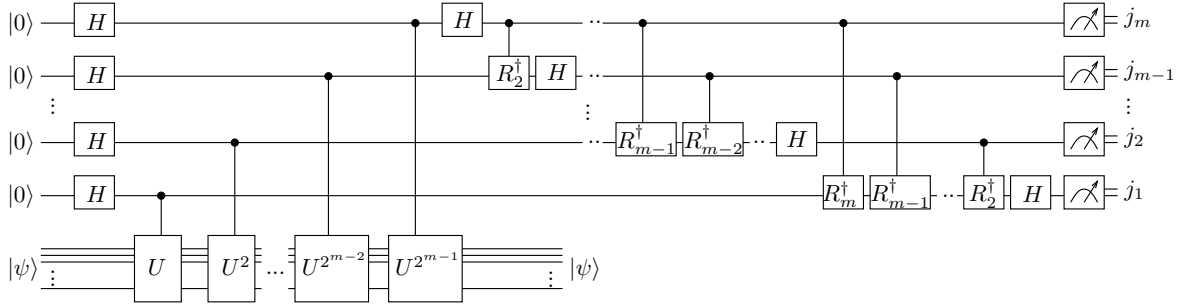}
 \end{center}
 \caption[Detailed scheme for advanced phase estimation]
         {Detailed scheme for advanced phase estimation.
          The state  $\ket{\psi}$ is promised to be an
	  eigenstate of $U$ with an eigenvalue 
	  $e^{2\pi i \phi}$. The output given by
          $\sum_{k=1}^m j_k\, 2^{-k}=j/2^m$ is a good
	  estimate of $\phi$, $\abs{\phi - j/2^m} < 1/2^m$,
	  with high probability.}
 \label{complete-scheme-for-pea}
\end{figure}

\subsection[Linear property of the phase estimation algorithm]
           {Linear property of the advanced scheme}
\label{sec:pea-linear-property}
Details of the advanced scheme for phase estimation including
the efficient measurement in the Fourier basis (implemented
by the Fourier transform + projection onto the computational
basis) is shown in the Figure~\ref{complete-scheme-for-pea}. 

Essentially the scheme transforms the input state 
$\ket{0,\psi}$ in the following way:
\begin{subequations}
\begin{align}
 &
 \ket{0,\psi}\\ 
 \overset{H\otimes I}{\longrightarrow}\quad &
 \frac{1}{\sqrt{2^m}} \sum_j \ket{j,\psi} \\
 \overset{\text{controlled-}U^j}{\longrightarrow}\quad &
 \frac{1}{\sqrt{2^m}} \sum_j \ket{j,U^j \psi}
 =
 \frac{1}{\sqrt{2^m}} \sum_j
 e^{2\pi i j \phi} \ket{j,\psi}\\
 \overset{QFT^\dagger \otimes I}{\longrightarrow}\quad &
 \sum_j
 \alpha_j \ket{j,\psi}\,,
\end{align}
\end{subequations}
where states $\ket{j,\psi}$ such that
$\abs{\phi - j/2^m} < 1/2^m$ have dominating probabilities
$\abs{\alpha_j}^2$
and thus one of them is observed with high probability.
Therefore the overall transform can be described as
\begin{equation}
 \ket{0,\psi}
 \quad \overset{\text{phase estimation}}{
	       \longrightarrow} \quad
 \ket{\blacktriangleleft\phi\blacktriangleright,\psi}\,,
\end{equation}
where $\blacktriangleleft\phi\blacktriangleright$ denotes a
good estimator. Importantly, since $U$ is a linear operator,
the same concept applies to an input superposition of 
eigenvectors of $U$. Thus we can avoid preparing a
(possibly unknown) eigenstate, at the cost of introducing
some additional randomness into the scheme. 

Let 
$\ket{\varsigma}=\sum_s \gamma_s \ket{\psi_s}$ denote an
arbitrary quantum state expanded in terms of eigenstates 
$\ket{\psi_s}$ of $U$. Then
\begin{equation}
 \label{eq:pea-linear-property}
 \ket{0,\varsigma} = \sum_s \gamma_s \ket{0,\psi_s}
 \quad \overset{\text{phase estimation}}
               {\longrightarrow} \quad
 \sum_s \gamma_s
 \ket{\blacktriangleleft\phi_s\blacktriangleright,\psi_s}\,,
\end{equation} 
where $\blacktriangleleft\phi_s\blacktriangleright$ denotes
a good estimator related to the eigenstate $\ket{\psi_s}$.
A subsequent measurement on the first register yields one
estimator $\blacktriangleleft\phi_s\blacktriangleright$ 
chosen at random with probability $\abs{\gamma_s}^2$. An
important side effect of the measurement is a collapse of
the second register to the corresponding eigenvector (or
a superposition of eigenvectors related to a particular
degenerative eigenvalue). Therefore the phase estimation
scheme can be used to prepare eigenstate(s). Note that
the eigenstate remains unknown to us.

\subsection{Representative instances of the phase
               estimation problem}
\label{sec:pea-representative-instances}

\subsubsection{Factoring casted as a phase estimation
               problem}
Kitaev \cite{Kitaev96, KitaevBook97} showed that the order
finding algorithm
(the only quantum part of the factoring algorithm) is
equivalent to the scheme where we estimate an eigenvalue of
the unitary operator $U$ given by 'modular multiplication by
$k$':
\begin{equation}
 U: \ket{y} \longrightarrow \ket{k\,y \!\!\pmod N}\,.
\end{equation}
Let $r$ denote the unknown order, i.e. the least  positive
integer such that $k^r = 1 \pmod N$ for given coprimes $k$
and $N$. The main observation  done by Kitaev is that states
defined as
\begin{equation}
 \ket{\psi_s}
 \equiv
 \frac{1}{\sqrt{r}}
 \sum_{j=0}^{r-1}e^{-2\pi i  j s/r}\ket{k^j \!\!\pmod N}\,,
 \label{modular-multiplication-eigenstates}
\end{equation}
for integer $0\leq s \leq r-1$ are eigenstates of $U$.
Additionally, since $U^r\ket{y} = \ket{k^r\, y \pmod N} = 
\ket{y}$ and thus $U^r = I$, the eigenvalues of $U$
must be of the form $\lambda_s=e^{2\pi i s /r}$. We can
verify that
\begin{equation}
 U\ket{\psi_s} =
 \frac{1}{\sqrt{r}}
 \sum_{j=0}^{r-1}
 e^{-2\pi i  j s/r} \ket{k^{j+1} \!\!\!\pmod N}
 =
 e^{2\pi i s/r}\ket{\psi_s}
 =\lambda_s\ket{\psi_s}\,.
\end{equation}
It is also useful to rewrite
(Eq.~\ref{modular-multiplication-eigenstates}) in the form
\begin{equation}
\frac{1}{\sqrt{r}}
 \sum_{s=0}^{r-1}
 e^{2\pi i j s/r}\ket{\psi_s}=
 \ket{k^j \!\!\!\pmod N}\,,
\end{equation}
and by setting $j=0$ to derive
\begin{equation}
 \frac{1}{\sqrt{r}}
 \sum_{s=0}^{r-1}
 \ket{\psi_s} = \ket{1}\,.
\end{equation}
Thus the state $\ket{1}$ trivially constitutes an equally
weighted superposition of eigenstates $\ket{\psi_s}$.

The process of a phase estimation applied to the 'modular
multiplication by $k$' operator $U$ can be described as
follows.
\begin{subequations}
\begin{align}
   &
  \ket{0}\ket{1} \\
  \overset{H \otimes I}{\longrightarrow}\qquad  &
  \frac{1}{\sqrt{2^m}}
  \sum_{j=0}^{2^m-1}\ket{j}\ket{1} \\
  \hspace{1.1cm}
  \overset{\text{controlled-}U^j}{\longrightarrow}\quad &
  \frac{1}{\sqrt{2^m}}
  \sum_{j=0}^{2^m-1}\ket{j}U^j\ket{1}
  \hspace{2.3cm}
  \left\vert\;\Rightarrow
  \frac{1}{\sqrt{2^m}}
  \sum_{j=0}^{2^m-1}\ket{j}\ket{k^j \!\!\!\pmod N}
  \right.
  \\
  =\quad &
  \frac{1}{\sqrt{2^m}}
  \sum_{j=0}^{2^m-1}\ket{j} U^j \Bigl(
    \frac{1}{\sqrt{r}}
    \sum_{s=0}^{r-1}\ket{\psi_s} \Bigr)
\end{align}
\begin{align}
  =\quad &
  \frac{1}{\sqrt{2^m}}
  \sum_{j=0}^{2^m-1}\ket{j} \Bigl( \frac{1}{\sqrt{r}}
   \sum_{s=0}^{r-1}e^{2\pi i j s/r}\ket{\psi_s} \Bigr)\\
  =\quad &
  \frac{1}{\sqrt{2^m}} \frac{1}{\sqrt{r}}
  \sum_{j=0}^{2^m-1} \sum_{s=0}^{r-1}
  e^{2\pi i j s /r} \ket{j}\ket{\psi_s}\\
  \overset{QFT^\dagger \otimes I}{\longrightarrow}\qquad  &
  \sim\frac{1}{\sqrt{r}}
  \sum_{s=0}^{r-1}
  \ket{\blacktriangleleft s/r \blacktriangleright}
  \ket{\psi_s}
\end{align}
\end{subequations}

Given the outcome 
$a=\;\blacktriangleleft s/r \blacktriangleright$ for some $s$,
the continued fractions algorithm efficiently produces numbers
$s'$ and $r'$ with no common factor, such that 
$s'/r' = s/r$. For $m=\Theta(\log(N^2))$ the number $r'$ is 
equal to the order $r$ with high probability.

\subsubsection*{Efficient implementation of 
                controlled-$U^{2^l}$ gates}
In order for the period finding algorithm to be practically
useful, we need to make sure that all its steps are 
performed efficiently with respect to the size of the input
$m=\Theta(\log(N^2))$. So far we know that  
initialization of the input state to $\ket{0}\ket{1}$ is
efficient, measurement in the Fourier basis is efficient
and the classical postprocessing is efficient. The
implementation of the controlled-$U^{2^l}$ gates, where 
$0\leq l \leq m-1$, based on the repeated modular squaring
can be performed efficiently as well. Naive implementation
consisting only of repeated application of the 
controlled-$U$ gate is obviously not efficient with respect
to $m$.

\begin{figure}[h!]
 \begin{center}
  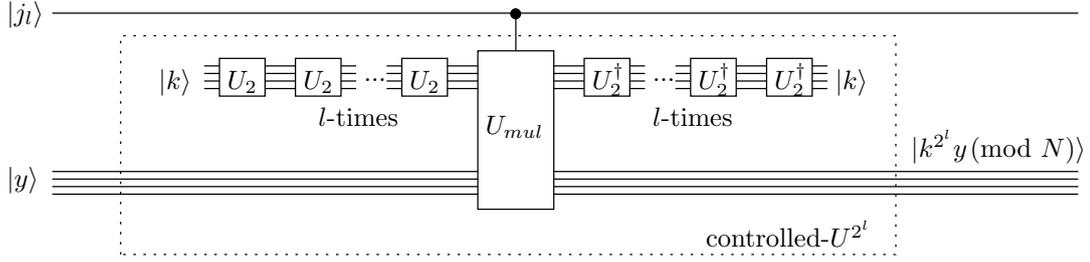
 \end{center}
 \caption[Modular exponentiation by modular squaring]
	 {Efficient implementation of the 
	  controlled-$U^{2^l}$ gate, where $U$ performs
	  a modular multiplication by $k$.}
 \label{efficient-implementation-u-2-k}
\end{figure}

Let us have gates for modular multiplication and modular 
squaring (these are manifestly efficient):
\begin{align}
 U_{mul} &: \ket{a}\ket{b} 
        \longrightarrow 
	\ket{a}\ket{ab \!\!\pmod N},
	\hspace{2cm} \text{(modular multiplication)}\\
 U_2 &: \ket{a}
        \longrightarrow 
	\ket{a^2  \!\!\pmod N}\,.
	\hspace{2.87cm} \text{(modular squaring)}
\end{align}
The implementation of the controlled-$U^{2^l}$ gate based
on the  modular squaring consists of four steps:
\begin{enumerate}
 \item Prepare two registers. The first is a scratch 
       register initially set to $\ket{k}$. The second 
       register contains an arbitrary state $\ket{y}$.
 \item Apply the squaring gate \underline{$l$-times} to the
       scratch register.
       \begin{equation}
         \ket{k,\,y}
	 \;\rightarrow\;
	 \ket{k^2 \!\!\!\pmod N,\, y}
	 \;\rightarrow\;
	 \ket{k^4 \!\!\!\pmod N,\,y} 
	 \;\rightarrow \cdots \rightarrow\;
	 \ket{k^{2^l} \!\!\!\pmod N,\, y}
       \end{equation}
 \item Conditioned on some bit $j_l$ perform the modular
       multiplication.
       \begin{equation}
         \ket{k^{2^l} \!\!\!\pmod N,\; y}
	 \;\longrightarrow\;
	 \ket{k^{2^l} \!\!\!\pmod N,\; k^{2^l}y \pmod N}
       \end{equation}
 \item Set the scratch register to $\ket{k}$ again, so it
       can be reused. This can be performed either by 
       'uncomputing' or a measurement based reset operation.
\end{enumerate}
The procedure is shown in the 
Figure~\ref{efficient-implementation-u-2-k}.

\subsubsection{Finding the energy eigenvalues of a local
               Hamiltonian}
In 1982, R. P. Feynman suggested that a quantum computer
might be capable of an efficient simulation of other 
quantum physical systems \cite{Feynman82}. S. Lloyd 
\cite{Lloyd96} showed
that an efficient simulation is indeed possible at least
for quantum systems with a local Hamiltonian. Systems with
a local Hamiltonian capture a large part of relevant 
physics. The question then arises, what kind of properties
can be efficiently extracted from the simulation. Abrams and
Lloyd \cite{Abrams-Lloyd99} focused on static properties and
showed that
the phase estimation procedure can be used to find 
efficiently the energy eigenvalues if approximate
eigenvectors are available.

Let $\mathcal{H}$ be a (local) Hamiltonian with energy
eigenvalues $\lambda_s$ and eigenstates $\ket{\psi_s}$. This
Hamiltonian generates a unitary evolution
$U=e^{-i\mathcal{H}t}$ during time $t$. Energy eigenvalues
are mapped to the phase of eigenvalues of $U$.
\begin{align}
 \mathcal{H}
 =
   \sum_s \lambda_s \ket{\psi_s}\bra{\psi_s}, \qquad
 U 
 =
   e^{-i\mathcal{H}t}
   = \sum_s e^{-i\lambda_s t}  \ket{\psi_s}\bra{\psi_s}
   = \sum_s e^{2\pi i \phi_s} \ket{\psi_s}\bra{\psi_s}\,.
\end{align}

Let $\ket{\varsigma}$ denote a promised approximate
eigenvector $\ket{\psi}$. In general, accessing an 
approximate eigenvector is a very hard problem 
\cite{Kempe-approx-eigvect06,Aharonov-on-a-line07}.
Therefore the Abrams-Lloyd algorithm is in practice
limited to cases where a rough approximate eigenvector
can be efficiently obtained by classical 'ab initio' methods
which result in a known wave function \cite{Abrams-Lloyd99}.
Another possibility is to exploit the adiabatic state 
preparation procedure
\cite{Aharonov-state-preparation03,Guzik05}. 
However, the power of adiabatic state preparation is yet not
fully explored and its further study will determine the
usefulness of the Abrams-Lloyd algorithm.

The phase estimation scheme applied to the operator $U$
and an approximate eigenvector $\ket{\varsigma}$ reveals a
good estimator of the phase $\phi$ with a probability
$\abs{\braket{\psi}{\varsigma}}^2$. The corresponding 
energy eigenvalue is then calculated as
\begin{equation}
 \lambda = 2\pi(1-\phi)/t\,.
\end{equation}
Clearly, time $t$ should be chosen such that 
$\phi \approx 1/2$ in order to avoid a long sequence
of initial zeroes in the binary expansion of $\phi$.
A satisfying value for time $t$ can be calculated using
classical approximation methods for rough  estimates of
$\lambda$. The dependence on a rough estimate of $\lambda$
is not that limiting as the dependence on an approximate
eigenvector. Potential initial zeroes contribute to the
overall complexity of the algorithm as a (large) hidden
constant.

\begin{figure}[h!]
 \begin{center}
 \hspace{-3.5mm}
 \scalebox{0.95}
          {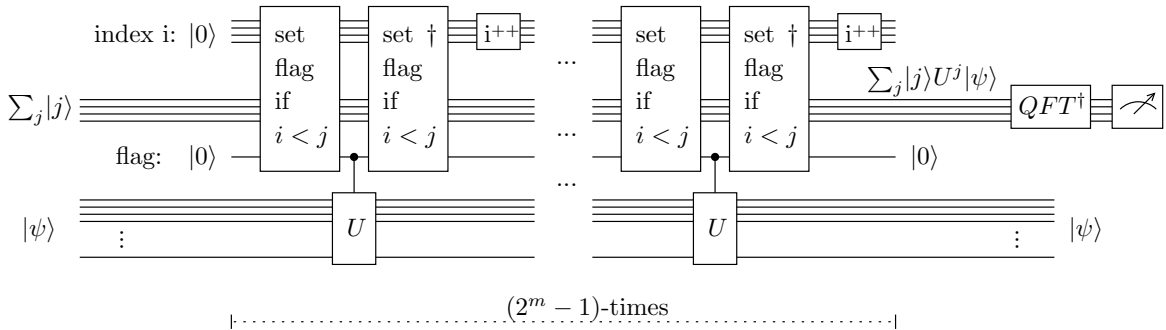}
 \end{center}
 \caption[The phase estimation scheme of Abrams and Lloyd]
         {The phase estimation scheme of Abrams and Lloyd.}
 \label{lloyd-phase-estimation}
\end{figure}

The scheme for phase estimation as proposed by Abrams and
Lloyd \cite{Abrams-Lloyd99} is shown in the 
Figure~\ref{lloyd-phase-estimation}. This scheme is 
functionally identical with the scheme in the 
Figure~\ref{complete-scheme-for-pea}. The only difference
is that an additional quantum logic is used to determine
how many times should be a controlled-$U$ gate applied to
a particular ancilla qubit $j_l$ in order to implement
a controlled-$U^{2^l}$ gate. This is intentional in order
to stress that the controlled-$U^{2^l}$ gates can be
implemented only by a long sequences of applications of 
a plain  controlled-$U$ gate. There is no  shortcut
similar to the one used in the period finding algorithm.

Assuming access to an approximate eigenvector and an
efficient implementation of $U$ (and  controlled-$U$
consequently) with respect to the size of
the problem $n$ (size of the system being simulated), the
algorithm for finding energy eigenvalues and preparing
eigenstates performs as follows. The scheme operates at
the Heisenberg limit and therefore the accuracy of 
order $1/2^m$ implies at most $2^m$ employments of the 
controlled-$U$ gate. The number $m$ is the amount of
ancillary qubits which are measured. In practice, the
precision is a fixed parameter and amounts to the overall
complexity as a hidden constant. For example, quantum
chemistry \cite{Guzik05} requires $m \approx 20$. 
This is in sharp contrast with the order finding algorithm
where the size of the ancilla register scales with the size
of the problem (order of few thousands of qubits for breaking
the RSA public-key cryptography). Considering $m$ a fixed
parameter, the algorithm is efficient by virtue of an
efficient circuit performing the controlled-$U$ gate. A
quantum simulation with $n>30$ outperforms classical
computers dramatically.

\subsubsection*{Efficient simulation of quantum systems
                with a local Hamiltonian}

A local Hamiltonian describing dynamics of a quantum 
system with $n$ particles can be written as a sum over
many local interactions
\begin{equation}
 \mathcal{H} = \sum_{k=1}^L \mathcal{H}_k\,,
\end{equation}
where $\mathcal{H}_k$ acts on at most a constant number
of particles, and $L$ is a polynomial in $n$. The key to 
quantum simulation algorithms is the convergence of the
Trotter formula which allows to approximate the evolution
\begin{equation}
 U=e^{-i \mathcal{H} t}\,.
\end{equation}

An exact simulation is not possible since the different
$\mathcal{H}_k$ do not have to commute with each other.
Therefore, the direct approach
\begin{equation}
  e^{-i\mathcal{H}t}
  =
   e^{-i\mathcal{H}_1 t}
   e^{-i\mathcal{H}_2 t}
   \ldots
   e^{-i\mathcal{H}_L t}
\end{equation}
is not correct.

{\theorem[Trotter formula]
 Let $A$ and $B$ be hermitian operators, then for any real
 $t$,
 \begin{equation}
  \lim_{q \rightarrow \infty}
  \left({e^{iAt/q}e^{iBt/q}}\right)^q
  = e^{i(A+B)t}.
 \end{equation}
 \sq
} 

By the definition of a matrix exponentiation and for some
$\delta < 1$, we have
\begin{align}
 e^{-i\mathcal{H}_A\delta }
 e^{-i\mathcal{H}_B\delta }
 &=
 \Bigl(I - i\mathcal{H}_A \delta  + O(\delta^2)\Bigr)
 \Bigl(I - i\mathcal{H}_B \delta  + O(\delta^2)\Bigr)\\
 \label{simulations-expansion1}
 &=
 I - i(\mathcal{H}_a+\mathcal{H}_b) \delta + O(\delta^2)\,,
\end{align}
and
\begin{equation}
 \label{simulations-expansion2}
 e^{-i(\mathcal{H}_A+\mathcal{H}_B)\delta}
 =
 I - i(\mathcal{H}_a+\mathcal{H}_b) \delta + O(\delta^2)\,.
\end{equation}
The asymptotic equations \eqref{simulations-expansion1}
and \eqref{simulations-expansion2} permit us to write
\begin{equation}
 \label{difference-of-order-delta-2}
 \hspace{2.1cm}
 e^{-i(\mathcal{H}_A + \mathcal{H}_B)\delta}
 =
 e^{-i\mathcal{H}_A\delta}
 e^{-i\mathcal{H}_B\delta}
  + O(\delta^2)\,.
 \hspace{1cm}
 \text{(for $\delta <1$)}
\end{equation}

Now, the equation \eqref{difference-of-order-delta-2} and
the convergence of the Trotter formula allow us to perform
the whole simulation by simulating the local evolution
operators over short discrete time slices and then repeat
the sequence of these local simulations a couple of times.
In particular, for $q>t$,
\begin{align}
  e^{-i\left(\mathcal{H}_1
  + \mathcal{H}_2
  + \cdots
  + \mathcal{H}_L\right)t}
  &=
  \left(e^{-i\mathcal{H}_1 (t/q)}
        e^{-i\mathcal{H}_2 (t/q)}
  \ldots
        e^{- i\mathcal{H}_L (t/q)}\right)^q
  + O\left(q\left(t/q\right)^2\right)\\
  &=
  \left(e^{-i\mathcal{H}_1 (t/q)}
        e^{-i\mathcal{H}_2 (t/q)}
  \ldots
        e^{- i\mathcal{H}_L (t/q)}\right)^q
  + O\left(t^2/q\right)\,.
\end{align}

Therefore by choosing $q \geq (t^2/\varepsilon)$, we
can implement the evolution $U=e^{-i\mathcal{H}t}$, where
$\mathcal{H}$ is a sum over local interactions, 
$\mathcal{H}=\sum_{k=1}^L \mathcal{H}_k$, within error
tolerance of $O(\varepsilon)$ using
$O(L\,(t^2/\varepsilon))$ 'basic' gates of the form
$e^{-i\mathcal{H}_k(t/q)}$. Since $L$ is polynomial in the
size of the problem $n$ and $\mathcal{H}_k$ acts on at 
most a constant number of particles, the whole simulation
requires an amount of gates of order
\begin{equation}
 O\Bigl(poly(n)\, \frac{t^2}{\varepsilon}\Bigr),
\end{equation}
and thus is efficient.

A similar result is known for Hamiltonians of
the form $\mathcal{H}= \bigotimes_{k=1}^L \sigma_{p(k)}$,
where $\sigma_{p(k)}$ are Pauli matrices.


 \newpage
 \section{Iterative phase estimation algorithm - IPEA}
 \label{sec:ipea}

\subsection{Motivation}
The Abrams-Lloyd algorithm is potentially the most important
quantum algorithm known so far. Even a small quantum
computer with as many as 100 qubits at total is expected to
outperform any classical computer in certain scopes of
quantum chemistry calculations. The issue of a relevance of
a small quantum computer for real chemistry problems has
been recently studied by Aspuru-Guzik {\it et. al.} 
\cite{Guzik05}.
In the work, the authors carried out a modified phase
estimation scheme where the size of the ancilla register is
fixed and a high accuracy of order $1/2^m$ is achieved via
$m$ iterations of the scheme. Using this scheme and an
adiabatic state preparation, simulated calculations of the
water and lithium hydride molecular ground-state energies
were performed. This required only 8 and 11 qubits for
storing an approximate eigenvector (the molecular
ground-state wave function) and four qubits in the
fixed-size ancilla register. These simulations strengthened
the conjecture of Abrams and Lloyd that quantum computers of
tens to hundreds of qubits can match and exceed classical 
calculations. The $k$-th iteration of the Aspuru-Guzik
scheme for phase estimation (recursively defined iterative
PEA) is shown in the Figure~\ref{gudzik-pea}.
\begin{figure}[h!]
 \begin{align*}
 &  
 \Qcircuit @R=1em @C=0.7em {
  \lstick{\ket{0}} 
  & \qw
  & \gate{H}
  & \qw
  & \qw
  & \qw
  & \ctrl{4} 
  & \multigate{3}{QFT^\dagger} 
  & \qw 
  & \meter 
  & \rstick{x_k^{(3)}} \cw \\
  \lstick{\ket{0}} 
  & \qw & \gate{H} 
  & \qw 
  & \qw 
  & \ctrl{3} 
  & \qw 
  & \ghost{QFT^\dagger} 
  & \qw 
  & \meter 
  & \rstick{x_k^{(2)}} \cw \\
  \lstick{\ket{0}} 
  & \qw 
  & \gate{H} 
  & \qw 
  & \ctrl{2} 
  & \qw 
  & \qw 
  & \ghost{QFT^\dagger} 
  & \qw 
  & \meter 
  & \rstick{x_k^{(1)}} \cw \\
  \lstick{\ket{0}} 
  & \qw 
  & \gate{H} 
  & \ctrl{1} 
  & \qw 
  & \qw 
  & \qw 
  & \ghost{QFT^\dagger} 
  & \qw 
  & \meter 
  & \rstick{x_k^{(0)}} \cw \\
  \lstick{\ket{\psi}} 
  & {/} \qw 
  & \qw 
  & \gate{V_k} 
  & \gate{V_k^2} 
  & \gate{V_k^{2^2}} 
  & \gate{V_k^{2^3}} & {/}\qw
  & \qw
 }\\[5mm]
 &  
 V_k = \Bigl( e^{-2\pi i \,\varphi_{k-1}} V_{k-1} \Bigr)^2
 \quad \text{for}\; k = 1,2,\ldots,m\,,\\
 & \text{where} \quad
 V_1 = U
 \quad \text{and} \quad
 \varphi_k = 0.x_k^{(0)} x_k^{(1)} x_k^{(2)} x_k^{(3)}\,.
\end{align*}
 \caption[The Aspuru-Guzik phase estimation scheme]
         {The Aspuru-Guzik phase estimation scheme.}
 \label{gudzik-pea}
\end{figure}

In the first iteration the operator $V_1$ is set to $U$,
where $U$ simulates the evolution of some physical
system. The input state $\ket{\psi}$ is promised to be an
eigenstate of $U$ with an eigenvalue $e^{2\pi i \phi}$. Let
us denote the binary expansion of 
$\phi$ as $\phi=0.\phi_1 \phi_2 \phi_3 \ldots$.
The result of the first iteration is a four bit
estimate $\varphi_1$ of the unknown phase $\phi$. With high
probability this estimate is accurate at least in order of
$1/2$. The estimate
$\varphi_1$ enters the next iteration with the operator 
$V_2=e^{-2\pi i (2 \varphi_1)}U^2$. The phase shift in the
operator $V_2$ effectively enables to measure a four bit
estimate $\varphi_2$ of a phase 
$ (2(\phi-\varphi_1) \bmod 1) =
 (0.\phi_2 \phi_3 \phi_4 \phi_5 \ldots
 \; - \;
 0.x_1^{(1)} x_1^{(2)} x_1^{(3)})$.
Now,
the number $(\varphi_1 + \varphi_2/2)$ constitutes an estimate
of $\phi$ which is accurate at least in order of $1/4$ with
high probability. Analogically, each additional iteration
yields one more bit of $\phi$. In the $k$-th iteration, we
work with a phase
\begin{equation}
\begin{array}{cccccccccccclcc}
 & \phi_1
 & \phi_2
 & \phi_3
 & \ldots
 & \ldots
 & \phi_{k-2}
 & \phi_{k-1}
 & .
 & \phi_{k}
 & \phi_{k+1}
 & \phi_{k+2}
 & \phi_{k+3}
 & \phi_{k+4}
 & \ldots \\ [1.5mm]
-& x_1^{(0)}
 & x_1^{(1)}
 & x_1^{(2)} 
 & x_1^{(3)} 
 & \\[1.5mm]
 \vdots
 &
 &
 &
 &
 & \vdots
 & \\
-&
 &
 &
 &
 & \ldots 
 & x_{k-2}^{(0)} 
 & x_{k-2}^{(1)} 
 & .
 & x_{k-2}^{(2)} 
 & x_{k-2}^{(3)}\\[1.5mm]
-&
 &
 &
 &
 & \ldots
 &
 & x_{k-1}^{(0)}
 & .
 & x_{k-1}^{(1)} 
 & x_{k-1}^{(2)}\
 & x_{k-1}^{(3)} 
 & ,
\end{array}
\end{equation}
measure a corresponding estimate $\varphi_k$,
and the final estimate of $\phi$ accurate to at least $m$
binary digits is given by
\begin{equation}
 \blacktriangleleft\phi\blacktriangleright
 \; = \,
 \sum_{k=1}^m \varphi_k/2^{k-1}\,.
\end{equation}
In conclusion, the Aspuru-Guzik scheme requires $O(m)$
iterations, $O(4m)=O(m)$ measurements, and the plain
controlled-U gate is employed $O(2^{m+3})=O(2^m)$ times in
order to create required operators controlled-$V_k^c$ for
$c=1,2,4,8$.

{\example[Aspuru-Guzik phase estimation]\quad\\
 Let $\phi=0.011101101\ldots$. The progress of
 obtaining gradually more accurate estimates of $\phi$ is
 captured in the following table.
 \begin{table}[h!]
 \begin{center}
 \begin{tabular}{c|l|l|l}
 iteration $k$  & working phase       &
 result $\varphi_k$ & final estimate \\ \hline
 $1$            & $0.011101101\ldots$ &
 $0.0111$           & $0.0111   \rightarrow 0.1$\\
 $2$            & $0.00001101\ldots$  &
 $0.0001$           & $0.01111  \rightarrow 0.10$\\
 $3$            & $0.1111101\ldots$   &
 $0.0000$           & $0.011110 \rightarrow 0.100$
 \end{tabular}
 \caption[Sample run of the Aspuru-Guzik phase estimation]
         {Sample run of the Aspuru-Guzik phase estimation.}
 \sq
 \end{center}
 \end{table}
}

\begin{figure}[h!]
 \[
 \Qcircuit @R=1em @C=0.7em {
   \lstick{\ket{0}}    
   & \gate{H} 
   & \ctrl{1} 
   & \qw 
   & \gate{H}
   & \meter 
   & \rstick{x_k} \cw \\
   \lstick{\ket{\psi}} 
   & {/} \qw      
   & \gate{U^{2^{k-1}}} 
   & \qw
   & {/} \qw
   & \qw }
 \]
 \caption[Kitaev's phase estimation circuit]
         {Kitaev's phase estimation circuit.}
 \label{kitaev-pea}
\end{figure}
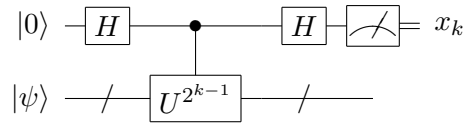

Another scheme for phase estimation which can be performed
in an iterative manner with a fixed-sized ancilla register
is the original procedure used by Kitaev in \cite{Kitaev96}.
Let $\phi$ denote the unknown phase of an eigenvalue
$e^{2\pi i \phi}$ related to some unitary operator $U$ and
its eigenstate $\ket{\psi}$. Kitaev observed that if one
knows estimates of the numbers 
$2^{k-1}\phi \!\pmod 1,$ for $k=1,2,\ldots m$, then there
exists a polynomial classical algorithm which can
reconstruct $\phi$ to at least $m$ binary digits. The 
circuit which determines an estimate of 
$2^{k-1}\phi \!\pmod 1$ is shown in the 
Figure~\ref{kitaev-pea}.

\begin{figure}[h!]
 \begin{center}
  \includegraphics[scale=0.65]
                  {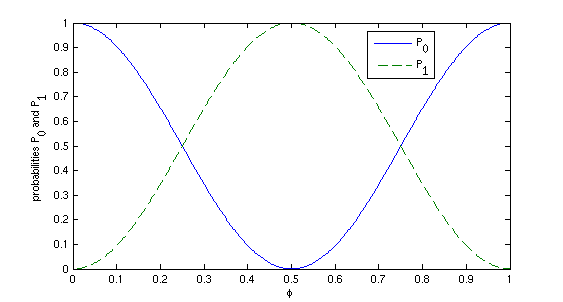}
 \end{center}
 \caption[A plot of outcome probabilities in the Kitaev PEA]
         {A plot of outcome probabilities in the Kitaev PEA.}
 \label{kitaev-probabilities-plot}
\end{figure}

The outcome $x_k=0$ is observed with probability 
$P_0=\cos^2(\pi \,2^{k-1} \phi)$. Accordingly, the
probability that $x_k=1$ is 
$P_1=1-P_0=\sin^2(\pi \,2^{k-1} \phi)$. See
Figure~\ref{kitaev-probabilities-plot}.

Exploiting independent trials of the circuit and using 
the Chernoff bound, it is possible to determine whether
$2^k\phi \!\pmod 1$ is closer to zero $("0.0")$, i.e. 
$P_0\geq P_1$, or to one half ("$0.1$") with error
probability
$\leq \varepsilon_k$ after $O(\log(1/\varepsilon_k))$
trials. By setting $\varepsilon_k \leq \varepsilon/m$,
all the values $x_k$, for $k=0,1,\ldots, m-1$, can be 
correctly determined with error probability
$\leq \epsilon$. Therefore this scheme requires at total
$O(m\log(m/\varepsilon))$ iterations (measurements) for 
estimating $\phi$ to $m$ binary digits. The plain
controlled-U gate is employed $O(2^m\log(m/\varepsilon))$ 
times at total in order to implement required operators
controlled-$U^{2^k}$.

Compared to the Aspuru-Guzik scheme, the Kitaev phase
estimation is asymptotically slower by a small logarithmic
factor.
However, in the context of estimating energy eigenvalues
where $m$ is a fixed constant between $20$ and $30$, the
performance of the Kitaev algorithm is actually better.
Besides, it requires only a single ancillary qubit and the
circuit is easier to implement.

The question arises, whether there exists an iterative
scheme for phase estimation (IPEA) which does not contain
the logarithmic factor (reaches the true Heisenberg limit)
and utilizes only a single ancilla qubit. Such a scheme
would be of utmost importance for experimental
implementation with a very limited number of qubits. The
answer is yes and such a scheme can be derived from the
textbook (QFT-based) phase estimation 
(Figure~\ref{complete-scheme-for-pea}). The derivation 
requires knowledge of how the entanglement is localized in
the circuit and understanding of the information flow in
the measurement in the Fourier basis. In particular, the
key point is a discovery of an adaptive implementation of
the measurement in the Fourier basis (so called 
semiclassical QFT) by Griffiths and Niu \cite{SemiQFT96}.
The
relevance of incorporating the semiclassical QFT into the
Shor's circuit for factoring was first noticed by 
C. Zalka \cite{Zalka98}, then by Mosca and Ekert
\cite{Mosca99} in the
context of solving the hidden subgroup problem by means of
the phase estimation approach, and by Childs {et. al.}
\cite{Childs00} for distinguishing Hamiltonians in some cases.

Regarding the problem of finding energy eigenvalues, Abrams
and Lloyd \cite{Abrams-Lloyd99} and Aspuru-Guzik {\it et. al.}
\cite{Guzik05} do
not employ IPEA although parts of their respective works
focus on a relevance of the algorithm for a quantum
computer with a very limited amount of qubits. Properties
of the IPEA in the context of finding energy eigenvalues
have been recently studied by Knill {\it et. al.} 
\cite{Knill07} and Dobšíček {\it et. al.} \cite{DobsicekPEA07}.

\subsection{IPEA derived from the Kitaev phase estimation}
\label{sec:ipea-derived-from-kitaev}
In contrast to the QFT-based PEA 
(Figure~\ref{complete-scheme-for-pea}), the Kitaev PEA
(Figure~\ref{kitaev-pea}) does not
explicitly use the quantum Fourier transform as such. 
Therefore, in some sense, the Kitaev PEA allows to solve
abelian HSP-like problems without using the relatively
extensive knowledge behind quantum Fourier sampling.
Moreover, it is possible to derive the IPEA scheme from
the Kitaev PEA by incorporating a classical postprocessing
algorithm used to construct the estimate 
$\blacktriangleleft\phi\blacktriangleright$ into the quantum
circuit. Such a result may be quite inspiring in the further
development of quantum algorithms as the usual approach is
to perform as much as possible of necessary calculations
classically, since quantum resources are considered 
'expensive'. 

\begin{figure}[h!]
 \begin{center}
  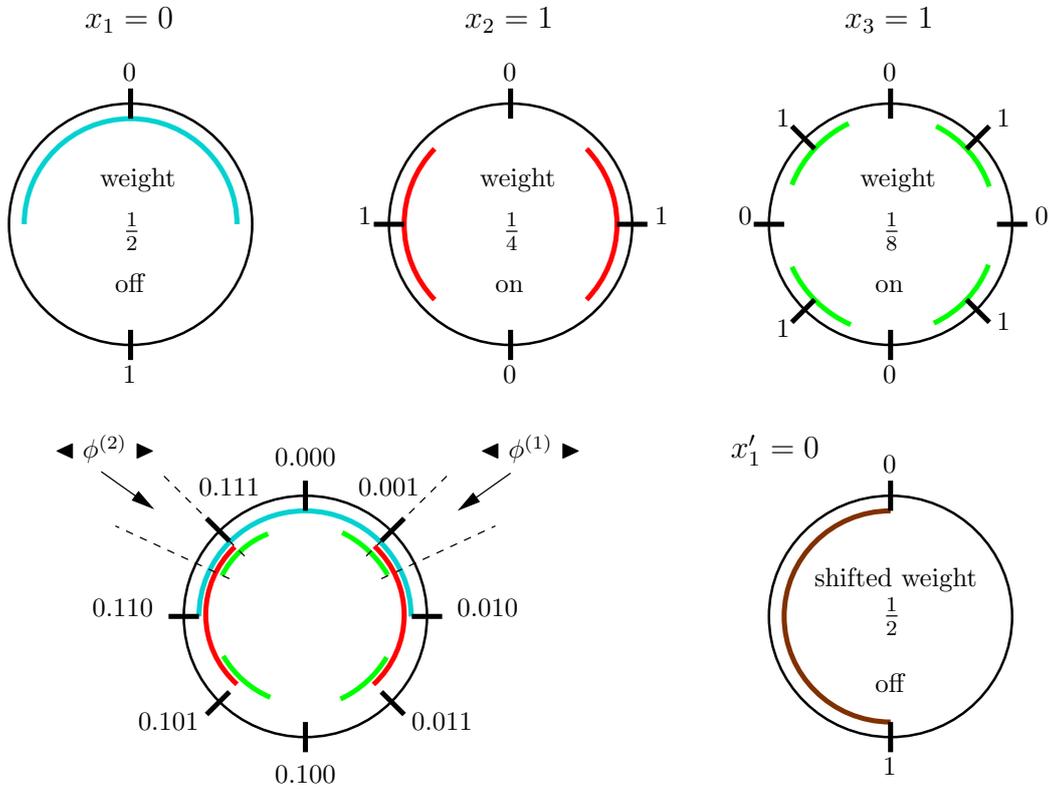
 \end{center}
 \caption[Weighted interval intersection method]
         {A weighted interval intersection method as a
	 postprocessing algorithm for the Kitaev PEA. In
	 general, there are two candidate estimates and
	 an intersection with additional shifted weight-1/2
	 interval determines the final estimate. The figure
	 shows a sample run for $\phi=0.110110\ldots$ which
	 gives the following estimates $x_k$ for values of 
	 $2^{k-1}\phi\pmod 1$: $x_1=0$, $x_2=1$, and
	 $x_3=1$.}
 \label{interval-intersection-method}
\end{figure}

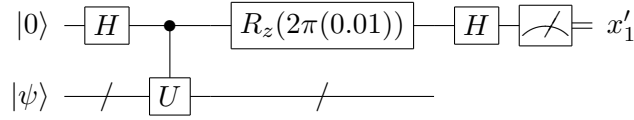
\begin{figure}[h!]
 \[
 \Qcircuit @R=1em @C=0.7em {
   \lstick{\ket{0}}
   & \gate{H}
   & \ctrl{1}
   & \qw
   & \gate{R_z(2\pi(0.01))}
   & \qw
   & \gate{H}
   & \meter
   & \rstick{x'_1} \cw \\
   \lstick{\ket{\psi}}
   & {/} \qw
   & \gate{U}
   & \qw
   & {/} \qw
   & \qw }
 \]
 \caption[Additional iteration in the Kitaev PEA]
         {Additional iteration in the Kitaev PEA.}
 \label{additional-iteration-kitaev}
\end{figure}

There are many possible efficient classical algorithms
suitable for postprocessing in the Kitaev PEA 
(no specific algorithm was given by Kitaev in \cite{Kitaev96}).
I 
propose to use an algorithm which can be described as a
weighted interval intersection method. The core of this
method is outlined in the 
Figure~\ref{interval-intersection-method}.
To each of the results $x_k$ is assigned an interval with
a weight $1/2^{k}$. The weight also determines the spanning 
range of the interval. By making intersections of the two
largest intervals per step, we end up with two small 
intervals of size $1/2^{m+1}$ after $m-1$ steps. Let us
associate a candidate estimate 
$\blacktriangleleft\phi^{(1)}\blacktriangleright$ to the 
value in the center of one of the two small intervals. This
ensures that the candidate contains information sufficient
to reveal $\phi$ to $m+2$ binary digits. Accordingly, 
there is a second candidate
$\blacktriangleleft\phi^{(2)}\blacktriangleright
\;=\;
1\;-\;\blacktriangleleft\phi^{(1)}\blacktriangleright$. The
last step is to decide which candidate is the right one. We 
perform one more intersection with an interval assigned to
the outcome $x'_1$ of an additional iteration of the Kitaev
scheme where to the phase $2\pi\phi$ (contained in the
ancilla qubit) is added a phase $2\pi(0.01)$. A circuit
corresponding to this additional iteration is shown in the
Figure~\ref{additional-iteration-kitaev}.

An important observation from the 
Figure~\ref{interval-intersection-method}
is that an interval with the smallest weight imposes the
largest constraints to the spanning range of the other
intervals. In other words, an interval with a small weight,
if determined at first, can help to keep the spanning range
of other intervals small. While this is not a good classical
strategy (there are too many small ranges to keep track 
of), nothing prevent us to obtain the least significant
bit $x_m$ at first and use this information to improve the
search for the bit $x_{m-1}$ in the quantum circuit. This is
a sort of feedback improvement similar to the one used in
the Aspuru-Guzik scheme.

\begin{figure}[h!]
 \[
 \Qcircuit @R=1em @C=0.7em {
   \lstick{\ket{0}}
   & \gate{H}
   & \ctrl{1}
   & \qw
   & \gate{R_z(\omega_k)}
   & \qw
   & \gate{H}
   & \meter
   & \rstick{x_k} \cw \\
   \lstick{\ket{\psi}}
   & {/} \qw
   & \gate{U^{2^{k-1}}}
   & \qw
   & {/} \qw
   & \qw }
 \]
 \caption[Iterative phase estimation scheme (IPEA)]
         {The $k$-th iteration of the iterative phase
          estimation (IPEA). The feedback angle depends
          on the previously measured bits through
          $\omega_k = -2\pi(0.0x_{k+1}x_{k+2}\ldots x_m)$,
	  and $\omega_m=0$. Note that $k$ is iterated
	  backwards from $m$ to $1$.}
 \label{ipea-circuit}
\end{figure}
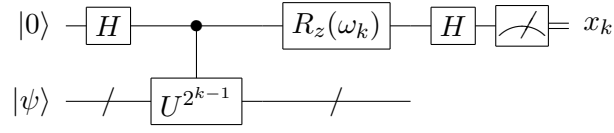

Let us write $\phi$ as $\phi=\tilde{\phi} + \delta2^{-m}$,
where $\tilde{\phi}=0.\phi_1\phi_2\ldots\phi_m$ denotes
the first $m$ bits of the binary expansion and 
$0\leq \delta <1$ is a reminder. In order to determine
the bit $\phi_m$, we perform an iteration of the Kitaev PEA
for $k=m$, i.e. the gate controlled-$U^{2^{m-1}}$ is used.
A single trial gives us the probability 
$P_0=\cos^2(\pi(0.\phi_m + \delta/2))$ to measure $x_m=0$
and the probability
$P_1=\sin^2(\pi(0.\phi_m + \delta/2))$ to measure $x_m=1$.
Thus for
\begin{equation}
\phi_m=\begin{cases}
 0 & \Rightarrow P_0=\cos^2(\pi\delta/2),\\
 1 & \Rightarrow P_1=\sin^2(\pi/2 + \pi\delta/2) 
        =\cos^2(\pi\delta/2)\,,
\end{cases}
\end{equation}
and therefore
\begin{equation}
 \label{intrinsic-error-from-delta}
 P_{m}
 = P(x_m=\phi_m) 
 = \cos^2(\pi\delta/2)\,.
\end{equation}
In the next iteration of the Kitaev PEA, $k=m-1$, the phase
of the ancilla qubit is 
$2\pi(0.\phi_{m-1}\phi_m + \delta/4)$. Assuming we know the
bit $\phi_m$, we can perform a Z rotation with angle
$\omega_{m-1} = -2\pi(0.0x_m)$. Then the conditional 
measurement probability becomes
$P_0=\cos^2(\pi(0.\phi_{m-1}0 + \delta/4)$ and
\begin{equation}
 P_{m-1}
 = P(x_{m-1}
 = \phi_{m-1} \;|\; x_m = \phi_m)
 = \cos^2(\pi\delta/4)\,.
\end{equation}
Note that $P_{m-1}$ is significantly larger than $P_m$.
Accordingly, we use bits $x_m$ and $x_{m-1}$ to improve
the probability that $x_{m-2}$ will be equal to 
$\phi_{m-2}$. The $k$-th iteration of this IPEA is shown
in the Figure~\ref{ipea-circuit}.
Generally, using the feedback angle 
$\omega_k=-2\pi(0.0x_{k+1}x_{k+2}\ldots x_m)$
the conditional probability $P_k$ for a bit $x_k$ to
be equal to $\phi_k$ is
\begin{equation}
 P_k = \cos^2(\pi\,2^{k-1}\,(\delta 2^{-m}))\,.
 \label{probability-p-k}
\end{equation}
Note that the equation \eqref{probability-p-k} shows an 
exponential increase of $P_k$ with decreasing $k$.
The overall probability for the IPEA to extract 
$\tilde{\phi}=0.\phi_1\phi_2\ldots\phi_m$, i.e. an $m$-bit
rounded down estimate of $\phi$, is then
\begin{align}
P_{down}(0\leq \delta < 1)
  =  P_m P_{m-1} \cdots P_1 
 &=  \prod_{k=1}^m  \cos^2(\pi \,2^{k-1}\, (\delta2^{-m}))
 \\[2mm]
 &=
 \label{probability-p-down-ipea}
 \begin{cases}
   1 
   & 
   \text{for } \delta = 0, \\[1.5mm]
   \frac{\sin^2(\pi\delta)}
        {2^{2m}\sin^2(\pi \delta 2^{-m})}\,,
   &
   \text{for } 0 < \delta < 1\,,
 \end{cases}
\end{align}
using the trigonometric identity
\begin{equation}
 \prod_{k=0}^{m-1} \cos(2^k \alpha) 
 = \frac{\sin(2^m \alpha)}{2^m \sin(\alpha)}
\end{equation}
for $\alpha \neq 0$. Using a similar derivation, the 
probability to extract $\tilde{\phi} + 2^{-m}$ (rounding
up) is
\begin{equation}
 \label{probability-p-up-ipea}
 P_{up}(0 \leq \delta < 1)
 = \frac{\sin^2(\pi(1-\delta))}
        {2^{2m}\sin^2(\pi (1-\delta) 2^{-m})}\,.
\end{equation}

Equations \eqref{probability-p-down-ipea} and
\eqref{probability-p-up-ipea} are identical to the QFT-based
PEA equations \eqref{probability-p-down-qft-based} and
\eqref{probability-p-up-qft-based}, respectively, 
see page \pageref{probability-p-down-qft-based}. Therefore,
the IPEA shares with the QFT-based PEA the same lower bound
on the probability to observe an estimate of $\phi$ accurate
to at least $m$ binary digits with error probability
$\varepsilon < 1 - 8/\pi^2$. In conclusion, the IPEA 
requires $O(m)$ iterations (measurements) in order to 
achieve the precision of order $1/2^m$ and the plain
controlled-$U$ gate is employed $O(2^m)$ times. Thus the
true Heisenberg limit is achieved utilizing only a single
ancilla qubit.

\begin{figure}[h!]
 \begin{center}
  \includegraphics[scale=0.65]
   {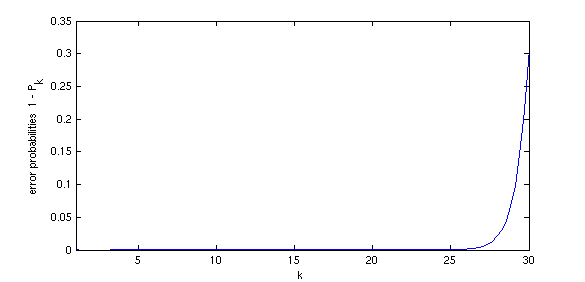}
 \end{center}
 \caption[Bitwise error probabilities in the IPEA]
         {Bitwise error probabilities 
          $\sin^2(\pi\,2^{k-1}\,(\delta 2^{-m}))$
	  for $m=30$ and $\delta=3/8$.}
 \label{least-significant-bits-error-probabilities}
\end{figure}

Similarly to the QFT-based PEA, success probability can by
amplified up to $1-\varepsilon$\, for $\varepsilon > 0$ by
extracting $m'=m+O(\log(1/\varepsilon))$ bits and keeping
only $m$ most significant ones. However, this  approach is
not favorable with respect to the costs of implementing
gates controlled-$U^{2^k}$ for $k>m$. The plain 
controlled-$U$ gate would be needed 
\begin{equation}
 O\Bigl(2^m \;2^{\log(1/\varepsilon)}\Bigr)
\end{equation}
times.
Per bit repetitions and majority voting is a better option.
First, we observe that the bare bitwise error probability 
\begin{equation}
 1-P_k=\sin^2(\pi\,2^{k-1}\,(\delta 2^{-m}))
\end{equation}
decreases exponentially with decreasing $k$. This implies
that only logarithmically many least significant bits have 
non-negligible error probabilities $1-P_k$; see 
Figure~\ref{least-significant-bits-error-probabilities}.

Additionally, the probability of a bit
$x_k$ to be correctly determined after $N_k$ independent
trials is
\begin{equation}
 \label{probability-amplification}
 P'_k 
 = \sum_{s=0}^{\lfloor \frac{N_k-1}{2} \rfloor}
   \binom{N_k}{s}
   (P_k)^{N_k-s}(1-P_k)^s \,,
\end{equation}
given that $x_k$ is chosen using majority voting from the
results. Thus the effective error probability decreases
exponentially with the number of repetitions, according to
the binomial distribution. Therefore repeating the
iterations for the $O(\log(1/\varepsilon))$ least
significant bits an $O(\log(1/\varepsilon))$ number of
times gives an error probability smaller than $\varepsilon$,
independently of $m$. The total amount of iterations
(measurements) then goes like $O(m+\log^2(1/\varepsilon))$
and the plain controlled-$U$ gate is utilized 
\begin{equation}
 \sum_{k=1}^m 2^{k-1} 
 \; + \;
 \lceil \log(1/\varepsilon) \rceil
 \sum_{t=1}^{\lceil \log(1/\varepsilon) \rceil}
 2^{m-t}
 \; = \;
 O(2^m\log(1/\varepsilon))
\end{equation}
times. The scaling $O(2^m\log(1/\varepsilon))$ is a
consequence of the fact that the least significant bits are the
most expensive bits in terms of constructing the corresponding
controlled gates. However, the factor $\log(1/\varepsilon)$
is more pleasing than $2^{\log(1/\varepsilon)}$.

\subsection{QFT-based PEA \,versus\, IPEA}
\label{sec:qft-based-pea-versus-ipea}
\begin{figure}[h!]
 \begin{center}
  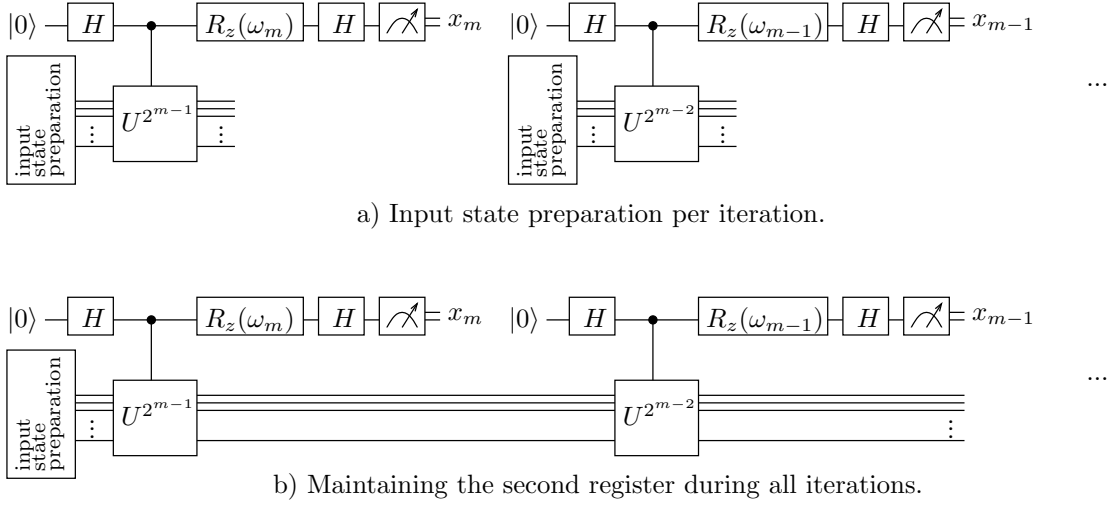
 \end{center}
 \caption[The IPEA and an input eigenstate]
         {The IPEA and an input eigenstate.}
 \label{maintaining-the-eigenstate}
\end{figure}

The IPEA can be considered as a direct viable alternative
to the QFT-based PEA, especially for experimental quantum
devices with a very limited amount of qubits. Next to a
smaller number of qubits the IPEA requires, an additional 
advantage for experiments is the possibility to exchange a
long coherence time of an input eigenstate for an input
state preparation per iteration (and/or repetition). See 
Figure~\ref{maintaining-the-eigenstate}. The possibility of
having a choice is valid for problems where we work with a
particular (approximate) input eigenstate since we can
prepare this eigenstate for each iteration.

Thanks to the linear property of any PEA scheme 
(Eq.~\ref{eq:pea-linear-property},
Section~\ref{sec:pea-linear-property}), it is not a problem
to work with an approximate eigenstate at the cost of some
additional randomness, however, this randomness now enters
(complicates) each iteration. Depending on the quality of 
the approximation of the eigenstate, repetitions of each
iteration are needed to make sure that the measured bit is
related to the eigenvalue corresponding to the desired
eigenstate. Otherwise, we may end up in a situation where
we work with a different eigenvalue in each iteration.

In problems where an eigenstate is completely unknown and
a state with special properties is used instead (e.g. a 
weighted superposition of eigenstates in the factoring 
problem), the corresponding register must be maintained
during all iterations and repetitions. This is due to the
fact that each measurement on the ancilla qubit results in
a collapse in the second register. The resulting 'collapsed'
state is unknown to us.

\subsubsection*{Optimal quantum circuits for general phase
                estimation}
The question of finding optimal quantum circuits for general
phase estimation has been addressed by Dam {\it et. al.} 
\cite{Dam-optimal-pea07}. It was shown that the QFT-based PEA
is indeed an
optimal circuit (so are derived iterative schemes) and
further optimization with respect to some cost function can
be done only by optimizing the input state in the ancilla
register. A common cost function is the window cost function
that allows any error smaller than
$\xi$:
\begin{equation}
 C_W^\xi(\phi,\blacktriangleleft\phi\blacktriangleright)
 =
  \begin{cases}
    0 
   &
    \text{if }
    \abs{ \phi
       \;- \blacktriangleleft\phi\blacktriangleright}
    < \xi,\\
    1
   & 
    \text{otherwise}.
  \end{cases}\
\end{equation}
Minimization of this cost function leads to the
ancilla input state
\begin{equation}
 \ket{ancilla_{opt_W}}
 = \frac{1}{\sqrt{2^m}}
   \sum_{j=0}^{2^m-1}
   \ket{j}\,.
\end{equation}
This state is created by the initial Hadamard gates in the
default QFT-based PEA setting. Since the state is a rather
simple product state, the initial Hadamard gate in each 
IPEA iteration serves as well. Thus the QFT-based PEA and
the IPEA are capable of producing optimal estimates with
respect to the window cost function. Another commonly used
cost function is the (in)fidelity cost function
\begin{equation}
 C_F (\phi,\blacktriangleleft\phi\blacktriangleright)
 = 
  \sin^2
  \Bigl(
   \frac{\phi\;- \blacktriangleleft\phi\blacktriangleright}
	{2}
  \Bigr)\,,
\end{equation}
and the minimum cost is achieved with the initial state
\begin{equation}
 \
 \ket{ancilla_{opt_F}}=
 \sum_{j=0}^{2^m-1}
 \sqrt{\frac{2}{2^m+1}}
 \sin\left(\frac{(j+1)\pi)}{2^m+1}\right)\ket{j}\,.
\end{equation}
The state $\ket{ancilla_{opt_F}}$ is derived in 
\cite{Bartlett-review07}.
An important observation is that this state is an entangled
state. In this case, an analytical calculation shows that
the information flow is such that it cannot be simulated by
the IPEA. This implies that there are cost functions for
which the IPEA is not able to yield optimal estimates. To
which degree entangled ancilla input states can be
approximated in the IPEA is an open question.


 \section{IPEA applications}
 \label{ipea-apps}
 

\subsection{Iterative phase estimation: A two-qubit test-bed
            application}

The ability to perform single qubit gates and two-qubit
(entangling) gates has been demonstrated for all major
quantum computation technology candidates such as 
superconducting quantum computation, cavity quantum 
electro-dynamic computation and ion trap quantum computation.
Having a very limited amount of not yet reliable qubits, the
question is what kind of test-bed applications can be
performed. For example, five to seven qubits are already
sufficient to run the smallest instances of Shor's or
Grover's algorithms, but experiments with only two qubits
have been so far limited only to testing Bell's inequality
\cite{Testing-bell-ineq04} or doing quantum state tomography
\cite{State-tomography06}. Phase
estimation in its iterative form (IPEA) shows up as a new
test-bed application with only two qubits.

Here, I deal with three different simulations of the IPEA
applied to a simple Z-rotation operator
\begin{equation}
 \label{simple-z-rotation-operator}
 U=\left(\begin{array}{cc}
   e^{-i\alpha} & 0 \\
   0 & e^{i\alpha} \end{array}\right)
  =\left(\begin{array}{cc}
   e^{-2\pi i \phi} & 0 \\
   0 & e^{2\pi i \phi} \end{array}\right)
\end{equation}

on a two qubit system. In the simulations, the underlying
Hamiltonians 
\begin{align}
 \label{xx-coupling-hamiltonian}
 \mathcal{H}_{xx}
  =&
    B_{x_1} X^{(1)}
  + B_{z_1} Z^{(1)}
  + B_{x_2} X^{(2)}
  + B_{z_2} Z^{(2)}
  + \gamma (X^{(1)} \otimes X^{(2)})\,,
    \quad \text{(XX coupling)}\\[1mm]
 \label{zz-coupling-hamiltonian}
 \mathcal{H}_{zz}
  =&
    B_{x_1} X^{(1)}
  + B_{z_1} Z^{(1)}
  + B_{x_2} X^{(2)}
  + B_{z_2} Z^{(2)}
  + \gamma (Z^{(1)} \otimes Z^{(2)})\,,
    \quad\; \text{(ZZ coupling)}
\end{align}
where $B_{z_1,z_2}, B_{x_1,x_2}, \gamma$ are tunable
parameters (absorbing the reduced Planck constant $\hbar$)
and $Z,X$ are Pauli matrices, and noise models were
chosen with respect to \cite{GJ07}. 
The Hamiltonians are related to superconducting 
quantum computation \cite{NanoHandbook06}.

The operator \eqref{simple-z-rotation-operator} has been
identified during analysis as especially suitable since:
(1) it is diagonal in the qubit eigenbasis, thus the initial
preparation of its eigenstate is straightforward, (2) the
phase to be measured can be controlled directly, and (3)
controlled powers of this gate can be done with a very
short circuit. These properties allow to keep the 
complexity of a real experiment low and isolate well the
performance of single qubit rotations and coupling terms.

\subsubsection*{Simulations with focus on single qubit
rotations performance and overall setup stability}

In general, any single qubit gate $U \in SU(2)$ can be
decomposed to at most three successive rotations about two
non-parallel axis. Given the form of Hamiltonians
\eqref{xx-coupling-hamiltonian} and 
\eqref{zz-coupling-hamiltonian}, the natural arising single
qubit rotations are $R_z$ and $R_x$ for both of them, and we
can write
\begin{equation}
 \label{u-z-x-decomposition}
 U = R_{z}(\alpha) \, 
     R_{x}(\theta) \,
     R_{z}(\beta)\,.
\end{equation}

Using the Z-X decomposition \eqref{u-z-x-decomposition}, a
decomposition of the gate $U^2$ can be obtained as
\begin{equation}
\begin{split}
 U^2 &= \Bigl(
        R_{z}\left(\alpha\right) \;
        R_{x}\left(\theta\right) \;
        R_{z}\left(\beta\right)  \;
	\Bigr)\Bigl(
        R_{z}\left(\alpha\right) \;
        R_{x}\left(\theta\right) \;
        R_{z}\left(\beta\right)
	\Bigr)
	\\
     &= R_{z}\left(\alpha\right) \;
        R_{x}\left(\theta\right) \;
        R_{z}\left(\alpha + \beta\right) \;
	R_{x}\left(\theta\right) \;
	R_{z}\left(\beta\right) \\
     &= R_{z}\left(\alpha + \nu_1\right) \;
        R_x  \left(\nu_2\right) \;
	R_z  \left(\beta + \nu_3\right)
\end{split}
\end{equation}
for $\nu_1, \nu_2$ and $\nu_3$ such that
\begin{equation}
  \label{angles-computation}
  R_x \left(\theta\right)      \;
  R_z \left(\alpha+\beta\right)\; 
  R_x \left(\theta\right)
= R_z \left(\nu_1\right)    \;
  R_x \left(\nu_2\right)    \;
  R_z \left(\nu_3\right)\,.
\end{equation}
Solving the equation \eqref{angles-computation} gives us
\begin{align}
 \nu_2 
 =&
    \left\{
     \begin{array}{ll}
        2 \arcsin 
	  \left(
	    \sin\theta\cos\frac{\alpha+\beta}{2}
	  \right),
     & 
        \text{ if (*) or (**), }
      \\[2mm]  
        2 \left(
	    \pi - \arcsin
	          \left(
		  \sin\theta\cos\frac{\alpha+\beta}{2}
		  \right)
	  \right),
     & 
        \text{ otherwise,}
     \end{array}
    \right.\\
  & 
    \begin{array}{ll}
      \nonumber
      \text{ (*) } 
      &
      \sin\left(\alpha+\beta\right)=0 \;\, 
      \text{ and } \;\,
      \cos\theta > 0 \; ,\\[0.5mm]
      \text{ (**) }
      &
      \cos\theta\sin\left(\alpha+\beta\right) > 0 \; ,
    \end{array}\\[1mm]
 \nu_1 
 =&
    \arcsin
    \left(
      \sin\frac{\alpha+\beta}{2}\cos\frac{\nu_2}{2}
    \right),\\[1mm]
 \nu_3
 =& 
    \nu_1.
\end{align}

Analogically, a decomposition of the gate $U^{2^k}$ is
easily calculated in the following loop:
\begin{equation}
\begin{tabular}{l}
 \textbf{repeat} $k$ \textbf{times}:\\
 \qquad$\;$\begin{tabular}{l}
 \begin{tabular}{l}
      calculate $\nu_2$\\
      calculate $\nu_1$
     \end{tabular}\\
     \begin{tabular}{lll}
      $\alpha$& $\Leftarrow$& $\alpha + \nu_1$\\
      $\theta$& $\Leftarrow$& $\nu_2$\\
      $\beta$ & $\Leftarrow$& $\beta +  \nu_1$
     \end{tabular}
 \end{tabular}\\[12mm]
 \textbf{return} \;
   $U^{2^k}
   \,\Leftarrow\,
   R_z\left(\alpha\right) \,
   R_x\left(\theta\right) \,
   R_z\left(\beta\right)$\;.
\end{tabular}
\end{equation}

Knowing the angles $\alpha,\theta,\beta$ corresponding 
to the single qubit gate $U^{2^k}$, we can implement
(Theorem~\ref{controlled-u-simulation}) its controlled
version with a circuit
$$
\Qcircuit @R=1em @C=0.7em {
 & \qw
 & \ctrl{1}
 & \qw
 & \ctrl{1}
 & \qw
 & \qw
 \\
 & \gate{C}
 & \targ
 & \gate{B}
 & \targ
 & \gate{A}
 & \qw
}
$$
where
\begin{equation}
\begin{split}
 A &= R_z \left(\alpha\right)    \;
      R_x \left(\theta/2\right)  \;
      R_z \left(-\pi/2\right),   \\
 B &= R_z \left(\pi/2\right)     \;
      R_x \left(-\theta/2\right) \,
      R_z \left(-(\alpha+\beta+\pi)/2\right),\\
 C &= R_z \left((\beta-\alpha+\pi)/2\right).
\end{split}
\end{equation}

\begin{figure}[h!]
  $$
  \scalebox{0.9}{
  \Qcircuit @R=1em @C=0.7em {
    &  \ctrl{1}
    & \qw
    &
    &
    & \qw
    & \ctrl{1} 
    & \qw 
    & \ctrl{1} 
    & \qw 
    & \qw \\
    & \gate{ \left( \begin{array}{cc}
                e^{-i\alpha} & 0 \\
                0            & e^{i\alpha}
              \end{array}            \right)^{2^k} } 
    & \qw
    & \push{\rule{.3em}{0em}=\rule{.3em}{0em}} 
    &
    & \gate{R_z( 2^k\alpha\!\!\!\!\pmod {2\pi})} 
    & \targ 
    & \gate{R_z(-2^k\alpha\!\!\!\!\pmod {2\pi})}
    & \targ 
    & \qw 
    & \qw
 }}
 $$
 \caption[Controlled powers of the simple Z-rotation
          operator]
         {Controlled powers of the simple Z-rotation
	  operator.}
 \label{ctrl-powers-z-rot-general-construction}
\end{figure}
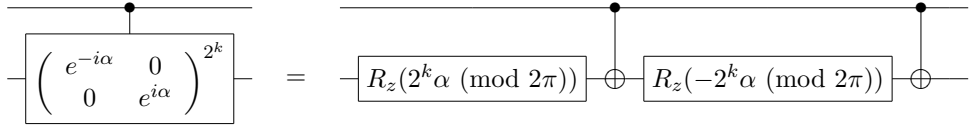

This construction can be pretty much simplified for
the operator \eqref{simple-z-rotation-operator}. See the
circuit in the 
Figure~\ref{ctrl-powers-z-rot-general-construction}.
Essentially, the circuit exploits the identity 
$X R_z(\alpha) X = R_z(-\alpha)$.

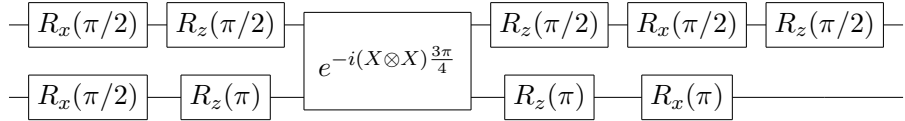
\begin{figure}[h!]
 $$
 \scalebox{0.96}{
 \Qcircuit @R=1em @C=0.7em {
  & \gate{R_x(\pi/2)}
  & \gate{R_z(\pi/2)}
  & \multigate{1}{e^{-i(X \otimes X)\frac{3\pi}{4}}}
  & \gate{R_z(\pi/2)}
  & \gate{R_x(\pi/2)}
  & \gate{R_z(\pi/2)}
  & \qw \\
  & \gate{R_x(\pi/2)}
  & \gate{R_z(\pi)}
  & \ghost{e^{-i(X \otimes X)\frac{3\pi}{4}}}
  & \gate{R_z(\pi)}
  & \gate{R_x(\pi)}
  & \qw
  & \qw
 }}
 $$
 \caption[Controlled-NOT implementation using the XX 
          coupling]
         {Controlled-NOT implementation using the XX
	  coupling; $\gamma t=3\pi/4$.}
 \label{cnot-implementation-xx-coupling}
\end{figure}

Now, the question is how to implement the Controlled-NOT
gate. The XX coupling term in the Hamiltonian
\eqref{xx-coupling-hamiltonian} produces during time $t$
an unitary evolution described by a matrix
\begin{equation}
 \label{xx-evolution-native}
 \hspace{1.5cm}
 e^{-i\gamma(X \otimes X)t}=
 \left(\begin{array}{cccc}
  \iota_1 & 0 & 0 & \iota_2 \\
  0 & \iota_1 & \iota_2 & 0 \\
  0 & \iota_2 & \iota_1 & 0 \\
  \iota_2 & 0 & 0 & \iota_1
 \end{array}\right)\,,
 \quad\text{where}\quad
 \iota_1 =  \cos \gamma t,\;\;
 \iota_2 =-i\sin \gamma t\;.
\end{equation}
Hadamard rotations modify 
\eqref{xx-evolution-native} to the diagonal form
\begin{equation}
 (H \otimes H) \;
 e^{-i\gamma(X \otimes X)t} \;
 (H \otimes H)
 =
 \diag(e^{-i\gamma t},
       e^{ i\gamma t},
       e^{ i\gamma t},
       e^{-i\gamma t})\,,
\end{equation}
and additional single qubit gates lead (after
simplification) to the Controlled-NOT implementation shown
in the Figure~\ref{cnot-implementation-xx-coupling}.

\begin{figure}[h!]
 $$
 \scalebox{0.74}{
   \Qcircuit @R=1em @C=0.7em {
          \lstick{\ket{0}}
	& \qw	
        & \gate{R_x(\pi/2)}
        & \multigate{1}{e^{-i(X \otimes X)\frac{3\pi}{4}}}
	& \qw
        & \gate{R_x(\pi/2)}
        & \multigate{1}{e^{-i(X \otimes X)\frac{3\pi}{4}}}
        & \gate{R_x(\omega_k)}
        & \meter
        & \rstick{x_k} \cw \\
          \lstick{\ket{0}}
        & \gate{R_z(2^{k-1}\alpha\!\!\!\!\pmod {2\pi})}
        & \gate{R_x(\pi/2)}
        & \ghost{e^{-i(X \otimes X)\frac{3\pi}{4}}}
        & \gate{R_z(2^{k-1}\alpha\!\!\!\!\pmod {2\pi})}
        & \gate{R_x(\pi/2)}
        & \ghost{e^{-i(X \otimes X)\frac{3\pi}{4}}}
        & \rstick{\ket{0}} \qw
   }
 }
 $$
 \caption[Benchmark circuit I]
         {The $k$-th step of the IPEA applied to the
	  operator \eqref{simple-z-rotation-operator}, 
	  implemented on a two-qubit system with XX coupling.
	  Note the block structure.}
 \label{first-simulation}
\end{figure}
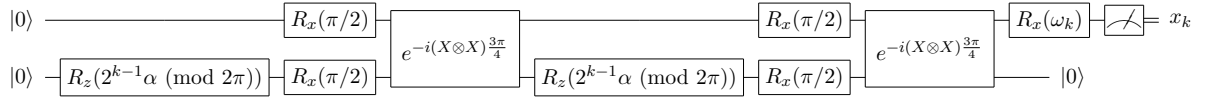

A complete minimized gate sequence implementing the $k$-th
step of the iterative phase estimation 
(Figure~\ref{ipea-circuit}) applied to the operator
\eqref{simple-z-rotation-operator} is shown in the 
Figure~\ref{first-simulation}. After $m$ 
iterations (measurements), we obtain an $m$-bit estimate of
a phase $\phi=\alpha/2\pi$ with high probability.

For simulations of a real experiment, we used a noise
model with accurate initialization and measurement, but
imperfect gates. The unitary operation of the $i$-th gate can
be parameterized as a rotation of a certain angle $\varphi_i$
around some axis (X, Z or XX). 
The angle is a product of the tunable strength of the
corresponding control or interaction term in the Hamiltonian
$\lambda_i$ and time $t_i$ the interaction is switched on,
$\varphi_i=\lambda_i t_i$. Assuming a precise timing, a
fluctuating interaction/control strength 
\begin{equation}
 \lambda_i
 \;\longrightarrow\;
 \lambda_i + \delta_{\lambda_i}
\end{equation}
leads to fluctuations in the rotation angle
\begin{equation}
 \varphi_i
 \;\longrightarrow\;
 \varphi_i \Bigl(
                 1+\frac{\delta_{\lambda_i}}{\lambda_i} 
           \Bigr) \,.
\end{equation}
The fluctuations are assumed to be evenly distributed in the
interval 
\begin{equation}
 -\frac{\Delta}{2}
 \,<\,
 \frac{\delta_{\lambda_i}}{\lambda_i}
 \,<\,
 \frac{\Delta}{2}\;,
\end{equation}
where $\Delta$ is a dimensionless parameter, indicating the
strength of the noise. As an example, for $\Delta=1$ a
$\pi/2$ X-rotation will be replaced with an X-rotation with
a random angle, evenly distributed between $\pi/4$ and 
$3\pi/4$. 

\begin{figure}[h!]
 \centering
 \subfloat[The success probability of the IPEA correctly
          determining $m=10$ binary digits of the phase
          $\phi=\alpha/2\pi$, as a function of the noise
          level $\Delta$. The curves shows errors in all
          Z, X, XX-coupling gates separately, as well as in
          all gates simultaneously.]
  {
    \includegraphics[scale=0.45]
    {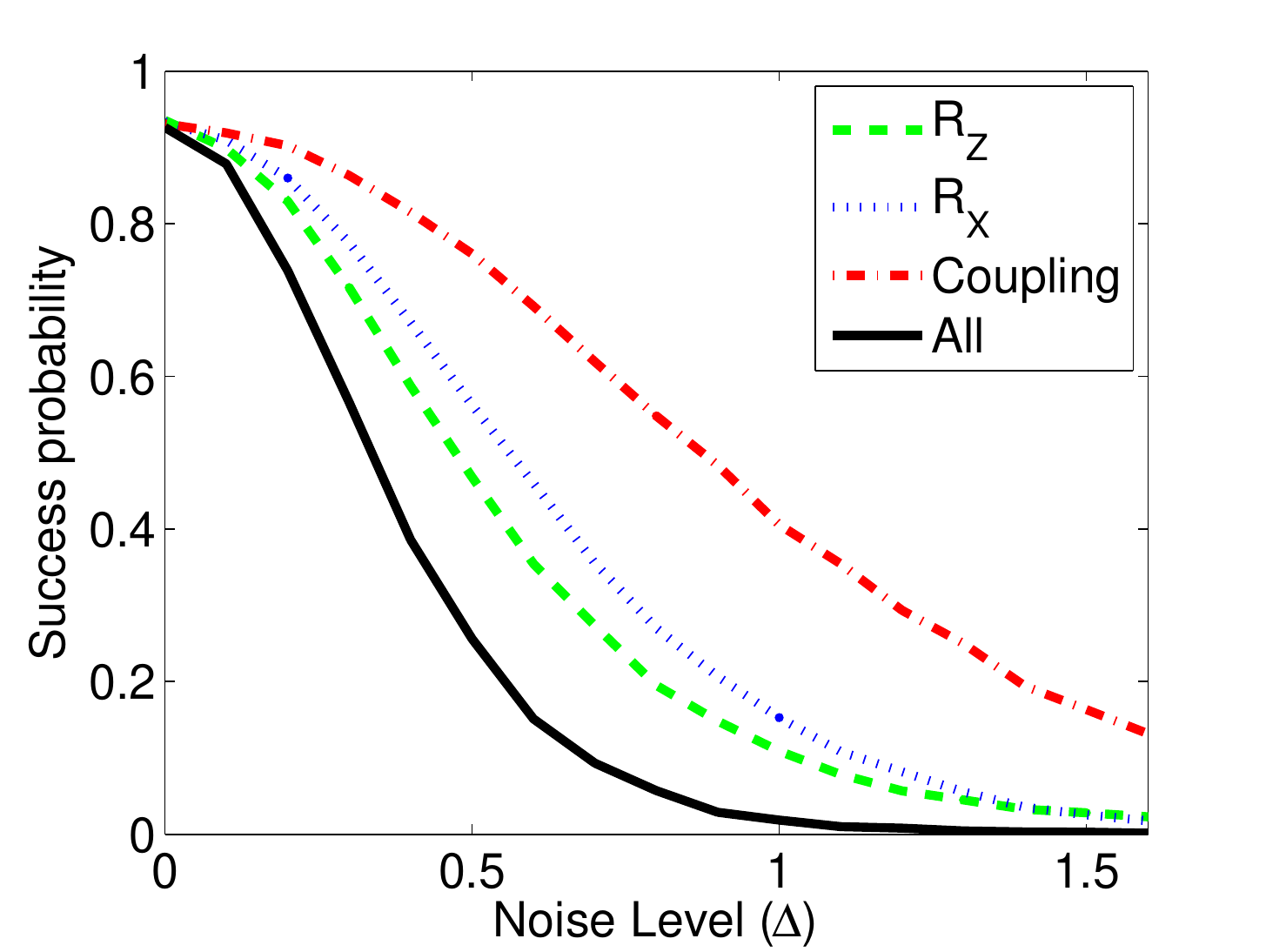}
    \label{xx-two-cnots-success-probability}
  } 
 \quad
 \subfloat[The total number of measurements as a function of
          the number of bits $m$ and noise level $\Delta$.
          The noise level $\Delta$ is the same for all
          gates.]
  {
     \includegraphics[scale=0.45]
     {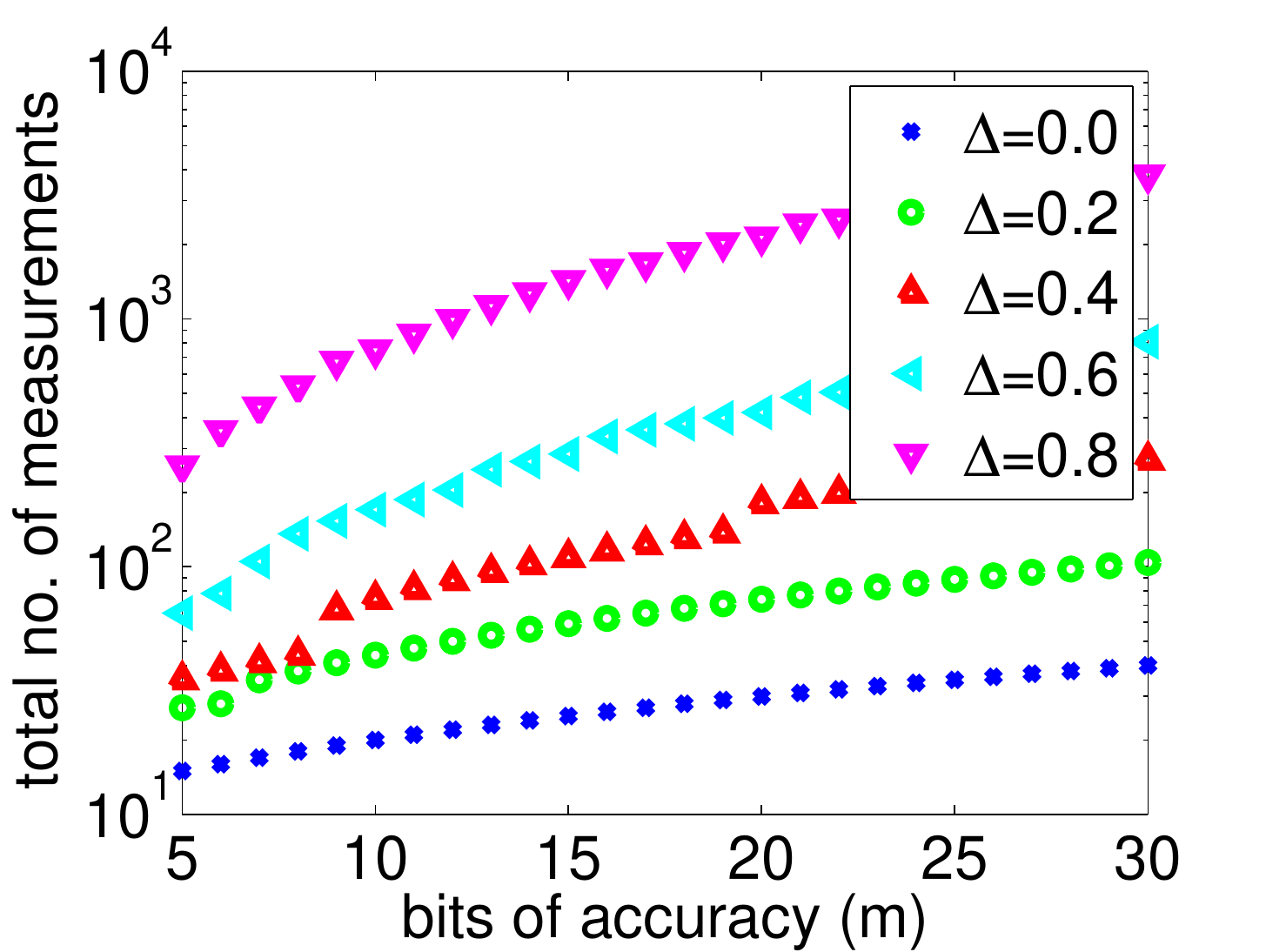}
     \label{xx-two-cnots-bits-of-accuracy}
  }
 \caption[Results of simulations I]
         {Results of simulations of the circuit shown in the
          Figure~\ref{first-simulation}.}
\end{figure}

The success probability of the circuit as a function of the
noise level $\Delta$ is shown in the
Figure~\ref{xx-two-cnots-success-probability}.

As can be seen the setup is quite robust towards errors up
to $\Delta \sim 0.3$ and the $R_z$ gates are the most 
sensitive gates. When the noise level is larger, repeated 
measurements on individual bits are necessary. To achieve
an overall success probability of $1-\varepsilon$, we need
to increase the per bit success probability to 
$(1-\varepsilon/m)$, using $N_k$ trials (repeated
measurements). The number $N_k$ can be determined in the
following way. We write down the estimates obtained from
few sample runs of the circuit. From these results it is 
possible to calculate per bit success probabilities $P_k$.
Next, we set the desired per bit success probability to
$P_k' = 1-\varepsilon/m$ and $N_k$ can be calculated using
the equation \eqref{probability-amplification} 
(see Page~\pageref{probability-amplification}). Essentially,
$N_k$ can be read out from the plot of the function
\eqref{probability-amplification}. A convenient way is to
express this cumulative distribution function in terms of
the regularized incomplete beta function
\begin{equation}
  P'_k
  = \sum_{s=0}^{ \frac{N_k-1}{2}}
    \binom{N_k}{s}
    (P_k)^{N_k-s}(1-P_k)^s 
  \;=\;
    \betainc
    \Bigl(P_k, \frac{N_k+1}{2}, \frac{N_k+1}{2}\Bigr)
    \quad
    \text{for } N_k=1,3,5,\ldots
\end{equation}
in order to avoid numerical problems with a huge binomial
coefficient, and calculate $N_k$ using the following 
Matlab$^{\text{\tiny \textregistered}}$ 6 compatible code:
\begin{center}
\begin{minipage}{140mm}
\begin{verbatim}
function nreps = NRepetitions (p0, perr)
   % nreps is the number of trials needed to determine a bit with
   % perr probability of error if
   % p0 is the single shot success rate

   nreps = ceil(fzero(@(x) myfun(x,p0,perr), [1 1000000]));

function f = myfun (x, p0, perr)
   f = betainc (1-p0, (x+1)/2, (x+1)/2) - perr;
\end{verbatim}
\end{minipage}
\end{center}
Figure~\ref{xx-two-cnots-bits-of-accuracy} shows the total
number of measurements needed to obtain the phase
$\phi=\alpha/2\pi$ with $m$ accurate bits under different
noise levels and an error probability $\varepsilon < 0.05$.
As one can see, repeated bitwise measurements works well up
to rather high noise levels. The total number of
measurements scales as $O(m\log(1/\varepsilon))$. However,
above a certain noise level, when the external noise
dominates over the intrinsic errors arising from the 
remainder $\delta$ 
(see Page~\pageref{intrinsic-error-from-delta}), the
error probability per bit $\varepsilon_k$ becomes almost
independent of the bit position $k$. Besides other problems,
the feedback improvement does not work any more. Then $N_k$
on average is proportional to 
$\log(1/\varepsilon_k)=\log(m/\varepsilon)$
and total number of measurements scales as 
$O(m\log(m/\varepsilon))$ with a relatively large hidden 
constant.

The circuit shown in the 
Figure~\ref{first-simulation} can be also
rearranged using the ZZ-coupling. The results of the 
simulations are very similar. However, the rearranged 
circuit does not have the nice block structure which might
be useful in a real experimental setup.

\subsubsection*{Simulations with focus on the coupling term}

\begin{figure}[h!]
 \centering
 \subfloat
   [Construction utilizing directly the ZZ-coupling.]
   {
   $$
   \scalebox{0.9}{
     \Qcircuit  @R=1em @C=0.7em {
	   \lstick{\ket{0}}
	 & \gate{R_x(\pi/2)}
	 & \multigate{1}
	   {ZZ\left(
	     \frac{1}{2} \, \alpha 2^{k-1}\!\pmod {2\pi}
	   \right)}
	 & \gate{R_z(\omega_k)}
	 & \gate{R_x(-\pi/2)}
	 & \meter
	 & \rstick{x_k} \cw \\
	   \lstick{\ket{0}}
	 & \qw
	 & \ghost
	   {ZZ\left(
	     \frac{1}{2} \, \alpha 2^{k-1}\!\pmod {2\pi}
	   \right)}
	 & \qw
	 & \qw
	 & \qw
	 & \rstick{\ket{0}} \qw
   }}$$
   \label{second-simulation-zz}
   } \\[7mm]
 \subfloat
   [Construction utilizing directly the XX-coupling.]
   {
   $$
   \scalebox{0.9}{
     \Qcircuit  @R=1em @C=0.7em {
	   \lstick{\ket{0}}
	 & \qw
	 & \qw
	 & \multigate{1}
	   {XX\left(
	     \frac{1}{2} \, \alpha 2^{k-1}\!\pmod {2\pi}
	    \right)}
	 & \gate{R_x(-\omega_k)}
	 & \gate{R_z(\pi/4)}
	 & \meter
	 & \rstick{x_k} \cw \\
	   \lstick{\ket{0}}
	 & \gate{R_x(\pi/2)}
	 & \gate{R_z(-\pi/2)}
	 & \ghost
	   {XX\left(
	     \frac{1}{2} \, \alpha 2^{k-1}\!\pmod {2\pi}
	   \right)}
	 & \qw
	 & \qw
   }}$$
   \label{second-simulation-xx}
   }
 \caption[Benchmark circuit II]
         {The $k$-th step of the IPEA applied to the
          operator \eqref{simple-z-rotation-operator},
	  implemented on a two-qubit system. 
	  Implementation makes direct use of the coupling
	  terms in the Hamiltonians 
	  \eqref{xx-coupling-hamiltonian} and
	  \eqref{zz-coupling-hamiltonian}.}
 \label{second-simulation}
\end{figure}
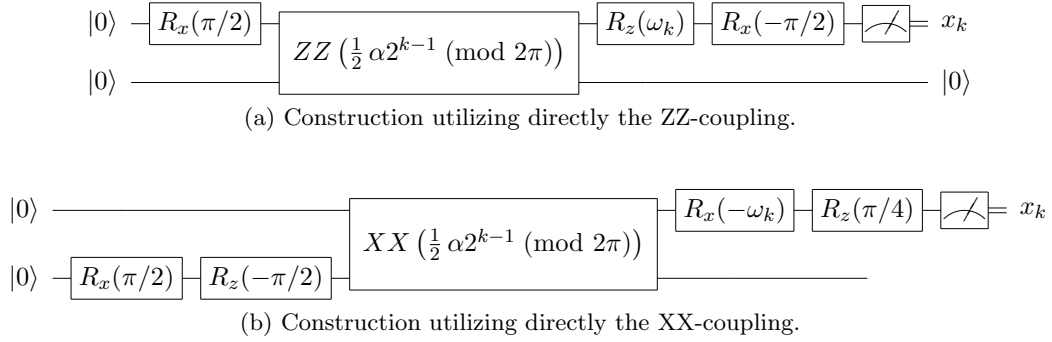

The construction used to
implement controlled powers of the operator 
\eqref{simple-z-rotation-operator} shown in the 
Figure~\ref{ctrl-powers-z-rot-general-construction} is not
the only possible construction given the underlying
Hamiltonians \eqref{xx-coupling-hamiltonian} and 
\eqref{zz-coupling-hamiltonian}. Unitary gates
\begin{align}
 \label{def-zz-gate}
 ZZ(\gamma t) 
  &= e^{-i \gamma (Z \otimes Z) t}
   = \diag(e^{-i\gamma t},\,
 	   e^{ i\gamma t},\,
	   e^{ i\gamma t},\,
	   e^{-i\gamma t})\\
 \hspace{-3.33cm}\text{and}\hspace{3cm}
 XX(\gamma t)
  &= e^{-i \gamma (X \otimes X) t}
\end{align}
arising from the ZZ and XX coupling terms, respectively,
can be used straightforwardly to achieve the same goal
without intermediate construction of the two controlled-NOT
gates. Figure~\ref{second-simulation-zz} shows a complete
minimized circuit implementing the $k$-th step of the IPEA
applied to the operator \eqref{simple-z-rotation-operator},
exploiting the ZZ-coupling directly. A circuit making use
of the XX-coupling is shown in the 
Figure~\ref{second-simulation-xx}.

\begin{figure}[h!]
 \centering
 \subfloat[The success probability of the IPEA correctly
           determining $m=10$ binary digits of the phase
	   $\phi=\alpha/2\pi$, as a function of the noise
	   level $\Delta$. The curves shows errors in all
	   Z, X, ZZ-coupling gates separately, as well as
	   in all gates simultaneously.]
  {
     \includegraphics[scale=0.45]
     {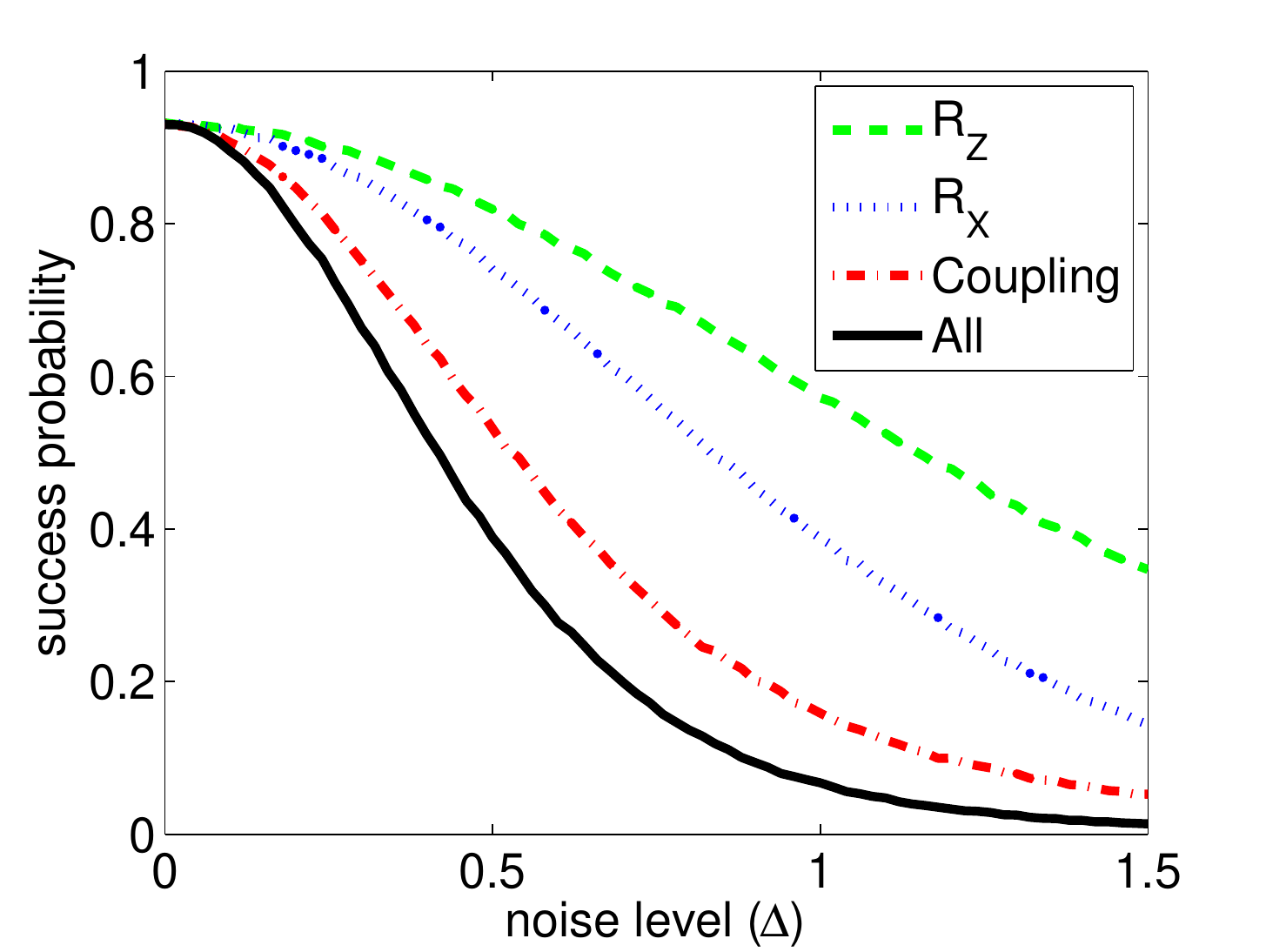}
     \label{zz-coupling-success-probability}
  }
 \quad
 \subfloat[The total number of measurements as a function
           of the number of bits $m$ and noise level 
	   $\Delta$. The noise level $\Delta$ is the same
	   for all gates.]
  {
     \includegraphics[scale=0.45]
     {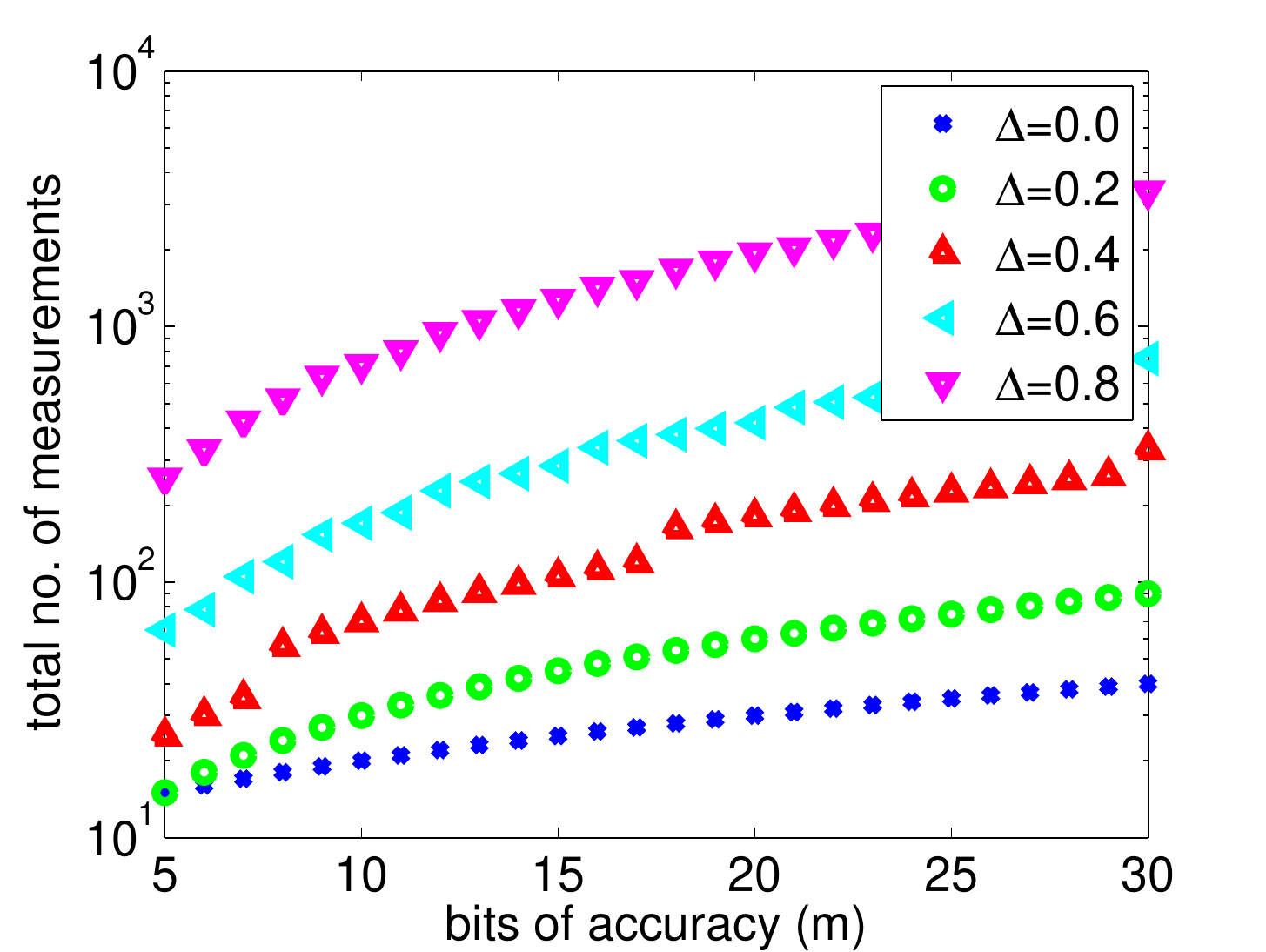}
     \label{zz-coupling-bits-of-accuracy}
  }
 \caption[Results of simulations II]
         {Results of simulations of the circuit shown in the
          Figure~\ref{second-simulation-zz}.}
 \label{second-simulation-results}
\end{figure}

In the simulations of a real experiment under the noise
model described in the previous section, both circuits
performed almost identically. 
Figure~\ref{zz-coupling-success-probability} shows the
success probability as a function of the noise level 
$\Delta$ for a circuit with the ZZ-coupling.

The success probability is somewhat better compared to the
result shown in the 
Figure~\ref{xx-two-cnots-success-probability} and
as expected the coupling gate is the most sensitive gate.
Importantly, the feedback gate $R_z(\omega_k)$ is quite
robust compared to the other gates. The total number of
measurements needed to obtain the phase with $m$ accurate
bits under different noise levels and an error probability
$\varepsilon < 0.05$ is then shown in the 
Figure~\ref{zz-coupling-bits-of-accuracy}. These results
are very much the same as results shown in the
Figure~\ref{xx-two-cnots-bits-of-accuracy}.

\subsubsection*{Simulations of a setup close to the 
                Abrams-Lloyd algorithm}
         
The two previous setups represented simple test-bed 
applications (benchmarks) where an input parameter $\alpha$
is plugged in as $\sim(2^k\alpha\pmod {2\pi})$ in each
iteration of the IPEA and the obtained outcome 
$\blacktriangleleft\phi\blacktriangleright$ is expected
to exhibit strong correlations with $\alpha$,
$\blacktriangleleft\phi\blacktriangleright
 \;\doteq\,
 \alpha/2\pi$.
A different (stronger) benchmark can be performed within a
scenario where the coupling strength $\gamma$ is supposed to
be unknown and the goal of the IPEA is to estimate $\gamma$.

\begin{figure}[h!]
 $$
  \scalebox{0.9}{
     \Qcircuit  @R=1em @C=0.7em {
           \lstick{\ket{0}}
         & \gate{R_x(\pi/2)}
         & \multigate{1}
           {ZZ\left(( \alpha/2)\, 2^{k-1}\right)}
         & \gate{R_z(\omega_k)}
         & \gate{R_x(-\pi/2)}
         & \meter
         & \rstick{x_k} \cw \\
           \lstick{\ket{0}}
         & \qw
         & \ghost
           {ZZ\left(( \alpha/2)\, 2^{k-1}\right)}
         & \qw
         & \qw
         & \qw
         & \rstick{\ket{0}} \qw
   }}
 $$
 \caption[Benchmark circuit III]
         {The $k$-th step of the IPEA applied to the 
          operator \eqref{simple-z-rotation-operator}, 
	  implemented on a two-qubit system using the
	  entangling $ZZ$ gate. The parameter $\alpha$
	  is considered to be unknown. The circuit
	  estimates the coupling strength $\gamma$ of
	  the $ZZ$ gate.}
 \label{third-simulation}
\end{figure}
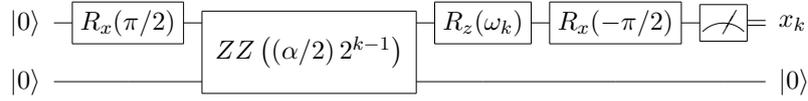

Let us consider a circuit shown in the 
Figure~\ref{third-simulation}. The ZZ gate is parameterized
similarly as in the Figure~\ref{second-simulation-zz}, but
since the parameter $\alpha$ is not known now, the modulo
operation is not present any more.

The elementary entangling gate in this circuit is the
$ZZ(\alpha/2)$ gate and it describes an unitary evolution
$e^{-i\gamma\left(Z \otimes Z\right)t}$ produced during a
time interval $t$, typically order of nanoseconds. As 
follows from the definition of the $ZZ$ gate 
(Eq.~\ref{def-zz-gate}), $\alpha~=~2\gamma t \pmod {2\pi}$.
Since the outcome of the IPEA is an estimate of the value
$\phi=\alpha/2\pi$, an estimate of $\gamma$ is then equal
to
\begin{equation}
 \blacktriangleleft\gamma\blacktriangleright
 \;\;=\,
 \frac{\pi \, \blacktriangleleft\phi\blacktriangleright}{t}
 \;.
\end{equation}

In the view of the fact that $\gamma$ is unknown, the only
possibility how to implement the powers of the 
$ZZ(\alpha/2)$ gate, as required by the steps of the IPEA,
is to create a long sequences of applications of the plain
$ZZ(\alpha/2)$ gate. This is equivalent to an evolution
during a time interval $(2^{k-1}\,t)$ :
\begin{equation}
 \left(ZZ\left(\alpha/2\right)\right)^{2^{k-1}}
 = ZZ\Bigl((\alpha/2)\,2^{k-1}\Bigr)
 = e^{-i\gamma\left(Z \otimes Z\right)2^{k-1}\,t} \,.
\end{equation}

In the simulations of this setup while considering
superconducting qubits, it is necessary to take into account
the effect of pure dephasing in the computational basis with
rate $\Gamma_\varphi$ as a result of the environmental noise
and long time intervals $(2^{k-1}\,t)$ 
\cite{GJ07,NanoHandbook06}.
Dephasing is a process in which a physical qubit without any
error corrections is losing its phase information. In our
setup, the accuracy of the $ZZ((\alpha/2) 2^{k-1})$ gate 
will be affected with a dephasing rate proportional to 
$\Gamma_\varphi/\gamma$. In addition, it is considered
imperfect X-rotations of the form 
$R_x(\pm \pi/2 \,+\, \delta_x)$, where $\delta_x$ is a
normally distributed random angle with variance $\Delta_x$.
Errors in the $R_z(\omega_k)$ gate are omitted, since as
compared to errors in other gates, they do not contribute to
the decrease of the success probability too significantly,
see Figure~\ref{second-simulation-zz}. 

\begin{figure}[h!]
 \centering
 \subfloat[The success probability of the IPEA to correctly
           determine the phase $\phi=\alpha/2\pi$,
	   with precision 
	   better than $2^{-5}$ (upper/green line) and 
	   $2^{-7}$ (lower/blue line) as a function of the
	   noise level. The three cases of pure X-gate 
	   errors, pure dephasing, and both types of noise
	   acting simultaneously are considered. The 
	   simulation was averaged over evenly distributed
	   $\alpha$.]
  {
     \includegraphics[scale=0.465]
     {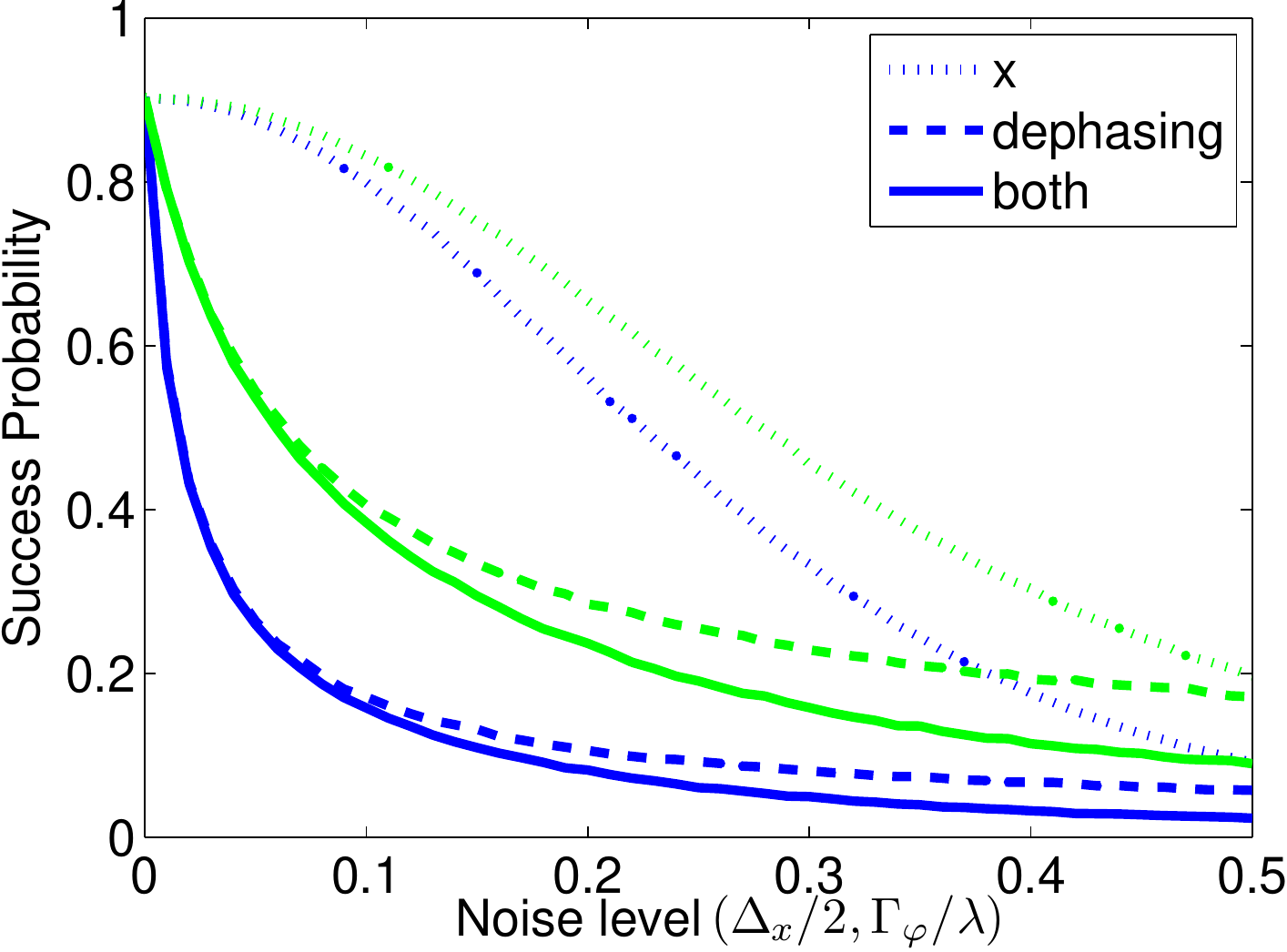}
     \label{third-simulation-results-a}
  }
 \quad
 \subfloat[The total number of measurements needed to obtain
           the phase $\phi=\alpha/2\pi$ with precision better
	   than $2^{-m}$ $(2\leq m \leq 11)$, with error 
	   probability $\varepsilon<0.05$.]
  {
     \includegraphics[scale=0.45]
     {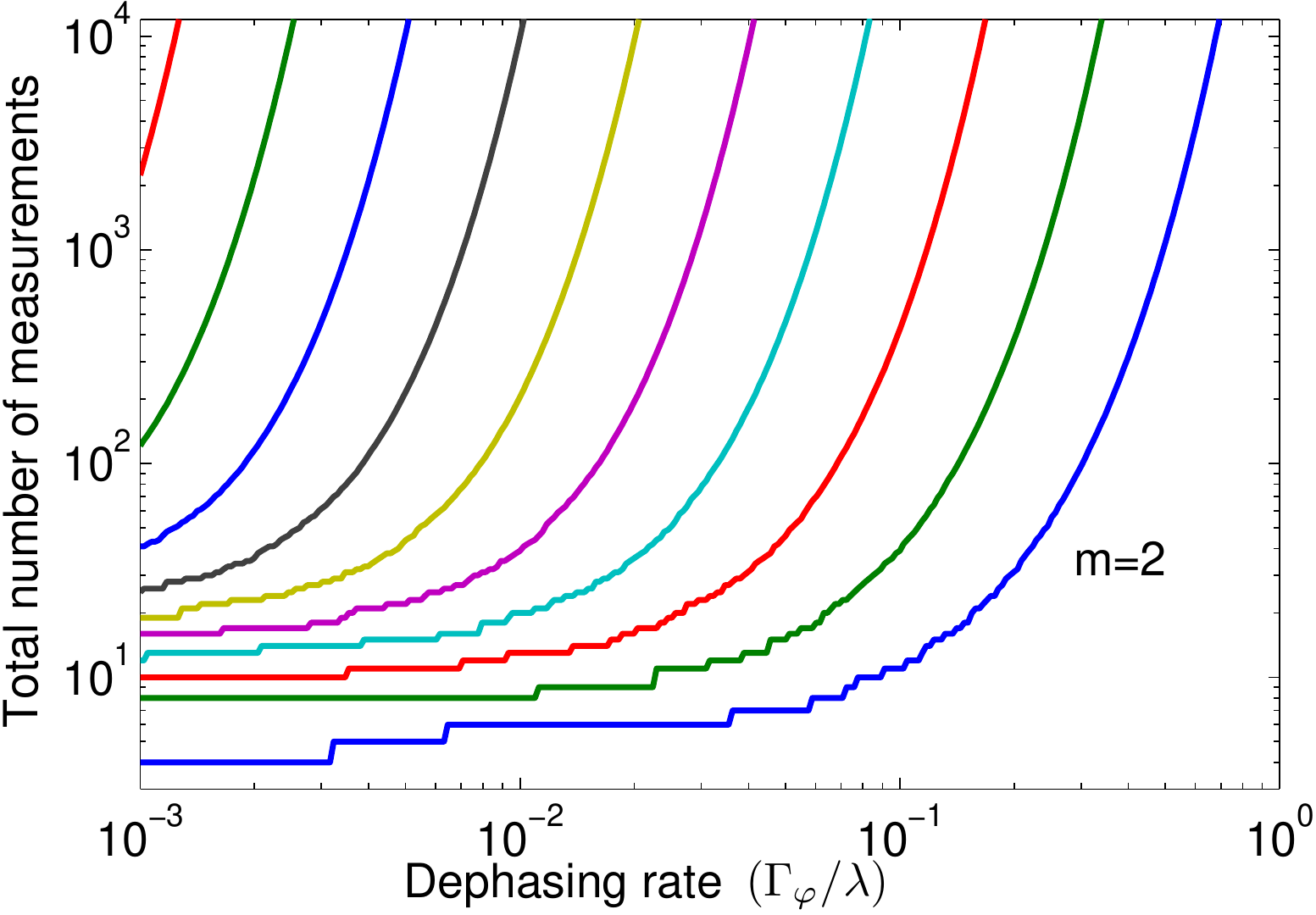}
     \label{third-simulation-results-b}
  }
 \caption[Results of simulations III]
         {Results of simulations of the circuit shown in 
          the Figure~\ref{third-simulation}.}
 \label{third-simulation-results}
\end{figure}

The success probability of the circuit as a function of
 dephasing rate $\Gamma_\varphi/\gamma$ and variance 
$\Delta_x$ is shown for $m=5$ and $m=7$ in the 
Figure~\ref{third-simulation-results-a}. The circuit is
rather robust against X-rotation errors, while being much
more sensitive to dephasing.

In order understand this sensitivity, an analytical
expression was carried out. The per bit success probability
without errors as derived in the 
Section~\ref{sec:ipea-derived-from-kitaev} is
\begin{equation}
 \label{p-k-formula}
 P_k = \cos^2(\pi\,2^{k-1}\,(\delta 2^{-m}))
     = \frac{1}{2}
       \left(
         1 + \cos\bigl(\pi\, 2^k (\delta 2^{-m})\bigr)
       \right) \;.
\end{equation}
The errors modify \eqref{p-k-formula} into
\begin{equation}
 \label{p-k-error-formula}
 P_k^{err} 
 = \frac{1}{2}
   \left(
      1
    +
      e^{  -   \Delta_x^2 
         \,-\, \abs{\alpha}\, 2^k (\Gamma_\varphi/\gamma)}
      \cos\bigl(\pi\, 2^k (\delta 2^{-m})\bigr)
   \right) \;.
\end{equation}
It is evident that the factor $2^k$ brought by dephasing
into the exponent of \eqref{p-k-error-formula} is
responsible for the circuit's increased sensitivity to the
environmental noise. As in the previous simulations, the
success probability can be improved by repeated 
measurements of each bit. The number of per bit repetitions
$N_k$ is proportional to 
$\exp\bigl(2^k (\Gamma_\varphi/\gamma)\bigr)$,
to neutralize the effect of the 'noise term' 
$\exp(-\Delta_x^2
      \,-\, \abs{\alpha}
      2^k (\Gamma_\varphi/\gamma))$.
Figure~\ref{third-simulation-results-b} shows the total
number of measurements needed to obtain the phase $\phi$
with $m$ accurate bits under different noise levels and
an error probability $\varepsilon < 0.05$. For a realistic
dephasing rate $0.01 < \Gamma_\varphi/\gamma < 0.1$ 
\cite{GJ07}, 
between $5$ and $8$ binary digits of $\phi$ can be
extracted with less than $10^4$ measurements.

\begin{center}\end{center}
\newpage
\subsection{Multiround protocols for reference frame 
            alignment}

The IPEA has been so far presented as a viable alternative
to the QFT-based PEA in situations where the number of
available qubits is a crucial limiting factor. Besides, the
IPEA and other iterative variants for phase estimation can
also be useful in communication tasks. In communication
between two parties, a common frame of reference is usually
needed. This is not less true for quantum communication
where e.g. to encode information into the horizontal or
vertical polarization of a photon, we have to agree on the
spatial direction of these axes. Think of ground to
satellite quantum cryptography \cite{QC-overview00}.
While some quantum
information processing tasks can be achieved completely 
without a shared reference frame by using entangled states
of multiple systems 
\cite{Bartlett-without-srf03, Bartlett-review07}, this is not
true in
general, and a shared reference frame is always a valuable
resource. The question is how to align (share) a reference
frame in the most efficient way, optimally at the Heisenberg
limit.

Protocols for reference frame alignment which do not require
transmission of large entangled states have been studied by
Rudolph and Grover \cite{RG03}, and de Burgh and Bartlett
\cite{BB05}.
This is of importance since large entangled states are 
difficult to be communicated faithfully. Essentially, the
protocols which have been discovered are instances of the
Kitaev phase estimation procedure. Due to its iterative 
(multiround) nature, these protocols are referred to as
multiround protocols. Protocols utilizing the QFT-based PEA
are called single round protocols.

\subsubsection*{Phase reference alignment}
Phase reference alignment corresponds to the problem of
synchronizing distant clocks. There are two classical
approaches to distant clock synchronization. The Einstein
light-pulse synchronization protocol and the Eddington
adiabatic clock transfer scheme. Einstein
synchronization uses a light pulse sent between the clocks,
in order to estimate the transmission time. In the Eddington
scheme, a wristwatch synchronized to the first clock is sent
to the place of the second clock.

Entanglement enhanced versions of both protocols have been
suggested \cite{GLM-clock01,Chuang-clock00}, as well as a new
protocol for clock
synchronization based on shared prior entanglement
\cite{Prior-entgl-clock00}. 
In the quantum version of the Eddington protocol 
\cite{Chuang-clock00}, the
crucial idea is to encode the time difference into the
relative phase of physical systems (qubits) having a 
well-defined energy difference, so called 'ticking' qubits.
Then the QFT-based PEA is used to estimate this phase. 

Burgh and Bartlett \cite{BB05} pointed out that the protocol 
\cite{Chuang-clock00} is, however, using quantum operations
in such a way
that the two parties would need an a priori shared phase 
reference (e.g. synchronized clocks), just to be able to
agree on the operations. The multiround protocol of Burgh
and Bartlett (BB) solves this problem and additionally does
not require entanglement. The framework for the BB protocol
is as follows.

Let us assume that Alice and Bob share an inertial 
reference frame (relativistic effect can be neglected) and
agree on a free Hamiltonian of two level atomic ticking 
qubit
\begin{equation}
 \mathcal{H}_{tick} 
 = \frac{\omega}{2}Z
 = \frac{\omega}{2}(\ketbra{0}{0} - \ketbra{1}{1}) \,.
\end{equation}
This means they can agree on the upper and lower energy
eigenstates and the frequency standard $\omega$. States
$\ket{0}$ and $\ket{1}$ can also be understood as defining
a common shared $Z$ axis of the Bloch sphere that describes
the two dimensional qubit (atomic) Hilbert space. The shared
$Z$ axis viewed as a resource is enough for Alice and Bob to
agree on a qubit state
\begin{equation}
 \ket{\psi}
 = \cos\theta\ket{0} + e^{i\xi}\sin\theta\ket{1} \,,
\end{equation}
up to the phase $\xi$. They would need a shared phase 
reference to agree on a qubit state completely. Let us assume
that both Alice and Bob define their local phase reference
relatively to their clocks that are ticking at the same 
frequency $\omega$. The time difference of their clocks is
given by $\Delta t = t\vert_B - t\vert_A$ (Bob's clock is
ahead). Then for example, a state described by Alice as 
$\ket{\psi}_A = \frac{1}{\sqrt{2}}(\ket{0}+\ket{1})$ 
corresponds to the state $\ket{\psi}_B=
(\ket{0}+e^{i(\omega\Delta t)}\ket{1})$ in Bob's frame.

The state of the ticking qubit evolving under the free
Hamiltonian $\mathcal{H}_{tick}$ can be pictured as the 
Bloch vector rotating anticlockwise about the $Z$ axis,
however, it is more convenient to work in the rotating 
frame in which the qubit no longer ticks. In this rotating
frame, it is assumed that Alice and Bob use the Rabi
pulses to interact with the qubit. The Rabi Hamiltonian is
defined as
\begin{equation}
 \label{rabi-hamiltonian}
 \mathcal{H}_{Rabi}(\alpha)
 = \frac{\Omega}{2}(X\cos\alpha + Y\sin\alpha),
\end{equation}
where $\alpha$ is the phase angle of the laser, defined
relatively to the local clock. The Rabi pulse applied for
time $t=\beta/\Omega$ corresponds to a unitary operator
\begin{equation}
 U(\alpha,\beta) 
 = e^{-i\alpha Z/2} \,
   e^{-i\beta  X/2} \,
   e^{ i\alpha Z/2} 
 = R_z(\alpha) \, R_x(\beta) \, R_z(-\alpha) \,.
\end{equation}
Using the operator $U(\alpha, \beta)$ both parties, Alice
and Bob, can relatively to their clock implement arbitrary
single qubit gates. For example, the NOT gate $X=U(0,\pi)$, 
the Hadamard gate $H=X\,U(\pi/2,\pi/2)$, and the $\pi/8$ 
gate $T=H\,U(0,\pi/4)\,H$. It follows that an operation 
$U(\alpha\vert_B,\beta)=U_B(\alpha\vert_B)$
performed by Bob corresponds to the operation 
$U_A(\alpha\vert_A - \phi)$ performed by Alice, where 
$\phi=\omega\Delta t$;
\begin{equation}
 \label{operations-relation}
 U_B(\alpha\vert_B) 
 = U_A(\alpha\vert_A-\phi)
 = R_z(-\phi)\,U_A(\alpha\vert_A)\,R_z(\phi) \,.
\end{equation}

\begin{figure}[h!]
 \centering
 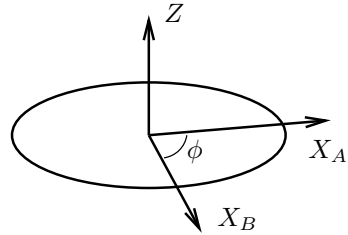
 \caption[The problem of phase reference alignment]
         {Alice and Bob share the $Z$ axis. However, their
          $X$-axes differ by an unknown angle $\phi$. 
	  Working in the rotating frame simplifies this 
	  picture by being static and not rotating 
	  anticlockwise about the $Z$ axis with the angular
	  frequency $\omega$.}
 \label{clock-synchro-basic-picture}
\end{figure}

The relation of operations $U_A$ and $U_B$ as described by
the equation \eqref{operations-relation} can be translated
into the language of the Bloch sphere as shown in the
Figure~\ref{clock-synchro-basic-picture}. Alice and Bob
share the $Z$ axis, but their respective X-axes $X_A$ and
$X_B$ do differ by an unknown angle $\phi=\omega\Delta t$.
The problem of clock synchronization is hereby reduced to 
estimating the angle $\phi$.

\begin{figure}[h!]
 \centering
 \hspace{-1.4cm}
 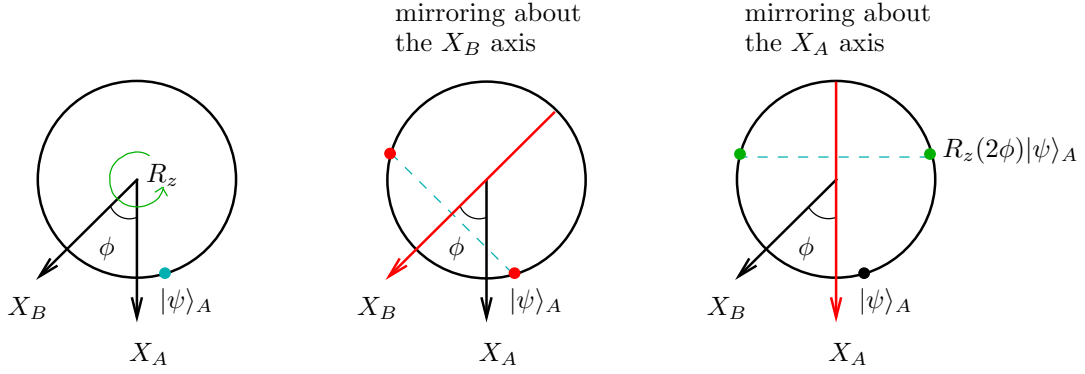
 \caption[The Grover amplification viewed as a mirroring]
         {Grover phase amplification method graphically
          viewed as a mirroring technique.}
 \label{mirroring-technique}
\end{figure}

The question is now how to encode an angle $\sim (2^k\phi)$ 
into the relative phase of a qubit to enable the phase
estimation procedure. The BB protocol uses a clever phase
amplifying technique known from the Grover search algorithm
\cite{Grover97,Nielsen-Chuang00} which can be graphically
viewed as mirroring a
qubit state about the $X_A$ and $X_B$ axes. Let
$\ket{\psi}_A
 = \cos\theta\ket{0}  + e^{i\xi}\sin\theta\ket{1}$
be a qubit state described in the Alice's phase reference
frame. The result of mirroring the state $\ket{\psi}_A$
about the $X_B$ axis and then about the $X_A$ axis is shown
in the Figure~\ref{mirroring-technique}.

In practice, this mirroring effect can be achieved by a
qubit exchange between Alice and Bob. Alice prepares a state 
$\ket{\psi}_A$ and sends the qubit to Bob, Bob applies his
NOT gate $X_B$, sends the qubit back, and  Alice applies
her NOT gate $X_A$. The net gain of this exchange
expressed in Alice's phase reference frame is that a
relative phase of the qubit is increased by an angle
$2\phi$,
\begin{equation}
 X_A \, X_B \ket{\psi}_A
 = \underset{R_z(\phi)}
            {\underbrace{       X_A               \,
	                 \Bigl( R_z(-\phi)        \,
			        X_A \Bigr. }}     \,
			 \Bigl. R_z(\phi) \Bigr)  \,
			 \ket{\psi}_A
 = R_z(2\phi)\ket{\psi}_A \,.
\end{equation}

It is now easy to see that 
 $X_A \, X_B \, H_A \, \ket{0}_A
 = \frac{1}{\sqrt{2}}(\ket{0}+e^{i (2\phi)}\ket{1})_A$
and the phase $\phi$ can be estimated using the Kitaev phase
estimation applied to the operator $(X_A \, X_B)$, see 
Figure~\ref{bb-protocol}. This is the BB protocol for  
phase reference alignment.
\begin{figure}[h!]
 $$
 \scalebox{1}{
  \Qcircuit @R=1em @C=0.7em {
	   \lstick{\ket{0}_A}
	 & \qw
	 & \gate{H_A}
	 & \gate{\left( X_A \, X_B \right)^{2^{k-1}}}
	 & \gate{H_A}
	 & \meter
	 & \rstick{x_k} \cw
 }}$$
 \caption[The Burgh and Bartlett protocol for phase
          reference alignment]
         {The Burgh and Bartlett protocol for phase
          reference alignment.}
 \label{bb-protocol}
\end{figure}

Utilizing the IPEA derived in the 
Section~\ref{sec:ipea-derived-from-kitaev}, the scaling of the 
BB protocol can be improved. In order to do so, Alice only 
needs to modify the scheme by inserting the $R_z(\omega_k)$
feedback. Given the Rabi Hamiltonian \eqref{rabi-hamiltonian},
the  $R_z(\omega_k)$ gate can be implemented by a sequence of 
pulses producing $(H_A \, U_A(0,\omega_k) \, H_A)$. The IPEA
applied to the operator $(X_A \, X_B)$ is shown in the
Figure~\ref{ipea-phase-reference-alignment}.
\begin{figure}[h!]
 $$
 \scalebox{1}{
  \Qcircuit @R=1em @C=0.7em {
           \lstick{\ket{0}_A}
         & \qw
         & \gate{H_A}
         & \gate{\left( X_A \, X_B \right)^{2^{k-1}}}
         & \gate{H_A}
	 & \gate{U_A(0,\omega_k)}
         & \meter
         & \rstick{x_k} \cw
 }}$$
 \caption[Improved protocol for phase reference alignment]
         {Improved protocol for phase reference alignment.}
 \label{ipea-phase-reference-alignment}
\end{figure}

\subsubsection*{Spatial axes alignment}

The problem of establishing either a single direction in 
space or an orthogonal trihedron (xyz axes) using quantum
enhanced methods was studied extensively in the past, see
\cite{RG03} for a list of references. A particular multiround
protocol making no use of entanglement is discussed by
Rudolph and Grover in \cite{RG03}. The framework for the RG 
protocol is as follows.

Let us assume that Alice and Bob define their spatial axes
relatively to their single qubit rotations axes. 
Additionally, let the rotation matrix describing the change
from Alice's to Bob's frame of reference be given by
\begin{equation}
 R = R_z(\xi)    \,
     R_y(\theta) \,
     R_z(\phi)   
   = e^{-i(\xi+\phi)/2}\left(\begin{array}{rr}
                         \cos \frac{\theta}{2} 
       &\; -e^{i\phi}    \sin \frac{\theta}{2} \\[3mm]
           -e^{i\xi}     \sin \frac{\theta}{2} 
       &\;  e^{\phi+\xi} \cos \frac{\theta}{2}
     \end{array}\right) \,.
\end{equation}
The goal is to estimate the Euler angles $\phi, \theta,
\xi$.

Rudolph and Grover showed explicitly that the $k$-th bit
describing the angle $\theta$ (Kitaev-like approach) can be
determined by applying the transform 
\begin{equation}
 U(k)
 = (Z_B \, Z_A)^{2^{k-1}} 
 = (Z_B \, R^\dagger \, Z_B \, R)^{2^{k-1}}
 = \left(\begin{array}{rr}
                    \cos (2^{k-1}\theta) 
     & -e^{ i\phi}  \sin (2^{k-1}\theta) \\[2mm]
        e^{-i\phi}  \sin (2^{k-1}\theta)
     &              \cos (2^{k-1}\theta)
   \end{array}\right)
\end{equation}
to a qubit prepared in the state $\ket{\psi}_B=\ket{0}$. 
The state $\ket{\psi}_B$ is expressed in the Bob's reference
frame. The protocol goes as follows. Bob prepares the state
$\ket{\psi}_B$ and sends the qubit to Alice, Alice applies
her $Z_A$ gate, sends the qubit back, and Bob applies his
$Z_B$ gate. The qubit exchange is repeated $2^{k-1}$ times
and then Bob performs the measurement to estimate the bit
$x_k$. The outcome '0' is observed with probability
\begin{equation}
 P_0 
 = \abs{\cos(2^{k-1}\theta)}^2
 = \cos^2(2^{k-1}\theta)
\end{equation}
and the outcome '1' with probability
\begin{equation}
 P_1
 = \abs{e^{-i\phi}\sin(2^{k-1}\theta)}^2
 = \sin^2(2^{k-1}\theta) \,.
\end{equation}

The other angle $\xi$ or $\phi$ needed to align the spatial
axes can be estimated by changing the transform that Alice
and Bob perform and/or the initial state Bob prepares. The
important point is that it is possible to make the outcome
probabilities dependent only on a specific angle at time.
Intriguingly, in the light of the IPEA, it is not clear how 
to improve the scaling of the RG protocol by using a 
feedback. That is to find a unitary transform $V(\omega_k)$
such that the outcome probabilities remain dependent only 
on $\theta$ and $\omega_k$, for example. In fact, I was not
able to find such transform, nor to rigorously prove that it
does not exist. One open direction I keep working on is to
focus on the order of angles to be estimated. 

\newpage

 \section{Quantum phase estimation conlusions}
 \label{sec:pea-conclusions}
 \begin{center}\end{center}

In this chapter, I discussed the broad issue of quantum phase
estimation algorithm as a practical approach for problem
solving on a quantum computer. Alternatively, it is possible
to use the framework of quantum Fourier sampling. An
example of a problem which is more naturally understood
using the quantum Fourier sampling is the Shor factoring
algorithm \cite{Shor-factoring97}. On the other side of
the problem spectrum is the Abrams-Lloyd algorithm 
\cite{Abrams-Lloyd99} which is rather well described using
the phase estimation.

In contrast to the modern quantum phase estimation 
based on the Fourier transform, the original Kitaev PEA 
\cite{Kitaev96} did not explicitly use the quantum Fourier
transform and Kitaev's approach looked completely different
from Shor's.
However, soon it was recognized that the Kitaev's approach
can be explained in terms of the Fourier transform over 
$\mathbb{Z}_2^n$ and that a transform over 
$\mathbb{Z}_{2^n}$ would be in fact more appropriate. Hence
Kitaev's approach was unified with Shor's QFT approach
for factoring and the QFT-based PEA was born 
\cite{Nielsen-Chuang00, Cleve98}. While this is perfectly
fine on the big scale of understanding the power of a quantum
computer, a few interesting aspects of the original Kitaev
work faded out.

Guzik {\it et. al.} \cite{Guzik05} building on the work of
Abrams and Lloyd even carried out their own iterative phase 
estimation procedure in order to reduce the number of
required qubits at time.
Paradoxically, the Kitaev's circuit would require less 
resources under given circumstances and it is more robust to
the environmental noise. When I started working on the phase
estimation algorithm, my task was to investigate how to
implement an instance of the Abrams-Lloyd PEA on three or
less qubits. The stepping stones were the QFT-based PEA and
the work of Guzik {\it et. al.}. As a first step, I 
rediscovered the Kitaev's circuit by reducing gates in the
Guzik PEA to absolute minimum. Hereupon, I focused on the
exigent classical postprocessing and incorporated it to the
quantum circuit. Later, the validity of the obtained IPEA
was verified using the adaptive measurement in
the Fourier basis \cite{SemiQFT96} (semiclassical QFT).

The QFT-based PEA was implemented experimentally a few 
times \cite{Lee-PEA-implement02, Zhang-clock-synchro-impl04}
on three qubits and two-bit phase estimates were obtained.
Using a benchmark circuit of similar complexity, I showed
that a much higher precision can be achieved using the IPEA
instead.

In the Section \ref{sec:qft-based-pea-versus-ipea},
differences between the QFT-based PEA and the IPEA were
discussed. Using the IPEA, repetitions needed due to the
environmental noise and/or a high success probability require
less resources compared to the QFT-based PEA. Next, in the
context of the Abrams-Lloyd algorithm, the IPEA allows to exchange
a long coherence time of an input eigenstate for a per 
iteration input state preparation. On the other hand, the
QFT-based PEA enables to employ entangled ancilla input states
which can be optimized with respect to a given cost function.

The question of yielding optimal estimates with respect to
the fidelity cost function without using entanglement was
discussed a lot in multiround protocols for reference frame
alignment \cite{RG03, BB05}. It was concluded that presented
protocols do not produce such optimal estimates. An existence
of better protocols was left as an open problem. It should be
clear from this thesis that the proposed protocols could not
produce desired optimal estimates, since they use the Kitaev
circuit which has a small logarithmic penalty compared to the
IPEA. Moreover, an exact simulation of information flow in the
QFT-based PEA cannot be achieved using small fixed-size ancilla
registers present in iterative schemes. Now, an open problem is
to which degree entangled ancilla input states can be approximated
in the IPEA.







 \chapter{Quantum cryptography}
 \label{chap:quantum-cryptography}
 \begin{center}\end{center}

Special properties of quantum physical systems such as
the no-cloning property of an unknown quantum state and the 
randomness of a wave function collapse allow for many 
quantum enhanced cryptographic primitives. These primitives
range from authentication, secret key 
generation/distribution to covered communication
(steganography) and more. Up to date, the BB84 protocol
\cite{BB84} and its variants \cite{B92,Ekert91} for quantum
key generation are the most celebrated quantum cryptography
primitives.

For myself, I am particularly interested in quantum
steganography and its applications related to watermarking
protocols for authentication. Similarly to the classical
steganography, hidden communication is achieved via
embedding a message into a redundant part of a (innocent
looking) cover medium. The main  difficulties for a
steganographer are to identify the redundant part of the
cover medium and insert a message in an unobtrusive way.
For watermarking, it is additionally desirable for the
message to be tied with the cover medium so well that the
message cannot be removed by a third party without damaging
the cover medium substantially.

Embedding methods may differ to a large extent in the medium
access level. The main levels are (1) physical carrier
level,
(2) error correction code level, and (3) data formats and 
protocol level. Steganography at the physical carrier level
is somewhat specific in the sense that the theory of
information coding does not play a big role, but instead it
is a clear race in the technology available to the
steganographer and steganalyst. To put it more to the
context, let us take a look at the first commercially
available products for quantum key generation 
\cite{Cerberis,Magiq}.
For these devices to work as              
securely as proposed by the theory, single photon emitters
and detectors are crucial. Nevertheless, single
photon sources are often substituted with an industry-standard
weak coherent pulse approach. Whilst practical,
this approach suffers from non-zero probability of
multiple-photons events. Apart from the introduced security
loophole \cite{Brassard00,QC-trojan02},
these extra photons can be, in priciple, used for
steganographic purposes while legitimate quantum key
generation seems to happen.

Regarding quantum steganography on the error correction
level, first steps along this line were accomplished
by J. Gea-Banacloche in \cite{Banacloche02}. It was proposed
to encode
the message to be hidden into the error syndrome. Compared
to the classical counterpart, quantum error correction codes
(QECC) are very suitable for steganographic purposes. This
is due to the fact that quantum errors are continuous and
quantum gates are very faulty, and therefore QECC schemes
require a rather big bunch of extra physical qubits to
create and preserve logical qubits. Especially, at higher
levels of concatenated QECC \cite{ConcatenatedQECC96}, it
is plausible to
assume that errors occur not too often, and the capacity of
the QECC scheme can serve as a steganographic channel.

\begin{figure}[h!]
 \centering
 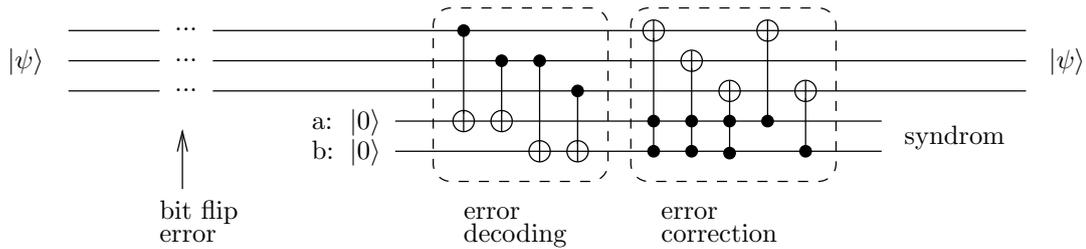
 \caption[A simple scheme for error correction]
         {A circuit to decode the error syndrome of the 
          3-qubit repetition code and correct the error
	  accordingly. Note that the correction part does
	  not require measurement on the ancilla qubits
	  $a$ and $b$.}
 \label{steganography-basic-scheme}
\end{figure}

An example of simple QECC scheme is shown in the 
Figure~\ref{steganography-basic-scheme}. Gea-Banacloche used
the same scheme to present his ideas.

The state 
$$
 \ket{\psi}
 = \alpha\ket{000} + \beta\ket{111}
 = \alpha\ket{0_L} + \beta\ket{1_L}
$$
represents a logical qubit
protected against a single bit flip error on any of the
three physical qubits, using the 3-qubit repetition code 
\cite{Shor-repetition-code95, Nielsen-Chuang00}.
After the decoding \& correction stage, the state of the
ancilla qubits $a$ and $b$ holds the information on what
had happened\\
\begin{tabular}{lll}
 \hspace{2cm} & &    \\[-3mm]
 & $\ket{00}_{ab}$ :
 & no error occurred, \\
 & $\ket{10}_{ab}$ :
 & bit flip error $X^{(1)}$ on the first physical qubit, \\
 & $\ket{11}_{ab}$ :
 & bit flip error $X^{(2)}$ on the second physical qubit,\\
 & $\ket{01}_{ab}$ :
 & bit flip error $X^{(3)}$ on the third physical qubit.
\end{tabular}
 
\begin{figure}[h!]
 \centering
 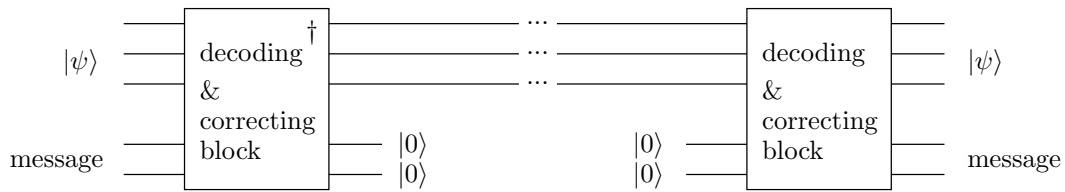
 \caption[Quantum steganography on the error correction
          level]
         {Quantum steganography on the error correction
	  level.}
 \label{steganography-inserting}
\end{figure}

Now, assuming that no accidental error will happen, it may
be the steganographer who deliberately introduces an error
and herewith transfers two bits of information. A convenient
method for introducing an error (inserting a message) is to
use the decoding \& correction block in a reversed mode, see
Figure~\ref{steganography-inserting}.

This steganographic channel can be also used for a simple
authentication. Alice and Bob a priori agree on a special
message (a tag). Then later, when Alice sends a state 
$\ket{\psi}$ to Bob, Bob considers the received state as
authentic only if the embedded message corresponds to the
 agreed tag. Additionally, it is required that a
third party (with an access to a transmission channel)
should not be able to change the state $\ket{\psi}$ without
affecting the tag as well. It was shown that this
requirement can be fulfilled only if the tag is a quantum
superposition of the form
$$
    \iota\ket{00}_{ab}
  + \kappa\ket{01}_{ab}
  + \mu\ket{10}_{ab}
  + \nu\ket{11}_{ab}
$$
and not a classical digital tag, e.g. '10'. Effectively, the 
operation (decoding \& correction block)$^\dagger$ encodes
the coefficients $\iota, \kappa, \mu, \nu$ together with the
coefficients $\alpha, \beta$ in such a way that the
transmitted state $\ket{\psi}$ becomes encrypted and it is
impossible for a third party to modify $\alpha, \beta$
without affecting the tag. In fact, the tag becomes a
watermark.

These results indicate that quantum authentication implies
encryption and that it is impossible to use digital 
signing to authenticate quantum states. Indeed, this was 
proven by Barnum {\it et. al.} \cite{Barnum02} in a more general 
study on authentication of quantum states. A limit case
study addressing authentication of a single qubit using a 
shared quantum key of minimum length has been conducted by
Curty {\it et. al.} \cite{Qubit-auth02}. Particular details
of this study
are discussed in the Section~\ref{sec:qubit-authentication}.

Finally, we discuss quantum steganography at the level of
data formats and protocols. In digital data formats, tens
of methods are known that exploit redundancy in JPEG
images \cite{JPEG-stego06}, MP3 music files \cite{MP3Stego06},
binary executable
files \cite{Hydan04} and so on. However, there are no such
formats
in the quantum domain by this time. It is too early to predict
how and when will the technology reach this stage. 
Nevertheless, some general methods such as modifying the
least significant bits can be ported to the quantum domain
quite easily \cite{Curty-conf-stego00}.

Let us assume that a classical information, e.g. bits 
representing a picture, is going to be transmitted using
qubits (a quantum channel). Each bit is mapped in some 
standard way to a qubit state. Additionally, since the
carried information is a classical information and the
mapping is publicly known, anybody with an access to the
quantum channel can make copies of the qubits and check
what is being transmitted.

Under this scenario, Alice and Bob want to establish a
steganographic channel. It is enough for them to merely
agree on a non-standard base encoding and measurement on
the qubits which corresponds to the least significant
bits. Hereby they set up a secret channel. Other parties
who are not aware of this deal treat all qubits in a
standard way. As a consequence, these parties may obtain
wrong bit values upon measurement, but due to the low
significance of corresponding bits it can hardly be
detected. Additionally, once the qubit carrying hidden
information is collapsed by a measurement, no later 
leakage of the stego key (the deal of Alice and Bob)
enables to obtain the past hidden information. There is
no classical analogue of this pretty striking property.
Of course, the difficult part for the steganographer
is to properly identify which bits are the least 
significant ones.
A variant of the above described method has been
successfully used by K. Martin \cite{Martin07} to
establish a
steganographic channel within the BB84 protocol. Arbitrary
third party sees only that a legitimate quantum key 
generation takes place.

Regarding quantum steganography with entanglement, it is
possible to build on schemes which are based on the
superdense coding \cite{Bennett-dense-coding92}. Basically,
Alice and Bob start with a shared entangled pair of qubits.
Then, (1) Alice applies a local unitary operation to her
qubit, (2) Alice sends the qubit to Bob, (3) Bob performs
joint measurement on both qubits, and (4) the result 
represents a delivered two-bit message. A third party
having an access to the sent off qubit while in traffic
cannot get even a single bit of that message.

\section{Qubit authentication with a quantum key of minimum
         length}
\label{sec:qubit-authentication}

Curty {\it et. al.} \cite{Qubit-auth02} studied the question
whether a
single qubit can be authenticated with a quantum key of
minimum length. It was shown that, unlike in the case of
quantum-authenticated classical bit value 
\cite{Curty-classical-auth01}, this is
not possible. Nevertheless, it is interesting to see how the
protocol works since all what is ultimately needed  to make
the protocol operating securely is to increase the length of
the key \cite{Perez-qubit-auth03}.

The protocol \cite{Qubit-auth02} is as follows. Alice wants
to send a 
general quantum state (possibly a mixed state) described by
the density operator $\rho_\mathsf{M}$ acting on a 
two-dimensional message space $\mathsf{M}$. As in a 
classical authentication, a tag must be attached to the
message, in order to allow to Bob to convince himself about
the authenticity of the message. Let the tag be given by a
density operator $\rho_\mathsf{T}$ acting on a two 
dimensional tag space $\mathsf{T}$. In principle, the tag
can be larger, however, this is advantageous only in a
presence of noise. No extra security is added. The tag
space $\mathsf{T}$ has to be divided into two orthogonal
subspaces. One subspace represents a valid tag, while the
other represents an invalid tag. Without loss of generality,
the state $\rho_{\mathsf{T}}=\ketbra{0}{0}$ can be fixed as
a valid tag.

The space of tagged messages is then defined as 
$\epsilon=\mathsf{M}\otimes\mathsf{T}$ and the tagged 
message is given by $\rho_\epsilon = \rho_\mathsf{M} \otimes
\rho_\mathsf{T}$. On the space $\epsilon$ a unitary coding
set $\{I_\epsilon, U_\epsilon\}$ is defined, where 
$I_\epsilon$ is the identity operation and $U_\epsilon$
a unitary transformation. A shared quantum key has the form
of a maximally entangled EPR pair, e.g. 
$\ket{\psi}_{ab} = \frac{1}{\sqrt{2}}(\ket{01} - \ket{10})$.
Alice and Bob each own one qubit of that EPR pair. The key
can be publicly known as well as the coding set. The state
of the global system (key + tagged message) is given by
\begin{equation}
 \rho_{ab\epsilon}
 \;=\; \ketbra{\psi}{\psi}_{ab} \otimes \rho_\epsilon
 \;=\; \ketbra{\psi}{\psi}_{ab} \otimes
       \rho_\mathsf{M}          \otimes
       \rho_\mathsf{T} \,.
\end{equation}

\begin{figure}[h!]
 \centering
 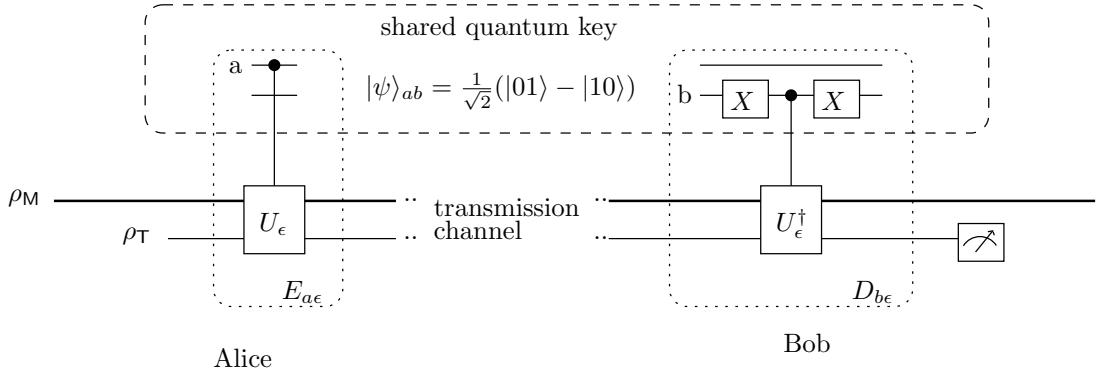
 \caption[A protocol for qubit authentication]
         {A protocol for qubit authentication with a quantum
          key of minimum length.}
 \label{qubit-authentication}
\end{figure}

{\it Encoding stage.} Alice wants to send an authenticated
message to Bob. In order to do so, Alice performs an
encoding operation $E_{a\epsilon}$ which acts on her part of
the key and the tagged message. With respect to the coding
set $\{I_\epsilon, U_\epsilon\}$, the encoding operation is 
defined as
\begin{equation}
 E_{a\epsilon}
 = \ketbra{0}{0}_a \otimes I_b \otimes I_\epsilon
   \;\;+\;\;
   \ketbra{1}{1}_a \otimes I_b \otimes U_\epsilon \,.
\end{equation}
Once the encoding operation is applied, Alice sends the
encoded tagged message to Bob. The state of the global
system after the encoding operation is given by
\begin{align}
 \rho_{ab\epsilon}^{enc}
 &= E_{a\epsilon}     \,
    \rho_{ab\epsilon} \,
    E_{a\epsilon}^\dagger \\
 &= \frac{1}{2}\left(
    \ketbra{01}{01} \otimes \rho_\epsilon
    \;\;-\;\;
    \ketbra{01}{10} \otimes \rho_\epsilon U_\epsilon^\dagger
    \;\;-\;\;
    \ketbra{10}{01} \otimes U_\epsilon \rho_\epsilon
    \;\;+\;\;
    \ketbra{10}{10} \otimes
                    U_\epsilon
		    \rho_\epsilon
		    U_\epsilon^\dagger
    \right) \,.
    \nonumber
\end{align}
{\it Decoding stage.} Bob receives the encoded tagged
message and decodes it using a decoding operation
\begin{equation}
 D_{b\epsilon}
 = I_a \otimes \ketbra{0}{0}_b \otimes U_\epsilon^\dagger
   \;\;+\;\;
   I_a \otimes \ketbra{1}{1}_b \otimes I_\epsilon \,.
\end{equation}
The state of the global system after the decoding
is described as
\begin{align}
 \rho_{ab\epsilon}^{dec}
 &= D_{b\epsilon}           \,
    \rho_{ab\epsilon}^{enc} \,
    D_{b\epsilon}^\dagger \\
 &= \frac{1}{2}\left(
    \ketbra{01}{01} \otimes 
                    \rho_\epsilon
    -
    \ketbra{01}{10} \otimes 
                   (\rho_\epsilon
		    U_\epsilon^\dagger)
		    U_\epsilon
    -
    \ketbra{10}{01} \otimes
                    U_\epsilon^\dagger
		   (U_\epsilon
		    \rho_\epsilon)
    +
    \ketbra{10}{10} \otimes
                    U_\epsilon^\dagger
		   (U_\epsilon
		    \rho_\epsilon
		    U_\epsilon^\dagger)
		   U_\epsilon
    \nonumber
    \right) \\
 &= \,\ketbra{\psi}{\psi}_{ab} \otimes \rho_\epsilon \,.
    \nonumber
\end{align}
{\it Verification.} Bob takes decoded tagged message, which
is formally given by 
$\rho_\epsilon^{dec}=\tr_{ab}(\rho_{ab\epsilon}^{dec})$, and
measures the tag portion. If the outcome belongs to a valid
tag subspace of space $\mathsf{T}$, then the extracted 
message is considered to be authentic. Using the gate model,
the protocol can be depicted as shown in the 
Figure~\ref{qubit-authentication}.

\subsubsection*{Message attack}
Let us now consider a 'message attack' performed by an 
adversary party called Eve. This party with a full access
to a public transmission channel sees a mixed state
\begin{align}
 \label{mixed-state-eve-sees}
 \rho_\epsilon^{E}
 &= \tr_{ab}(\rho_{ab\epsilon}^{enc}) \\
 &= \frac{1}{2}\left(
    \rho_\mathsf{M}  \otimes \ketbra{0}{0}_\mathsf{T}
    \;\; + \;\;
    U_\epsilon  \,
    (\rho_\mathsf{M} \otimes \ketbra{0}{0}_\mathsf{T}) \,
    U_\epsilon^\dagger
    \right) \,.
    \nonumber
\end{align}
Her task is to find a transformation $Q_E$ which, applied
to $\rho_\epsilon^{E}$, will modify the $\rho_\mathsf{M}$
keeping the tag portion intact. The authors of
\cite{Qubit-auth02} gave
an existential proof that such a transformation always 
exists regardless of the choice of $U_\epsilon$, thus the
protocol is not secure. However, the form of $Q_E$ with 
respect to $U_\epsilon$ and the range possible damages was
not discussed. In a later paper \cite{Perez-qubit-auth03},
it was shown that
if the key is longer, then there exists a larger coding set
such that Eve's desired $Q_E$ does not exists.

An interesting point is what exactly happens when the
operation $U_\epsilon$ is a separable gate. Classically,
the tag must be a function of both the key and the message. 
Now, since the mixed state \eqref{mixed-state-eve-sees} 
is achievable even if Alice and Bob had shared only a 
classical one bit key, this indicates that $U_\epsilon$
should be an entangling gate to limit Eve's actions, at
least. However, the shared key is a quantum one and the
question is whether entanglement present in the key can
compensate for non-entangling operation $U_\epsilon$.

Let $U_\epsilon$ be a separable gate of the form
$U_\epsilon = U_\mathsf{M} \otimes U_\mathsf{T}$, then
\begin{equation}
 \rho_\epsilon^{E}
 = \frac{1}{2}\left(
   \rho_\mathsf{M} \otimes \ketbra{0}{0}_\mathsf{T}
   \;\; + \;\;
   (U_\mathsf{M} \rho_\mathsf{M} U_\mathsf{M}^\dagger)
   \otimes
   (U_\mathsf{T}
    \ketbra{0}{0}_\mathsf{T}
    U_\mathsf{T}^\dagger)
   \right) \,.
\end{equation}
For $Q_E = R \otimes I$, where $Q_E \in \epsilon$, $R \in
\mathsf{M}$ and $I \in \mathsf{T}$, we have
\begin{align}
 \rho_\epsilon^{Q_E}
 &= Q_E \rho_\epsilon^E Q_E^\dagger \\
 &= \frac{1}{2}\left(
    R \, \rho_\mathsf{M} \, R^\dagger
    \otimes
    \ketbra{0}{0}_\mathsf{T}
    \;\; + \;\;
    \left(R\,(U_\mathsf{M}
          \,\rho_\mathsf{M}
	  \,U_\mathsf{M}^\dagger)
          \,R^\dagger\right)
    \otimes
    \left(U_\mathsf{T}
          \,\ketbra{0}{0}_\mathsf{T}
	  \,U_\mathsf{T}^\dagger\right)
    \right) \,.
    \nonumber
\end{align}
And after Bob's decoding operation we get
\begin{equation}
 \rho_\epsilon^{Q_E, dec}
 = \frac{1}{2}\left(
   R \, \rho_\mathsf{M} \, R^\dagger
   \otimes
   \ketbra{0}{0}_\mathsf{T}
   \;\; + \;\;
   \left(U_\mathsf{M}^\dagger
   \, R \, U_\mathsf{M} \, \rho_\mathsf{M}
   \, U_\mathsf{M}^\dagger \,
   R^\dagger \, U_\mathsf{M} \right)
   \otimes
   \left(U_\mathsf{T}^\dagger\,U_\mathsf{T}\,
   \ketbra{0}{0}_\mathsf{T}\,
   U_\mathsf{T}^\dagger\,U_\mathsf{T}\right)
   \right)
\end{equation}
and $\tr_\mathsf{M}\left( \rho_\epsilon^{Q_E, dec} \right)
= \ketbra{0}{0}_\mathsf{T}$.

Hence the adversary party is always able to change the 
message $\rho_\mathsf{M}$ keeping the tag portion intact,
and there are {\bf no constraints} on the operation $R$.
Having some statistics of usually sent messages, $R$ can
be even optimized to cause maximal damage. This result
also implies that there is no added security in the
protocol by using quantum key instead of a classical key.

\subsubsection*{Secret-key discussion}
Authenticating quantum data makes sense only in a scenario
where a reliable technology for quantum information 
processing is available. This means that there are no extra
costs in using quantum keys instead of classical ones. 
Moreover, quantum keys have better key-management properties
thanks to the no-cloning theorem, and, strikingly, in the
above described protocol the EPR pair does not collapse
so it can be in principle reused. 

But anyway, reliable storing of an EPR pair requires a 
periodic entanglement purification \cite{Entanglement-purf-b}.
In between two
successive refreshments the EPR pair should be inevitably
considered as corrupted to some degree. A well designed
protocol has to be insensitive to a small corruption 
(entanglement purification guarantees to output only almost
pure states), and a non-negligible corruption should
result in a high probability of rejecting received
messages as authentic ones. 

Let us correlate the corruption of the EPR pair with the
quantity $p$ of a maximally mixed state in the mixture. The 
state of the quantum key is then described by a density
operator
\begin{equation}
 \rho_{\psi}^p
 = (1-p)\ketbra{\psi}{\psi}_{ab}
   \;\; + \;\;
   p\,\frac{I_{ab}}{4} \,.
\end{equation}
When the protocol is executed with this mixed key the
resulting global state of the system is described as
\begin{equation}
 \rho_{ab\epsilon}^{dec}
 = D_{b\epsilon} \left( E_{a\epsilon} \left(
   \rho_{\psi}^p
   \otimes \rho_\mathsf{M}
   \otimes \ketbra{0}{0}_\mathsf{T}
   \right) E_{a\epsilon}^\dagger \right) 
   D_{b\epsilon}^\dagger \,,
\end{equation}
and the decoded message after tracing out the key is 
\begin{align}
 \rho_\epsilon^{dec}
 &= \tr_{\rho_{\psi}^p}
    \left( \rho_{ab\epsilon}^{dec} \right) \\
 &= \frac{2-p}{p}
    \left( \rho_\mathsf{M}
           \otimes
	   \ketbra{0}{0}_\mathsf{T} \right)
    \;\; + \;\;
    \frac{p}{4}
    \left( U_\epsilon^\dagger 
           \left(
	           \rho_\mathsf{M}
		   \otimes
		   \ketbra{0}{0}_\mathsf{T}
	   \right) 
           U_\epsilon
           \; + \;
           U_\epsilon 
	   \left( 
	           \rho_\mathsf{M}
		   \otimes
		   \ketbra{0}{0}_\mathsf{T}
           \right)
	   U_\epsilon^\dagger
    \right) \,.
    \nonumber
\end{align}
Here, we can see that the probability of 
$\rho_\epsilon^{dec}$ passing Bob's verification procedure
is unpleasantly high even for apparently corrupted key. 
Therefore, in this protocol and its variants a classical key
should be preferred.







 \chapter{Conclusions}
 \label{chap:conclusions}
 \begin{center}\end{center}

In this thesis, I have addressed the emerging field of 
quantum computing. Since at the time of starting my PhD
studies I was a complete newcomer to the field, a 
relatively large part of the thesis deals with portions of 
linear algebra essential to understanding the framework
of quantum mechanics, and with comparison of deterministic,
probabilistic and quantum Turing machines in order to 
emphasize the fundamental differences. The quantum gate
model and the thin borderline between quantum evolutionary
universal sets of gates and not even classically universal
sets of gates are discussed as well. At this point, the 
thesis hopefully provides a self contained description of
the main issues related to quantum computing.

The major part of the thesis concerns the quantum phase 
estimation algorithm. This algorithm presents a fundamental
approach for problem solving on a quantum 
computer and it can be used to well explain most of known
quantum algorithms which yield an exponential speed-up
compared to their classical counterparts. The two most
prominent quantum algorithms are the Shor factoring algorithm
and the
Abrams-Lloyd algorithm for finding energy eigenvalues. In
Section~\ref{sec:pea-representative-instances}, it was
shown how to cast both algorithms as a phase estimation
problem and it was stressed why both algorithms are efficient
with respect to the input size of the problem.

The novel work I did encompasses a new simple proof of the
lower bound of the PEA success probability, a derivation of
an iterative scheme for quantum phase estimation (IPEA) from
the Kitaev phase estimation, a study of robustness of the
IPEA utilized as a few-qubit testbed application and an
improved protocol for phase reference alignment. There is
also a chapter concerning quantum steganography and 
authentication which provides additional insights into
quantum enhanced protocols, however, no substantial
contribution in this field is presented.

The design and study of robustness of small testbed
applications currently represents one of the chief goals in
the quantum computing field. The famous theorem for
fault-tolerant quantum computation uses a very general
noise model and it is not clear at all whether the theorem
actually holds in an experimental setting. Testbed 
applications provide an important way how to realize the
potential of many-body quantum systems as quantum computers
and give understanding to the effects of environmental noise.
It might happen that our universe is not generous enough to
allow for robust large-scale quantum computing, however,
quantum information processing plays an eminent role in
forthcoming quantum sensors and similar nanoscale single
purpose devices. Think of the energy transfer in plants
during photosynthesis \cite{Photosynthesis07}.

Regarding future work, it is planned to work out a complete
example of a quantum chemistry calculation (the Abrams-Lloyd
algorithm). That is to choose a small quantum system such as
the Helium atom, map a corresponding truncated Hamiltonian
to the states of two to three qubits, simulate the evolution, 
adiabatically prepare an approximate ground state and 
utilize the IPEA to estimate the ground state energy.

\newpage
\markboth{BIBLIOGRAPHY}{}

\bibliography{phd}
\bibliographystyle{unsrt}

\end{document}

%
%
%
%
%